\documentclass[preprint,aps]{revtex4}
\usepackage{graphicx}
\usepackage{dcolumn}
\usepackage{bm}

\begin{document}
\preprint{T02/178}
\title{The  Gross-Neveu model at finite temperature \\ at next to leading order in the $1/N$
expansion}

\author{Jean-Paul Blaizot}
  \email{blaizot@spht.saclay.cea.fr}
  \affiliation{Service de Physique Th\'eorique, CEA/DSM/SPhT,
  91191 Gif-sur-Yvette Cedex, France.}

\author{Ram\'on M\'endez Galain}
  \email{mendezg@fing.edu.uy}
  \affiliation{Instituto de F\'{\i}sica, Facultad de Ingenier\'{\i}a, J.H.y Reissig 565, 11000 
  Montevideo, Uruguay}
  
\author{Nicol\'as Wschebor}
  \email{wschebor@spht.saclay.cea.fr}
  \affiliation{Service de Physique Th\'eorique, CEA/DSM/SPhT,
  91191 Gif-sur-Yvette Cedex, France\\ and \\
  Instituto de F\'{\i}sica, Facultad de Ingenier\'{\i}a, J.H.y Reissig 565, 11000 
  Montevideo, Uruguay}  

  \date{\today}
\begin{abstract}

We present new results on the Gross-Neveu model at finite temperature and at next-to-leading
order in the $1/N$ expansion. In particular, a new expression is obtained for the
effective potential which is explicitly invariant under renormalization group transformations.
The model is used as a playground to investigate various features of field theory at finite
temperature. For example we 
verify that, as expected from general arguments, the cancellation of ultraviolet divergences
takes place at finite temperature without the need
for introducing counterterms beyond those of  zero-temperature. As 
well known, the discrete chiral symmetry of the 1+1 dimensional model is spontaneously
broken at zero temperature and restored, in leading order,  at some temperature
$T_c$; we find that the $1/N$ approximation breaks down for temperatures  below $T_c$: As the temperature
increases, the fluctuations become eventually too large to be treated as corrections, and
a Landau pole invalidates the calculation 
of the effective potential in the vicinity of its minimum.   Beyond
$T_c$, the $1/N$ expansion becomes again regular: it predicts that in leading order
the system behaves as a free gas of massless fermions and that, at the next-to-leading order,
it remains weakly interacting. In the limit of large temperature, the pressure coincides with
that given by perturbation
theory with a coupling constant defined at a scale of the order of the temperature,
as expected from asymptotic freedom.


\end{abstract}

\pacs{Valid PACS appear here}
\maketitle

\newcommand \beq{\begin{eqnarray}}
\newcommand \eeq{\end{eqnarray}}
\newcommand{\ba}{\begin{eqnarray}}
\newcommand{\ea}{\end{eqnarray}}

\input epsf


\def\square{\hbox{{$\sqcup$}\llap{$\sqcap$}}}
\def\grad{\nabla}   
\def\del{\partial}  

\def\frac#1#2{{#1 \over #2}}
\def\smallfrac#1#2{{\scriptstyle {#1 \over #2}}}
\def\half{\ifinner {\scriptstyle {1 \over 2}}
   \else {1 \over 2} \fi}

 
\def\bra#1{\langle#1\vert} 
\def\ket#1{\vert#1\rangle}


\def\simge{\mathrel{%
   \rlap{\raise 0.511ex \hbox{$>$}}{\lower 0.511ex \hbox{$\sim$}}}}
\def\simle{\mathrel{
   \rlap{\raise 0.511ex \hbox{$<$}}{\lower 0.511ex \hbox{$\sim$}}}}


\def\buildchar#1#2#3{{\null\!               
   \mathop#1\limits^{#2}_{#3}               
   \!\null}}                                
\def\overcirc#1{\buildchar{#1}{\circ}{}}


\def\slashchar#1{\setbox0=\hbox{$#1$}          
   \dimen0=\wd0                             
   \setbox1=\hbox{/} \dimen1=\wd1              
   \ifdim\dimen0>\dimen1                       
      \rlap{\hbox to \dimen0{\hfil/\hfil}}     
      #1                                       
   \else                                       
      \rlap{\hbox to \dimen1{\hfil$#1$\hfil}}  
      /                                        
   \fi}


\def\real{\mathop{\rm Re}\nolimits}     
\def\imag{\mathop{\rm Im}\nolimits}     

\def\tr{\mathop{\rm tr}\nolimits}       
\def\Tr{\mathop{\rm Tr}\nolimits}       
\def\Det{\mathop{\rm Det}\nolimits}     

\def\mod{\mathop{\rm mod}\nolimits}     
\def\wrt{\mathop{\rm wrt}\nolimits}     


\def\TeV{{\rm TeV}}                     
\def\GeV{{\rm GeV}}                     
\def\MeV{{\rm MeV}}                     
\def\KeV{{\rm KeV}}                     
\def\eV{{\rm eV}}                       

\def\mb{{\rm mb}}                       
\def\mub{\hbox{$\mu$b}}                 
\def\nb{{\rm nb}}                       
\def\pb{{\rm pb}}                       

%
%

\def\picture #1 by #2 (#3){
  \vbox to #2{
    \hrule width #1 height 0pt depth 0pt
    \vfill
    \special{picture #3} 
    }
  }

\def\scaledpicture #1 by #2 (#3 scaled #4){{
  \dimen0=#1 \dimen1=#2
  \divide\dimen0 by 1000 \multiply\dimen0 by #4
  \divide\dimen1 by 1000 \multiply\dimen1 by #4
  \picture \dimen0 by \dimen1 (#3 scaled #4)}
  }

\def\centerpicture #1 by #2 (#3 scaled #4){
   \dimen0=#1 \dimen1=#2                                                
    \divide\dimen0 by 1000 \multiply\dimen0 by #4        
    \divide\dimen1 by 1000 \multiply\dimen1 by #4
         \noindent                                                                    
         \vbox{                                                                      
            \hspace*{\fill}                                                  
            \picture \dimen0 by \dimen1 (#3 scaled #4)      
            \hspace*{\fill}                                                       
            \vfill}}

\def\figun{\scaledpicture 40.57mm by 37.75mm
 (fermion_det scaled 800)}

\def\figunter{\scaledpicture 88.55mm by 32.81mm
 (ring scaled 800)}

\def\figunbis{\scaledpicture 143.5mm by 32.81mm
 (figure1 scaled 800)}

\def\figdeux{\scaledpicture 165.4mm by 29.28mm
 (figure2 scaled 750)}

\def\figcondensat{\centerpicture 125.2mm by 43.74mm
 (condensat scaled 750)}

\def\figfermass{\centerpicture 122.4mm by 32.46mm
 (fermass scaled 750)}

\def\figcinq{\centerpicture 118.8mm by 100.5mm
 (veff scaled 750)}

\def\figfunctionF{\centerpicture 117.4mm by 75.14mm
 (functionF scaled 750)}

\def\figfunctionfK{\centerpicture 118.1mm by 76.20mm
 (functionfK scaled 750)}

\def\figmasscorr{\centerpicture 117.8mm by 75.85mm
 (masscorr scaled 750)}

\def\figmodeg{\centerpicture 116.7mm by 74.79mm
 (modeg scaled 750)}

\def\figrhodez{\centerpicture 99.48mm by 61.38mm
 (rhodez scaled 800)}
\def\figphaseshift{\centerpicture 99.48mm by 61.38mm
 (phaseshift scaled 800)}

\def\fighuit{\scaledpicture 151mm by 167mm
 (new2 scaled 500)}

\def\figneuf{\scaledpicture 152mm by 165mm
(new4 scaled 450)}
\def\figdix{\scaledpicture 151mm by 91.7mm
(new5 scaled 500)}
%
\section{\label{sec:intro}Introduction}

This paper presents a detailed and pedagogical study of the Gross-Neveu model \cite{Gross:jv} at
finite  temperature at next-to-leading order in the $1/N$ expansion. As well known, the model
describes a system of interacting fermions in one spatial dimension. It is
renormalizable and asymptotically free, and, in the vacuum, exhibits chiral
symmetry breaking.  These properties, chiral symmetry breaking and asymptotic freedom,
mimic those of Quantum Chromodynamics (QCD) and the Gross-Neveu model constitutes an ideal
playground to study these questions in  a much simpler context than that of non abelian
gauge theories in four dimensions. Because of this connection, we shall often use the
language of QCD in this paper and refer to the fermion as a ``quark'',
and to the $N$ fermion species as ``flavors''.

Aside from being  perturbatively renormalizable, the model is also renormalizable in the $1/N$
expansion  for any dimension smaller than four \cite{Gross:vu,Rosenstein,Parisi,Shizuya,Zinn-Justin:yn}.
It is one of the main motivations
of the present work to study the details of the renormalization in a scheme which is not restricted
to perturbation theory. In particular, it is expected on general grounds \cite{Collins84,Landsman:uw,LeBellac96}
that  the short distance singularities leading to ultraviolet divergences are not
affected by the temperature. That is, infinities which occur in finite temperature calculations can
all be removed by the zero temperature counterterms. It is interesting to see explicitly, on a non
trivial example, how these cancellations of infinities take place.

Because the calculations are simpler in one dimension we restrict ourselves here to this situation. But, our
main interest is not  the one-dimensional physics and we leave aside aspects of the model which
are specific to one dimension, at the cost of obscuring perhaps 
the physical interpretation of some of our results. In particular, the 
$1/N$ expansion is built on field configurations which make the action an extremum, and in this paper we consider only
static, uniform, such configurations.  However the action is extremal also for configurations which are not uniform, namely
configurations commonly called ``kinks'', which interpolate  between the two degenerate minima of the leading order
effective potential.
The role of these kinks at finite temperature has been qualitatively discussed for
the Gross-Neveu model in
Ref.~\cite{Dashen:xz}. The kinks have energies typically of order
$\varepsilon\sim NM_f$ where
$M_f$ is the fermion mass (arising from symmetry breaking), and their number  is 
$\sim L\,{\rm e}^{-\varepsilon/T}$, where $L$ is the length of the system. Their role
depends then of the order of the two limits $N\to \infty$, $L\to\infty$. If one takes the
limit $N\to \infty$ first, then the kinks play no role since their number is exponentially small: this is the
situation described by the mean field approximation, or the leading order in the $1/N$ expansion, and where symmetry
breaking occurs. When going to the next-to-leading order in the
$1/N$ expansion, one has to take the thermodynamic limit $L\to\infty$ at fixed
$N$; in this case the entropy associated with the positions of the kinks  eventually overcome the cost in energy for 
producing them. These configurations are then expected to dominate, preventing symmetry breaking  at any non-zero
temperature, in agreement with general results about infinite systems in one-dimension
\cite{LL58}. Note also that kinks play a major role in the exact solution of the model found recently
\cite{Fendley:2001mw}.

In our calculations, kinks are ignored. Then, in  the leading order of the $1/N$ expansion,
chiral symmetry is restored at some finite temperature $T_c$. But the calculation of the corrections of order $1/N$
reveals that the expansion breaks down as one approaches $T_c$: The fluctuations become eventually too large to be treated as
corrections, which may be related to the existence of a Landau pole invalidating the calculation of the effective potential for small values
of the field. It is unclear to us whether such difficulties with the $1/N$ expansion are consequences of  the one
dimensional character of the system, and in particular whether they could be cured for instance by taking kinks into account. Since
treating kinks explicitly would represent a major effort beyond the scope of this paper, we leave this question open
and focus  on another interesting regime, that of high temperature. In this regime, because of asymptotic freedom, one
expects the system to become a weakly interacting gas of fermions, somewhat similar to the quark-gluon plasma of QCD.
As we shall see,  verifying explicitly how such properties emerge in the $1/N$ expansion is instructive.

There has been many studies of the Gross-Neveu model,
or of the related  Nambu Jona-Lasinio model \cite{Nambu:fr}, at finite temperature. Most of these studies however
concentrated on the mean field physics and chiral symmetry breaking
\cite{Jacobs:ys,Harrington:tf,Dittrich:nq,Wolff:av,Barducci:cb,Bernard:ir,Hatsuda:1986gu,Asakawa:bq}. More recently, 
$1/N$ corrections where calculated in a version of the Gross-Neveu model with continuous symmetry, focusing on the dominant role
of the soft pion modes at low temperature \cite{Barducci:1996jv,Modugno:qm}.  Investigations of the effect  of  the
$1/N$ corrections in the three-dimensional  Nambu Jona-Lasionio
model were also  presented in Refs.~\cite{Hufner:ma,Florkowski:1996wf}. However, none of these studies
provide systematic answers to  the set of questions that we address in the present work.

The outline of this paper is as follows. The next section is a general introduction to the model; we recall the
construction of the effective potential at finite temperature and of its $1/N$ expansion. In
Sect.~\ref{sec:renorpotential}, we calculate explicitly the zero temperature effective potential at next-to-leading
order in the $1/N$ expansion and carry out completely its renormalization.  Although there exists many calculations
of the effective potential at zero temperature \cite{Root:1974zr,Root:1974cs,Schonfeld:us,Haymaker:1978kp}, the 
expressions that we obtain are new and exhibit explicit 
invariance  under renormalization group
transformations. In particular, we include a correction of order $1/N$ to the fermion mass which has been left out in 
some previous analysis, but which is needed to ensure the explicit renormalization group invariance of the effective
potential. In Sect.~\ref{sec:renoreffpot}, we extend the calculation of the effective potential to finite temperature.
We analyze in particular the cancellation of ultraviolet divergences, and verify that indeed these cancellations
take place without the need for  counterterms other than the zero temperature ones. In
Sect.~\ref{sec:thermo} we present a physical discussion of the thermodynamical properties of the system. We first
analyze the mean field, or large
$N$, approximation and recover known results concerning  chiral symmetry breaking,  and its restoration at a finite
temperature
$T_c$. Then we show how the $1/N$ expansion breaks down for temperatures below $T_c$. This is signaled in particular by 
the appearance of a Landau pole which invalidates the calculation of the effective potential for small values of
the field. Finally  we turn to the high temperature regime, where we find that the $1/N$ expansion becomes again a
regular expansion. There chiral symmetry is restored and no Landau pole occurs. In this regime,  the
system behaves as a system of weakly interacting massless fermions, the
$1/N$ corrections providing interactions which decrease logarithmically with increasing temperature. This is  as expected
from  asymptotic freedom. However, since the coupling constant does not appear as an explicit parameter in our expression of
the effective
potential, this result is obtained only after a detailed analysis of the high temperature behavior of the effective
potential obtained in the $1/N$ expansion, and a comparison with the first orders of ordinary perturbation theory. We
show that both approaches yield identical results when the coupling is the running coupling at a scale of the order of
the temperature.

\section{The Gross-Neveu model. Generalities}
\label{sec:generalities}

The  lagrangian of the Gross-Neveu model \cite{Gross:jv}
\beq\label{GrossNeveu}
{\cal L}(\bar \psi, \psi) = \bar\psi \, i \slash \hskip -.22 cm  \partial
\,
\psi +\frac{g^2}{2N} (\bar\psi \psi )^2\, ,
\eeq
describes $N$ interacting massless fermions in one spatial dimension.
The summation over the
$N$ flavors is implicit in Eq.~(\ref{GrossNeveu}), e.g.
$\bar \psi \psi\equiv \sum_{a=1}^N \bar\psi_a
\psi_a$.  As usual,
$\slashchar{\del}=\gamma^\mu \del_\mu$, where the
$\gamma$ matrices are 2$\times$2 matrices satisfying
$\gamma_\mu\gamma_\nu+\gamma_\nu\gamma_\mu=2g_{\mu\nu}$, with
$g_{\mu\nu}={\rm diag}(1,-1)$, and 
$\gamma_5=\gamma_5^\dagger=\gamma^0\gamma^1$.

  The Lagrangian
(\ref{GrossNeveu}) is invariant  under the discrete chiral transformation
\beq\label{discretechiral}
\psi \to \gamma_5 \psi,  \qquad
\bar\psi\to - \bar\psi\gamma_5.
\eeq
Since $\bar\psi\psi\to - \bar\psi\psi$ under this transformation, 
while the Lagrangian (\ref{GrossNeveu})
remains invariant, the quark condensate
$\langle
\bar\psi
\psi\rangle$ plays the role of an order parameter: its non
vanishing indicates spontaneous chiral symmetry breaking, a situation 
met in the vacuum state \cite{Gross:jv}.

In order to study the thermodynamics, we use the imaginary time 
formalism \cite{BR86,Kapusta,LeBellac96}, and write the
partition function ${\cal Z}$ as the following path integral:
\beq\label{partfunct}
{\cal Z} ={\cal N}^{-1} \int {\cal D} \bar \psi {\cal D} \psi \, \exp\left\{-
\int_0^\beta {\rm d}\tau\int {\rm d} x \, \left[\psi^\dagger 
\left(\del_\tau+h_0\right)\psi-
\frac{g^2}{2N} (\bar\psi
\psi )^2\right]\right\},
\eeq
where $\beta=1/T$ is the inverse
of the temperature. We shall sometimes denote the space-time volume 
by  $\int {\rm d}^2
x=\beta L$, with
$L$ the length of the system (eventually to be taken infinite). In 
Eq.~(\ref{partfunct})
$h_0=-\gamma^0\gamma^1\del_x$ is the hamiltonian of a free Dirac 
particle.  The fields $\psi
(\tau,x)$ and $\bar\psi (\tau,x)$ are antiperiodic, with period $\beta$:
$\psi (\beta,x)=-\psi (0,x)$. The (infinite)   normalization
constant
${\cal N}$ in Eq.~(\ref{partfunct}), which depends on the temperature 
but not on the coupling constant, can be eliminated
when necessary (see e.g.
Eq.~(\ref{Vsigmac}) below)
  by dividing  ${\cal Z}$ by the (known)  partition
function
${\cal Z}_0$ of free fermions at the same temperature:
\beq\label{freeZ}
\frac{1}{\beta L}\ln{\cal Z}_0= \frac{2N}{\beta}\int_{-\infty}^\infty 
\frac{{\rm
d}p}{2\pi}
\ln(1+{\rm e} ^{-\beta e_p}) + \left( N\int \frac{{\rm d}p}{2\pi}\, 
e_p\right)\,
\eeq
with $e_p=|p|$, and the factor 2 in the first term accounts for quarks and
antiquarks. The last term (in parenthesis) is infinite, but 
independent of the temperature; it
contributes only to the vacuum energy, and can be discarded. This is 
a trivial part of the
renormalization which will be discussed at length in 
Sect.~\ref{sec:renorpotential}.

At this point, let us briefly digress on the notation that we shall 
use throughout to evaluate momentum integrals
in the imaginary time formalism. Such integrals
involve actually a sum over Matsubara frequencies and a true integral over the
one-dimensional momentum $p$. They will be denoted by:
\beq\label{measure_fermion}
\int \{ {\rm d}^2 P\}\equiv T\sum_{n,\, odd}\int_{-\infty}^{\infty} 
\frac {{\rm d}{p}}{2\pi}\qquad\qquad
\int [ {\rm d}^2 P]\equiv T\sum_{n,\, even}\int_{-\infty}^{\infty} 
\frac {{\rm d}{p}}{2\pi}.
\eeq
  where $P$ denotes a 2-momentum, and  the notation
$\sum_{n,\, odd}$ ($\sum_{n,\, even}$) indicates a sum over fermionic 
(bosonic) Matsubara frequencies:
$\omega_n=(2n+1)\pi T$ for fermions,
$\omega_n=2n\pi T$ for bosons. Depending
upon the context, the 2-momentum $P$ will be either a Minkowski momentum or an
Euclidean one. Whenever ambiguities may arise  we shall use the more
explicit notation
$P_M=(\omega,p)$ for a Minkowski momentum, and $P_E=(p_0,p)$ for an 
Euclidean one; we have
$P_M^2=\omega^2-p^2$, and  $P_E^2=p_0^2+p^2$.  Functions of the 
2-momentum $P$ will be considered in general as functions
of the Minkowski variables $\omega,p$ and, with a slight abuse in 
notation, denoted by either
$f(P)$ or $f(\omega,p)$; when Euclidean variables are appropriate, we 
shall write $f(P_E)$ or $f(ip_0,p)$. In doing finite
temperature calculations  one is led to set
  $\omega=i\omega_n$, where $\omega_n$ is a Matsubara frequency.
  Zero
temperature contributions may be obtained from a finite temperature 
calculation by replacing
$T\sum_{n}$ by
$\int_{-i \infty}^{i\infty} {\rm d}\omega/(2\pi 
i)=\int_{-\infty}^\infty {\rm d}p_0/(2\pi)$, leading
to an Euclidean integral $\int {\rm d}^2 p/(2\pi)^2$. The way to 
handle the ultraviolet
divergences will be described in Sect.~\ref{sec:renorpotential}; in 
the present section we assume that all
integrals are properly regularized, without making the regularization explicit.

\subsection{\label{Sec:eff_pot}Effective potential and the $1/N$ expansion}

We now return to  the partition function ${\cal Z}$ of 
Eq.~(\ref{partfunct}). Following a standard
procedure
\cite{Gross:jv}, we introduce an auxiliary scalar field
$\sigma$ and write:
\beq\label{ZZsigma}
{\cal Z} = {\cal N}^{-1}_\sigma\int {\cal D}\sigma\,{\rm 
e}^{-\frac{1}{2}\int{\rm d}^2x
\,\,\sigma^2}\,{\cal Z}_\sigma ,
\eeq
where
$
  {\cal N}_\sigma= \int {\cal D}\sigma\,{\rm e}^{-\frac{1}{2}\int{\rm 
d}^2x \,\,\sigma^2},
$
and the integration runs over periodic fields: $\sigma (\beta,x)=\sigma (0,x)$.
The quantity
  ${\cal Z}_\sigma$:
\beq\label{Zsigma}
{\cal Z}_\sigma ={\cal N}^{-1}\int {\cal D} \bar \psi {\cal D} \psi\, 
{\rm e}^{-\int{\rm
d}^2x\,
\psi^\dagger
\left(\del_\tau +h(\sigma)\right)\psi},
\eeq
where $h(\sigma)=h_0+\frac{g\sigma}{\sqrt N}\gamma_0$, may be viewed 
as  the partition function for a system of massless
fermions in an ``external field''
$\sigma$.
We denote by
$S_\sigma$  the   fermion propagator in this external field; it obeys
the equation
\beq\label{eqGfiniteT}
\left( \del_{\tau_1} +h(\sigma_1)\right) S_\sigma(\tau_1, x_1;\tau_2, x_2)
=\delta(\tau_1-\tau_2)\delta(x_1-x_2),
\eeq
where $\sigma_1=\sigma(\tau_1,x_1)$. For a given field $\sigma$, the 
Gaussian integral in
Eq.~(\ref{Zsigma}) can be calculated in terms of $S_\sigma$. By 
taking the ratio ${\cal Z}_\sigma/{\cal Z}_0$ one eliminates
the infinite normalization constant and gets:
\beq\label{Vsigmac}
\ln\left(\frac{{\cal Z}_\sigma}{{\cal Z}_0}\right)={\rm Tr} \ln
S_{\sigma}^{-1} -{\rm Tr} \ln S_0^{-1},
\eeq
where the propagator $S_0$ satisfies Eq.~(\ref{eqGfiniteT}) with 
$g=0$, and the symbol Tr implies a trace over the Dirac
matrices and an integration over space-time coordinates.

The  discrete chiral
symmetry of the massless Dirac hamiltonian, 
Eq.~(\ref{discretechiral}), entails the following
property: $\gamma_5
h(\sigma)\gamma_5=h(-\sigma)$, from which it follows that ${\cal Z}_\sigma$ is
invariant under the transformation $\sigma \to -\sigma.$ Since the 
weight function in Eq.~(\ref{ZZsigma}) is
also an even function of $\sigma$, one expects the average value of 
$\sigma$ to vanish, unless there is
spontaneous symmetry breaking.

Whether symmetry breaking occurs or not can be deduced from the 
effective potential for the field
$\sigma$. To get this effective potential, we evaluate  first the 
partition function in the presence of
an external source
$j$ coupled to
$\sigma$:
\beq\label{partition_j}
{\cal Z}[j]= e^{-W[j]} = {\cal N}^{-1}_\sigma\int {\cal D} \sigma 
\,{\cal Z}_\sigma \, e^{-
\int {\rm d}^2x  \,(\frac{1}{2} \sigma^2+j\sigma)}.
\eeq
The expectation value of $\sigma$ in equilibrium, that is, in the 
state of the system corresponding to the
minimum of the free energy in the presence of the source $j$,  is given by:
\beq
\label{eq10}
<\sigma(x)> _j= \frac{\delta W[j]}{\delta j(x)} \equiv \bar\sigma_j (x) .
\eeq
A  Legendre
transform allows us to eliminate the
source $j$ in favor of the expectation value of the field:
\beq\label{legendre}
\Gamma [\bar\sigma_j] =  W[j]-\int {\rm d}^2 x\, \bar\sigma_j(x)j(x) .
\eeq
When
$\bar\sigma_j$ is constant in space and time, we define (dropping now 
the subscript $j$):
\beq\label{Gamma}
\Gamma[\bar\sigma] = V(\bar\sigma) \,\int d^2x\,=\beta L V(\bar\sigma) ,
\eeq
where $V (\bar\sigma)$ is the effective potential. Note that while 
$\Gamma$  is dimensionless,
$V$ has the dimension of an energy density:  it is the free energy 
per unit length (or minus the pressure) for a prescribed
value of $\bar\sigma$.  Note also that the discrete chiral symmetry 
implies that $V$
is an even function of $\bar\sigma$, i.e., $V(\bar\sigma)=V(-\bar\sigma)$.

The effective potential allows a simple determination of the 
equilibrium state. Indeed, since  by
construction
$\delta
\Gamma/\delta
\bar\sigma(x)=-j(x)$, the equilibrium state in the absence of the 
source (i.e, for $j=0$) is determined by the
equation
\beq\label{minimumcondition}
\frac{{\rm d} V}{{\rm d} \bar\sigma}= 0.
\eeq
Furthermore, since the effective
potential is a convex function (this follows from the general 
properties of the Legendre transform and the
convexity of $W[j]$),
  the solutions to Eq.~(\ref{minimumcondition})  correspond to minima 
of the effective potential. In the absence
of symmetry breaking, $V(\bar\sigma)$ has a unique minimum, 
$\bar\sigma=0$. When spontaneous symmetry   breaking
occurs, two degenerate minima appear, at nonzero values of
$\bar\sigma$, say $\pm\bar\sigma_{min}$.

Strictly speaking, the technique of Legendre transforms allows us to
construct  $V(\bar\sigma)$ as indicated above for all values of 
$\bar\sigma$ outside the interval
$[-\bar\sigma_{min},+\bar\sigma_{min}]$, and yields  a constant value 
$V(\bar\sigma)=V(\bar\sigma_{min})$ inside this
interval  (for a pedagogical discussion of this point see e.g.
\cite{Brown:db}). Note that this construction is 
somewhat formal; in particular the values of
$\bar\sigma$ in the interval $[-\bar\sigma_{min},+\bar\sigma_{min}]$
are not  reached as minimal free energy 
solutions in the presence of a constant source $j$.
Now,  in the approximations to be developed shortly, we shall find that
the relation between the source $j$ and
$\bar\sigma$ is multivalued: one solution corresponds to the minimal 
free energy for a given $j$, while other
solutions yield higher free energies and values  of
$\bar\sigma$  inside the interval $[-\bar\sigma_{min},+\bar\sigma_{min}]$. By 
keeping all solutions, we define a
continuation of the effective potential which is not a constant in the interval
$[-\bar\sigma_{min},+\bar\sigma_{min}]$. It is this continuation of the effective
potential that is used in particular to  calculate  the
fluctuations of the field $\sigma$.

In this paper we consider the first two orders in the $1/N$ expansion of the
effective potential $V$. These are obtained by
  evaluating
the path integral (\ref{partition_j}) in a saddle point approximation 
\cite{Zinn-Justin:mi}, and then performing a
Legendre transform.  The leading contribution  will come from the 
saddle point itself, the $1/N$
correction
  from integrating the fluctuations about the saddle point.

The value of the field
$\sigma$ at the saddle point, i.e., the value $\sigma_c$ for which 
the exponent in Eq.~(\ref{partition_j}) (with ${\cal
Z}_\sigma$ of Eq.~(\ref{Zsigma})) is extremum,  is the solution of 
the following self-consistent equation (commonly referred
to as the ``gap equation''):
\beq\label{gap}
\sigma_c+j=\frac{1}{\beta L}\left.\frac{\del \ln {\cal Z}_\sigma}{\del
\sigma}\right|_{\sigma_c}=-\frac{g}{\sqrt{N}}
\langle\bar\psi\psi\rangle_{\sigma_c}, 
\eeq
where we have used Eq.~(\ref{Zsigma}) to express the derivative 
$\left.{\del \ln {\cal Z}_\sigma}/{\del
\sigma}\right|_{\sigma_c}$ in terms of the quark condensate
$\langle\bar\psi\psi\rangle_{\sigma_c}$, and we restrict ourselves to 
solutions $\sigma_c$ which are constant in
space and time. The  quark condensate
$\langle\bar\psi\psi\rangle_{\sigma_c}$ is easily calculated with the 
help of  the propagator $S_{\sigma}$ (see
Eq.~(\ref{eqGfiniteT})), whose Fourier transform for constant $\sigma$ reads:
\beq
S_{\sigma}(P)=\frac{1}{(-\slashchar{P} + M)}.
\eeq
This is the propagator of a quark with mass $M$ defined as:
\beq\label{M}
M=\frac{g \sigma}{\sqrt{N}}.
\eeq
For a given (constant) value of $\sigma$, the quark condensates reads then:
\beq\label{G(0)}
  \langle\bar\psi\psi\rangle_{\sigma} = - \int \{d^2P\} {\rm Tr} 
\left[ \frac{1}{
-\slashchar{P} + M}\right]= -2N\int \{d^2P\}\frac{M}{-P^2+M^2},
\eeq
from which we get  $\langle\bar\psi\psi\rangle_{\sigma_c}$ by setting 
$\sigma=\sigma_c$.
 From the equations above,  we
observe that
  $\langle\bar\psi\psi\rangle_{\sigma_c}$ is
of order $N$, so that $\sigma_c$ is of order $\sqrt{N}$ (the source
$j$ is to be considered as a constant of order
$\sqrt{N}$); as for the mass parameter
$M$, it is independent of $N$.

At this level of approximation, the free energy in the presence of 
the source, $W[j]=-\ln {\cal Z}[j]$, is
obtained from the value at the saddle point of the integrand in 
Eq.~(\ref{partition_j}):
\beq\label{W0}
W^{(0)}[j]=\left(\frac{1}{2}\sigma_c^2+j\sigma_c\right) \beta L-\ln 
\frac{{\cal Z}_{\sigma_c}}{{\cal Z}_0},
\eeq
and, using Eq. (\ref{eq10}), the expectation of the field is simply $\bar\sigma=\sigma_c$. 
Note that the gap equation (\ref{gap}) may
have multiple constant solutions
$\sigma_c$ corresponding to the same constant $j$ (this can be seen explicitly 
from the expressions given in
subsection ~\ref{sec:T0LO}). Keeping, as discussed above,  all solutions 
(i.e., not only that with the lowest free
energy), and eliminating the source $j$ according to 
Eq.~(\ref{legendre}), one obtains the following
``continuation'' of the effective potential:
\beq
V(\bar\sigma)=\frac{1}{2}\bar\sigma^2-\frac{1}{\beta L}\ln 
\frac{{\cal Z}_{\bar\sigma}}{{\cal
Z}_0}.
\eeq
As we shall see in subsection~\ref{sec:T0LO}, this function has two 
degenerate minima, but is not flat between the
two minima: the values of $\bar\sigma$ within the two
minima correspond to the solutions  of the gap equation  which, for a 
given value of $j$, accompany the
solutions with minimal free energy.

In order to get the next order in the $1/N$ expansion, we expand the 
field $\sigma$ around the  value $\sigma_c$,
i.e., we set $\sigma=\sigma_c+\tilde\sigma$, and do the corresponding 
change of variable in
the functional integral (\ref{partition_j}). The terms linear in $\tilde\sigma$
cancel because $\sigma_c$ satisfies Eq.~(\ref{gap}). There remains 
then in (\ref{partition_j})
a gaussian integral over the fluctuations of the field $\sigma$. This 
is easily evaluated, with the result:
\beq\label{integrale_gaussienne}
\frac{1}{{\cal N}}_\sigma\int {\cal D} \tilde \sigma \,\exp
\left\{ - \frac{1}{2}\int [d^2Q]\,
\tilde \sigma (Q) \,\left(1+g^2\Pi(Q) \right) \,\tilde \sigma (-Q) 
\right\}\nonumber\\
& & 
\!\!\!\!\!\!\!\!\!\!\!\!\!\!\!\!\!\!\!\!\!\!\!\!\!\!\!\!\!\!\!\!\!\!\!\!\!\!\!\!\!\!\!\!\!\!\!\!\!\!\!\!\!\!\!\!
=\exp\left\{ -\frac{\beta L}{2}  \int
[d^2Q] \,\ln \left(1+g^2\Pi(Q)\right)\right\}.
\eeq
The quantity  $\Pi(Q)$ that appears in this equation is  the 
contribution of the one loop diagram:
\beq\label{Pi_oneloop}
\Pi(Q) = \frac{1}{N} \int \{d^2K\} {\rm Tr} \left[ \frac{1}{
\slashchar{K} - M} \,\frac{1}{ \slashchar{K} +  \slashchar{Q} - M}
\right],
\eeq
which is calculated in App.~A; it plays the role of a self-energy
for the $\sigma$ meson, whose propagator $D(Q)$ can be read from 
Eq.~(\ref{integrale_gaussienne})
above:
\beq\label{sigmapropagator}
D(Q)=\frac{1}{1+g^2\Pi(Q)}.
\eeq

\begin{figure}
\includegraphics[width=10cm]{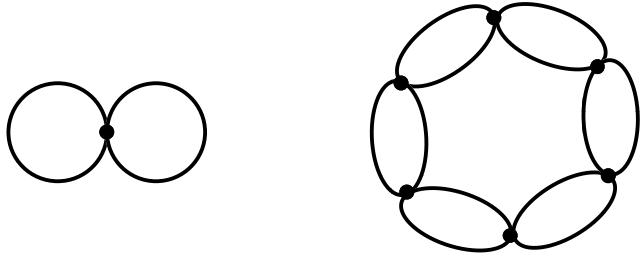}
\caption{\label{fig:free_energy}The contributions to the free energy 
in the $1/N$ expansion:
the left diagram represents the Hartree contribution; the right 
diagram is one of the infinite set of ``ring
diagrams''.}
\end{figure}

Combining the value $W^{(0)}$ obtained above, Eq.~(\ref{W0}),  with 
the gaussian integral
(\ref{integrale_gaussienne}),  we obtain $W[j]$ at order  $1/N$:
\beq\label{logZdej}
W[j]= \left(\frac{1}{2}\sigma_c^2+j\sigma_c\right) \beta L-\ln 
\frac{{\cal Z}_{\sigma_c}}{{\cal Z}_0}+\frac{\beta L}{2}\int
[d^2Q] \,\ln \left(1+g^2\Pi(Q)\right)\equiv W^{(0)}+W^{(1)} .
\eeq
By taking the derivative of $W[j]$ in Eq.~(\ref{logZdej}) with 
respect to $j$, one obtains
$\bar\sigma$, the expectation value of the field
$\sigma$. (Note that $\sigma_c$ depends implicitly on $j$ through the 
gap equation (\ref{gap}).) We can write
$\bar\sigma=\sigma_c+\delta\bar\sigma$,  and it follows immediately 
from  Eqs.~(\ref{logZdej}) and (\ref{gap}) that
$\delta\bar\sigma$ is of order
$1/N$ relative to $\sigma_c$. In computing the Legendre transform 
$\Gamma [\bar\sigma] =W^{(0)}+W^{(1)}-j \bar\sigma
\beta L$ (see Eq.~(\ref{legendre})),  we may simply replace
$\sigma_c$ by
$\bar\sigma$, since the error made is of order $1/N^2$ (the terms linear in
$\delta\bar\sigma$ drop in $W^{(0)}$ because $\sigma_c$ satisfies the 
gap equation (\ref{gap}); similarly, one
makes an error of order $1/N^2$ in ignoring $\delta\bar\sigma$ in
$W^{(1)}$ which is already a quantity of order $1/N$). The 
elimination of $j$ is then
straightforward
  and we finally obtain the  effective potential in the form:
\beq\label{Vdesigmac}
V(M)=N\frac{M^2}{2g^2}- N \int\{d^2P\} 
\ln\left(1-\frac{M^2}{P^2}\right)+\frac{1}{2}
\int [d^2Q] \,\ln \left(1+g^2\Pi(Q)\right),
\eeq
where $M$ is given by Eq.~(\ref{M}) with $\sigma$ replaced by $\bar\sigma$ .
We shall
write  this effective potential as
\beq\label{V0V1}
V(M)=NV^{(0)}(M)+V^{(1)}(M),
\eeq
where $NV^{(0)}(M)$ is the leading order contribution and
$V^{(1)}(M)$ the next-to-leading order one. Before renormalization,
$NV^{(0)}(M)$ is the sum of
the first two terms in Eq. (\ref{Vdesigmac}), and $V^{(1)}(M)$ is the last term.
After renormalization, as we shall see, the first term  in
Eq.~(\ref{Vdesigmac}) contributes also at order $1/N$.  For this 
reason, we shall often refer to the
first two terms in Eq.~(\ref{Vdesigmac}) as the ``fermionic'' 
contribution, and to the last term
as the ``bosonic'' one: the first two terms represent indeed the free 
energy density of  massive fermions in the ``Hartree
approximation'' \cite{BR86}, while the last term may be viewed as a 
the one-loop contribution of the bosonic degrees of
freedom associated with the $\sigma$ field.   In terms of the 
fermionic variables of the original lagrangian, the
corresponding contributions to the free energy can be given a simple 
diagrammatic interpretation: the first diagram in
Fig.~\ref{fig:free_energy} corresponds to the Hartree contribution, 
the second is one of the family of ``ring diagrams"
representing  the bosonic contribution.

Before closing this subsection, let us recall that the effective 
potential is the generating functional of the irreducible
$n$-point functions for the $\sigma$ field at zero momentum. In 
particular, the following, leading order, relation for the
2-point function can be easily established at zero temperature:
\beq\label{V0andD}
g^2\frac{{\rm d}^2V^{(0)}}{{\rm d}M^2}=D^{-1}(Q=0;M)
\eeq
where $D^{-1}=1+g^2\Pi$ is the inverse of the $\sigma$ propagator 
(see Eq.~(\ref{sigmapropagator})). The same relation
holds at finite temperature, but care must be exerted in specifying 
the limit $Q\to 0$ which is in general
  non analytic (it depends on which limit $\omega\to 0$ and $q\to 0$ 
is taken first). Thus, at finite temperature,
the right hand side of the relation (\ref{V0andD}) has to be 
understood as $\lim_{q\to
0}D^{-1}(\omega\!=\!0,q;M)$  (see subsection~\ref{sigmaexcitation} and App. A).

\subsection{\label{Sec:gap_eqn} Gap equation and quark condensate}

As we have
argued before, the equilibrium state is obtained by minimizing the 
effective potential with
respect to $M$. One gets then the gap equation in the form:
\beq\label{condensatedef}
0=\left.\frac{dV}{dM}\right|_{M_{min}}=\frac{NM_{min}}{g^2}+\langle\bar\psi 
\psi\rangle_{M_{min}},
\eeq
where we have used the fact that $\langle\bar\psi \psi\rangle$ may be 
obtained as the derivative with respect to
$M$ of the last two terms in Eq.~(\ref{Vdesigmac})  (this may be seen 
by going back to the
expression (\ref{ZZsigma}) of the partition function and using the 
fact that $\langle\bar\psi \psi\rangle$ can be obtained
by differentiating ${\cal Z}_\sigma$ in Eq.~(\ref{Zsigma}) with 
respect to $\sigma$; see also Eq.~(\ref{gap}) ).  Thus, the
value of
$M$ at the minimum of the potential and the value of the 
quark-condensate in equilibrium  are proportional:
\beq\label{condensatedef2}
M_{min}=-\frac{g^2}{N}\langle\bar\psi \psi\rangle_{M_{min}},
\eeq
which indicates that $\bar\sigma$ (or $M$) and 
$\langle\bar\psi \psi\rangle$ are equivalent order
parameters for the discrete chiral symmetry.

\begin{figure}
\includegraphics[width=10cm]{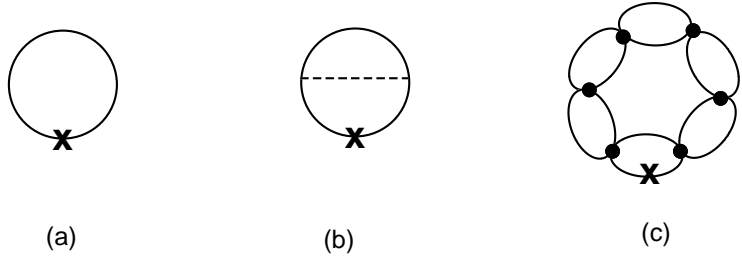}
\caption{\label{fig:condensat}The fermion condensate $\langle 
\bar\psi\psi\rangle$: (a) leading order; (b) $1/N$
correction, the dashed line represents the $\sigma$ propagator; (c) 
same as (b) in the
fermionic language. }
\end{figure}

We shall solve Eq.~(\ref{condensatedef}) order by order in the
$1/N$ expansion. To do so, we write:
\beq\label{Mexpanded}
M_{min}=M_{min}^{(0)}+\frac{1}{N}M_{min}^{(1)}.
\eeq
In leading order, we have:
\beq
0=\left.\frac{{\rm d}V^{(0)}}{{\rm 
d}M}\right|_{M_{min}^{(0)}}=\frac{M_{min}^{(0)}}{g^2}+\frac{\langle
\bar\psi
\psi\rangle_{M_{min}^{(0)}}}{N},
\eeq
where $\langle
\bar\psi
\psi\rangle_{M_{min}^{(0)}}$ is obtained from  Eq.~(\ref{G(0)}) with 
$M_{min}^{(0)}$
substituted for $M$. This gap equation needs to be solved exactly in 
order to get $M_{min}^{(0)}$.  The correction
$M_{min}^{(1)}$ is obtained by solving Eq.~(\ref{condensatedef}) to
$1/N$ accuracy. We get then:
\beq\label{Mmin1gen}
M_{min}^{(1)}=-\frac{   \left.  \frac{ {\rm d}V^{(1)} }{ {\rm d}M } 
\right|_{M_{min}^{(0)}}   }
{\left. \frac{ {\rm d}^2V^{(0)} }{ {\rm d}M^2 }\right|_{M_{min}^{(0)}} }.
\eeq

\begin{figure}
\includegraphics[width=10cm]{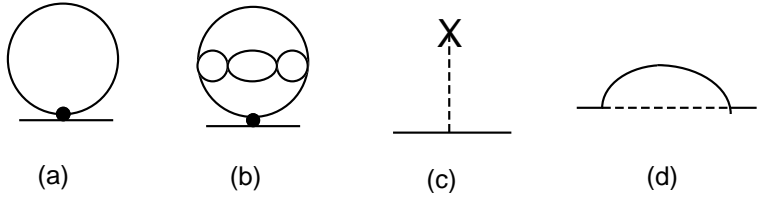}
\caption{\label{fig:fermion_mass}Contributions to the fermion mass. 
(a) leading order; (b) $1/N$ correction
to  the minimum of the effective potential; equivalently, the sum of 
the contributions (a) and (b) can be represented by
diagram (c) where the insertion with a cross represents $M_{min}$. 
Diagram (d) represents an additional
$1/N$ correction coming from
$\Sigma$. As in Fig.~\ref{fig:condensat}, the dashed line in (d) represents 
the $\sigma$ propagator}
\end{figure}

To the expansion (\ref{Mexpanded}) of $M_{min}$ corresponds,
according to Eq.~(\ref{condensatedef2}), a similar expansion of the 
quark condensate:
\beq
\langle\bar\psi \psi\rangle=N\langle\bar\psi \psi\rangle^{(0)}+\langle\bar\psi
\psi\rangle^{(1)}.
\eeq
The two contributions to the quark condensate correspond to the 
diagrams displayed in Fig.~\ref{fig:condensat}.
The leading order $\langle\bar\psi
\psi\rangle^{(0)}$ corresponds to the contribution of the first 
diagram evaluated for $M_{min}=M_{min}^{(0)}$.
The order
$1/N$ correction
$\langle\bar\psi
\psi\rangle^{(1)}$ contains the contribution
\beq\label{Picondensat}
\frac{d V^{(1)}}{dM}=\frac{1}{2}\int [d^2Q] \,\frac{ \del
\Pi/\del M}{1+g^2\Pi(Q)}
\eeq
associated with either of the last two diagrams in 
Fig.~\ref{fig:condensat}. In addition, as explicitly indicated in
Eq.~(\ref{Mmin1gen}),
$\langle\bar\psi
\psi\rangle^{(1)}$ receives also a contribution from the first 
diagram in Fig.~\ref{fig:condensat}
evaluated at $M_{min}\ne M_{min}^{(0)}$.

Before closing this subsection, it is worth emphasizing that neither 
the minimum $M_{min}$ of the effective potential, nor
the quark condensate are physical observables: after renormalization, 
and beyond leading order
in the $1/N$ expansion,
their values will depend on the renormalization scale.

\subsection{\label{Sec:quark_mass}The quark mass $M_f$}

As we have mentioned, chiral symmetry is spontaneously broken in the ground
state, and because of their coupling to the condensate  the quarks 
acquire  a mass. This mass is a
physical quantity  that we shall denote throughout as
$M_f$. That is,
$M_f$  will consistently refer to the mass of the quark in the 
vacuum, {\it a quantity to be kept constant at all
orders of our approximations}.

In leading order in the $1/N$ expansion, the coupling to the 
condensate is the only contribution to the fermion
mass and we can identify
$M_{min}^{(0)}$ with $M_f$. This identification between $M_{min}$ and 
$M_f$ does not hold in higher order.
Starting at order $1/N$, there are other contributions to the fermion 
mass besides
$M_{min}$.
Quite generally, the fermion mass
is given by the pole of the fermion propagator at $P_E^2=-M_f^2$.
The  fermion propagator itself, $S(P)$, can be written as
\beq\label{sigmadef}
S^{-1}(P)=S^{-1}_\sigma(P)+\Sigma(P)=-\slashchar{P}+M_{min}+\Sigma(P).
\eeq
The various Feynman diagrams contributing to the 
fermion mass are displayed in
Fig.~\ref{fig:fermion_mass}.  In leading order, $\Sigma(P)=0$. At
next-to-leading order,
$\Sigma(P)$ can be written as:
\beq\label{selffermion}
\Sigma(P)=a\slashchar{P}+b,
\eeq
where $a$ and $b$ are functions of $P$ given in App. C. The equation
$S^{-1}(\slashchar{P}=M_f)=0$ yields then  $M_f$ as a sum of two contributions:
\beq\label{totalM}
M_f=M_{min}+M_\Sigma,
\eeq
where $M_\Sigma$ (which is of order $1/N$) is calculated in App. C.

At finite temperature, we shall define a temperature dependent mass $M_f(T)$ by
the equation:
\beq
S^{-1}(\omega=M_f(T),p=0;T)=0.
\eeq
where $S(\omega,p;T)$ is the fermion propagator at finite temperature.
At leading order in the $1/N$ expansion $\Sigma(\omega,p;T)=0$, and one can simply
identify the fermion mass $M_f(T)$ with the minimum of the finite temperature effective
potential. At next-to-leading order, the evaluation of the $M_f(T)$ would require the
calculation of $\Sigma(\omega,p;T)$. But, as we shall see in the following sections, this
is not needed to evaluate the pressure at next-to-leading order.

\section{\label{sec:renorpotential}The renormalized effective potential at $T=0$}
As we have already mentioned, the formulae given in the previous section contain ultraviolet divergent integrals.
We  discuss  now
the procedure of renormalization which allows us to get rid of the  infinities. First
we
need to specify our regularization:  this will consist in a simple cut-off $\Lambda$ on the
length of the
Euclidean momenta (for a discussion of more sophisticated regularizations
in this type of models, see e.g.
Ref.~\cite{Ripka:zb}).  We then obtain the effective
potential as a function $V_B(M_B;g_B,\Lambda)$,
referred to as the ``bare'' potential; the variables
$\sigma_B$ (or $M_B$) and $g_B$, are the bare field (or 
mass) and coupling constant, respectively.

\begin{figure}
\includegraphics[width=10cm]{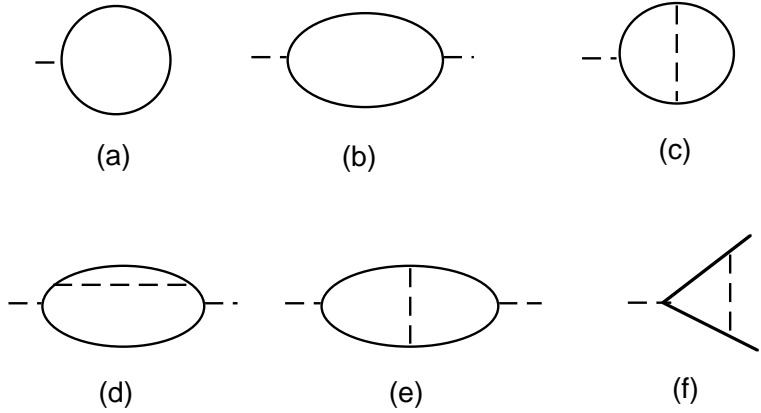}
\caption{\label{fig:Npoint} One-point
(a,c) and two-point (b,d,e) functions
for the
$\sigma$ field, in leading order in the $1/N$ expansion
(a,b), and at order $1/N$
(c,d,e). Diagram (f) is the first  correction to the quark-sigma vertex; it is  of
order $1/N$. The dashed lines represent $\sigma$ propagators.}
\end{figure}

Divergent integrals appear in ${\rm d}V_B/{\rm d} M_B$ and
${\rm d}^2V_B/{\rm
d} M_B^2$, i.e., in the one and
two-point functions of the $\sigma$-field.  The corresponding diagrams at
leading
order and at next-to-leading
order are displayed in Fig.~\ref{fig:Npoint}.
  Besides,  there are
divergences
which
are independent of $M_B$; these  are  eliminated
by subtracting from
$V_B(M_B;g_B,\Lambda)$ a constant $C(g_B,\Lambda)$.

In order to construct the renormalized effective
potential, we
define a renormalized field $\sigma$,
 a renormalized coupling constant $g$, and a renormalized
mass $M$ (related to $\sigma$ and $g$
by Eq.~(\ref{M})). We set:
\beq\label{renormalizations}
\sigma_B =
\sqrt{Z} \sigma, \hskip 2 cm g_B = \sqrt{\frac{Z'}{Z}} g,
\qquad\qquad M_B =\sqrt{Z'} M.
\eeq
where the
renormalization
constants $Z$ and $Z'$  are
dimensionless functions of the renormalized coupling $g$, and the
cut-off $\Lambda$.
 These
functions may be expanded in powers of $1/N$:
\beq\label{Z01}
Z =
Z^{(0)}+\frac{1}{N}Z^{(1)},  \qquad\qquad \qquad
Z' = Z'^{(0)}+\frac{1}{N}Z'^{(1)} ,
\eeq
and  chosen so
as to absorb
the ultraviolet divergences at each order. Because they are dimensionless, $Z$ and $Z'$
can
actually depend only on the ratio of $\Lambda$ to another mass scale $M_0$, to be referred to as the renormalization
scale,  at which the renormalized $n$-point
functions are specified (see below). In fact, we shall find convenient to
consider $Z$ and $Z'$ as functions of both $\Lambda/M_f$, where $M_f$ is the fermion mass, and
$M_0/M_f$ (the latter quantity being eventually a
finite function of the renormalized coupling):
the dependence in
$\Lambda$ will be fixed by the elimination
of ultraviolet divergences, that on $M_0$ 
will be determined by  renormalization conditions that we now
present.

To this aim, we note first that the  correction to the
vertex
$\sigma\bar\psi\psi$ is
of order $1/N$ (see Fig.~\ref{fig:Npoint}~f).
In  leading order, this
vertex  is therefore not renormalized and we may
choose to balance the renormalization of the field with that of
the 
coupling constant, i.e.,
set $g_B\sigma_B=g\sigma$  or equivalently,
  according to
Eq.~(\ref{renormalizations}), set $Z'^{(0)}=1$.
As for the constant $Z^{(0)}$, it will be determined by the
following condition:
\beq\label{renormalisation_condition_U}
\left.\frac{{\rm d}^2V^{(0)}}{{\rm
d}M^2}\right|_{M_0}=\frac{1}{g^2}.
\eeq
According to the relation (\ref{V0andD}) (extended to
the
corresponding renormalized quantities) this is equivalent to the condition
$D(Q^2=0;M_0)=1$ for the renormalized
$\sigma$ propagator.   At next-to-leading order we
shall impose
\cite{Coleman:jx} that the first and second
derivatives of the potential
at
$M=M_0$ are not changed by the $1/N$ corrections, that
is:
\beq\label{renormalisation_conditions}
   \left.\frac{{\rm d}V^{(1)}}{{\rm
d}
M^2}\right|_{M_0}=0,\qquad\qquad\left.\frac{{\rm d}^2V^{(1)}}{{\rm
d}
M^2}\right|_{M_0}=0.
\eeq

Summarizing the procedure that we have outlined, we may write the
renormalized
effective potential $V$ as
follows:
\beq\label{renormV}
V(M;g,M_0)=V_B(M_B;g_B,\Lambda)-C(g_B,\Lambda).
\eeq
By construction, the
right hand side has a finite limit when the
cut-off
$\Lambda$ is sent to infinity. For large $\Lambda$, the
renormalized
potential
becomes therefore independent of $\Lambda$, i.e., it satisfies
the
renormalization
group equation $
\Lambda{{\rm d}V}/{{\rm d}\Lambda}=0,
$
where the derivative is
taken at fixed values of the renormalized
quantities,
and fixed $M_0$.  This can be written explicitly
as
\beq\label{RG1}
\left[ \Lambda\frac{\del}{\del \Lambda}+\beta(g_B)\frac{\del}{\del
g_B}
-\gamma(g_B)M_B\frac{\del }{\del M_B}
\right]\left(V_B(M_B;g_B,\Lambda)
-C(g_B,\Lambda)\right)=0,
\eeq
where
\beq\label{betabare}
\beta(g_B)\equiv\Lambda\frac{
\del g_B}{
\del\Lambda}=\frac{g_B}{2}\,\frac{\del
\ln(Z'/Z)}{\del \ln
\Lambda}\,,\qquad\qquad\gamma(g_B)\equiv
-\frac{\Lambda}{M_B}\frac{\del
M_B}{\del
\Lambda}=-\frac{1}{2}\frac{\del\ln
Z'}{\del\ln\Lambda}.
\eeq

Alternatively,  expressing
the fact that the renormalized potential
in Eq.~(\ref{renormV}) does not depend on the scale
$M_0$, one can
write the renormalization group equation as $
M_0\,{{\rm d}V}/{{\rm d}M_0}=0,$
where the derivative is now
taken at fixed bare quantities, and fixed
cut-off
$\Lambda$. Explicitly:
\beq\label{RGM0}
\left[
M_0\frac{\del}{\del M_0}+\beta(g)\frac{\del}{\del g}
-\gamma(g)M\frac{\del }{\del
M}
\right]V(M;g,M_0)=0,
\eeq
where
\beq\label{renormM0}
\beta(g)\equiv M_0\frac{ \del g}{ \del
M_0}=\frac{g}{2}\,\frac{\del
\ln(Z/Z')}{\del
\ln M_0}\,,\qquad\qquad
\gamma(g)\equiv
-\frac{M_0}{M}\frac{\del M}{\del
M_0}=\frac{1}{2}\frac{\del \ln Z'}
{\del\ln  M_0}.
\eeq

These
various functions will be calculated in subsection \ref{sec:RGanal},
where
we shall also discuss some
consequences of the renormalization group
equations. We note here a general feature of the model that  is useful
to keep in mind in
order to understand the logic of the construction of the effective potential  in the
next
subsections. The model depends initially on two parameters, the bare coupling
$g_B$ and the
cut-off
$\Lambda$. Since these will be adjusted so as to reproduce the fermion mass $M_f$,
$g_B$ will become effectively a function of
$\Lambda$. This relation between $g_B$ and $\Lambda$ depends on the accuracy with
which
$M_f$ is calculated, and is therefore modified at each order in the
$1/N$ expansion. The same property
holds in the renormalized theory: there exists a relation
between the renormalized coupling $g$ and the scale $M_0$,
which is redefined at each
order so that the calculation of the fermion mass at that order reproduces the value
$M_f$.

We shall, in the next two subsections, construct the leading and
next-to-leading order
contributions to the renormalized  effective potential. We shall see
that it is possible to write these in terms
of a variable $x_M$ related to $M$,
but which is  invariant under renormalization group transformations ($M$ is
not invariant
beyond leading order). This choice of variable will make it obvious that the potential 
 is
independent of
$M_0$, and it follows from  the relation between
$M_0$ and $g$ alluded to above, that it
will be also independent of the renormalized
coupling.

\subsection{\label{sec:T0LO}The leading order
contribution}

The zeroth order in the $1/N$ expansion of the potential, $V^{(0)}_B$, is the
fermionic
contribution in Eq. (\ref{Vdesigmac}). At T=0 it can be written as:
\beq\label{potential_U}
NV^{(0)}_B (M_B)
&=& N\frac{1}{2g_B^2} M_B^2 - N\int \frac{d^2 p}{(2\pi)^2}
\ln\left(1-\frac{M_B^2}{P^2}\right).
\eeq
This
expression is divergent. Introducing a cut-off $\Lambda$
on the length of the  Euclidean momentum ($\sqrt
{P_E^2}<\Lambda$), one
obtains:
\beq\label{potUL}
  V^{(0)}_B
(M_B)=\frac{M_B^2}{2g_B^2}+\frac{M_B^2}{4\pi}
\left(\ln\frac{M_B^2}{\Lambda^2} -1\right) +
{\cal
O}\left(\frac{M_B^2}{\Lambda^2}\right).
\eeq

We proceed now to the renormalization of the
potential along the 
lines described in the previous section. Since at
leading order
$Z'=Z'^{(0)}$=1 (see
the discussion after Eq.~(\ref{Z01})), 
we 
can simply replace $M_B$ by $M$, and  we need only to
renormalize the coupling constant. After this, there will be no 
need for further subtraction.
By
expressing
$g_B$ in terms of
$g$ in  Eq. (\ref{potUL}), using Eq.~(\ref{renormalizations}) with
$Z=Z^{(0)}$, we get:
\beq\label{potUR}
  V^{(0)} (M)=\frac{Z^{(0)} M^2}{2g^2}+\frac{M^2}{4\pi}
\left(
\ln {\frac{M^2}{\Lambda^2}} -1\right) .
\eeq
The  $\Lambda$-dependent term is eliminated by
choosing
\beq\label{ZMf}
Z^{(0)} = \frac{g^2}{2\pi} \ln \frac{\Lambda^2}{M_f^2}+ \bar
Z^{(0)},
\eeq
where the finite
constant $\bar Z^{(0)}$ is fixed by the renormalization condition
(\ref{renormalisation_condition_U}): 
\beq\label{barZero}
\bar
Z^{(0)}(M_0,g)=1-g^2/\pi+(g^2/\pi)\ln(M_f/M_0).
\eeq 
The renormalized potential at leading order can then be
written as:
\beq\label{renormU}
V^{(0)} (M) = \frac{M^2}{2g^2}\bar Z^{(0)}
+ \frac{M^2}{4\pi}\left(\ln\frac{M^2}{M_f^2}-1\right)
=\frac{M^2}{4\pi}\left(\ln\frac{M^2}{M_0}-3+\frac{2\pi}{g^2}\right).
\eeq
It has two minima, at $\pm M_{min}^{(0)}$,
with:
\beq\label{Mmin0}
M_{min}^{(0)}(M_0,g) = M_f\,{\rm e}^{-(\pi/g^2)\bar Z^{(0)}}
= M_0\,{\rm e}^{1-\pi/g^2}.
\eeq
The
existence of such non trivial minima of the effective potential 
indicates spontaneous symmetry breaking
with,
according to  Eq.~(\ref{condensatedef2}), a non vanishing value  of 
the quark condensate.

It
remains to verify that $V^{(0)}$ does not
depend on $M_0$. To do so, we remind first that, at this order,
we can set $ M_{min}^{(0)}=M_f$, or equivalently (see Eq. (\ref{Mmin0})) $\bar Z^{(0)}=0$.
Then, one gets:
\beq\label{URindepM0}
V^{(0)}(M) =
\frac{M^2}{4\pi}\left(\ln\frac{M^2}{M_f^2}-1\right) =
   \frac{M_f^2}{4\pi}\, x_M(\ln x_M-1),
\eeq
where
we have set 
\beq\label{xMdef0}
x_M\equiv M^2/M_f^2.\eeq
Since at this order $M$ is not
renormalized, the variable $x_M$ is clearly
a renormalization group invariant, and so is the effective
potential. At next-to-leading
order, we shall see that it is still  possible to express the effective potential
in terms
of a renormalization group  invariant $x_M$ whose definition generalizes that in
Eq.~(\ref{xMdef0})
so as to include an $M_0$-dependent correction of order 
$1/N$ (this correction will compensate  the
$M_0$-dependence of $M$; see Eq.~(\ref{xm2})).
We note also that, as expected from the discussion at the end of
the previous subsection,
the  expression (\ref{URindepM0}) of the effective potential does not depend on the
coupling
constant anymore: by eliminating the explicit dependence on
$M_0$, we have  also eliminated the
dependence on $g$. 

In order to get the expression (\ref{URindepM0}) of the effective potential, we did not
need
the explicit expression of $\bar Z^{(0)}$ given by Eq.~(\ref{barZero}). This is needed however
to
specify the  explicit relation between
$M_0$ and
$g$. From the  equation  
$M_f=M_{min}^{(0)}(M_0,g)$, or
equivalently $\bar Z^{(0)}(M_0,g)=0$, one gets:
\beq\label{running_g}
g^2(M_0)=\frac{\pi}{1+\ln (M_0/M_f)} \,
.
\eeq
This formula indicates that $g$ becomes vanishingly small as $M_0\to \infty$, 
as expected from asymptotic freedom. It also shows that $g$ diverges when 
$M_0\to M_f/e$, a property that we shall discuss further shortly.

Before we do that, we note that the main  results obtained so far in this subsection 
could be derived by working solely with the bare
quantities. In terms of the bare parameters and the cut-off 
$\Lambda$, the minima of
$V^{(0)}_B(M_B)$ are given
by  the solutions of the gap
equation:
\beq\label{minbare}
\frac{M_B}{g_B^2}=\frac{M_B}{4\pi}\ln\left(\frac{\Lambda^2}{M_B^2}\right).
\eeq
The
right hand side of this equation is $-\langle \bar\psi \psi\rangle/N$, i.e.,
it is proportional to the quark
condensate (see Eq.~(\ref{condensatedef})). Aside
from the trivial solution
$M_B=0$, the gap equation has two
degenerate  solutions at $M_B=\pm M_{min}^{(0)}$ with:
\beq\label{minimbare}
M_{\rm
min}^{(0)}(\Lambda,g_B)=\Lambda {\rm e}^{-\pi/g^2_B}.
\eeq
By identifying one of the non trivial solutions
with the fermion mass, 
as we did above, e.g. setting $M_f=M_{\rm
min}^{(0)}(\Lambda,g_B)$, we get the relation
between the bare 
coupling and the cut-off:
\beq\label{running_gB}
g_B^2=\frac{\pi}{\ln \Lambda/M_f}
\,.
\eeq
This may be used to eliminate $\Lambda$ from the expression 
(\ref{potUL}) of the potential and recover
Eq.~(\ref{URindepM0}) (recall that at this order $M_B=M$).

\begin{figure}
\includegraphics[width=13cm]{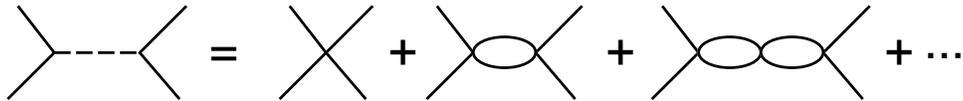}
\caption{\label{figscattampl}The quark-quark scattering amplitude corresponding to the exchange of a $\sigma$ meson
(left-hand side), and its expression in terms of the fermionic variables.}
\end{figure}
We end this subsection by considering the quark-quark scattering amplitude 
${\cal T}(\omega,q;M)$, where the quark mass $M$ is considered as an 
independent 
parameter. In leading order (in perturbation theory) this scattering 
amplitude is simply the bare coupling constant squared
$g_B^2$. Including the  exchange of the $\sigma$ meson, which contributes at the same order as the bare
coupling in the $1/N$ expansion,  one obtains:
\beq\label{scatteringamplitude0}
{\cal T}(Q_E;M)=\frac{g_B^2}{1+g_B^2
\Pi_0(Q_E;M)} .
\eeq
A diagrammatic interpretation of ${\cal
T}(Q_E;M)$ is given in Fig.~\ref{figscattampl}.
The scattering amplitude can also be expressed in terms of renormalized quantities (see App. A):
\beq\label{scatteringamplitude01}
{\cal T}(Q_E;M)=
g^2\,D_0(Q_E;M),
\eeq
where $g^2$ is the square of the renormalized coupling constant at the scale $M_0$, 
i.e., $g^2=g^2(M_0^2)$ (given by  Eq.~(\ref{running_g})) and $D_0$ is 
the renormalized $\sigma$ propagator (given by  Eq.~(\ref{Dinverse})). Note that the explicit $g^2$ in the r.h.s. 
of Eq.~(\ref{scatteringamplitude01}) cancels with the factor $1/g^2$ 
contained in
$D_0$ (see  Eq.~(\ref{Dinverse})) so that ${\cal T}(Q_E;M )$ is 
independent of the
renormalisation scale $M_0$, i.e., the scattering amplitude is a
renormalization group invariant. Using the explicit expression of 
$D_0(Q_E;M )$ given in App. A, one easily shows that ${\cal T}(Q_E=0;M_f)=g^2(M^2)$, and that at large
$Q_E^2$ ($Q_E^2\gg M_f^2$, $M\simle M_f$), ${\cal T}(Q_E;M_f)\approx 2\pi/\ln(Q_E^2/M_f^2)\approx
g^2(Q_E^2)$. This  logarithmic decrease of the
interaction strength at large $Q_E^2$ reflects of course the property of 
asymptotic freedom. 

As explained in App. A  (see
Eq.~(\ref{Dinverse}) and the discussion that follows), the 
$\sigma$ propagator develops a pole at finite $Q_E$ when $M< M_f/e$ and, according 
to  Eq.~(\ref{scatteringamplitude01}), so does 
the quark-quark scattering amplitude. This pole which does not correspond to a physical excitation of
the system is commonly referred to as a Landau pole; as we shall see, it is
responsible for several difficulties in our next-to-leading order calculation at finite temperature.
We shall denote by $M_*$ the value of $M$ at which the Landau pole appears at $Q_E=0$. Here $M_*=M_f/e$. As we
have seen before, at $Q_E=0$ the scattering amplitude 
is simply ${\cal T}(Q_E=0;M)= g^2(M^2)$ which diverges when $M\to M_*$ (see Eq.~ (\ref{running_g})). Since
${\cal T}(Q_E=0;M)$ is nothing but the inverse of the second derivative of the leading order effective potential $V^{(0)}$
(see Eqs.~(\ref{V0andD}) and (\ref{scatteringamplitude01}), and 
also  (\ref{Dmoins1})), $M_*$ is also the zero of the second derivative of (\ref{URindepM0}).
At zero temperature, $M_*$ is far from the physical quark mass $M_f$, and the quantum fluctuations of the
$\sigma$ field do not probe values of $M$ close to $M_*$; the Landau pole is then harmless. We shall see that this is
no longer the case at finite temperature.

\subsection{\label{sec:T0NLO}The
$1/N$ contribution}

We consider now the next-to-leading order contributions to the effective
potential.
There are two such contributions that we call $V^{(1)}_f$ and
$V^{(1)}_b$. The latter is the bosonic
contribution, i.e.,
the last term in Eq.~(\ref{Vdesigmac}). In terms of bare quantities, 
this
reads:
\beq\label{V1/Nbare}
V^{(1)}_{b,B} (M_B;g_B,\Lambda) = \frac{1}{2} \int^\Lambda
\frac{d^2q}{(2\pi)^2} \ln
\left[1+g_B^2\Pi_0(Q_E;M_B,\Lambda)\right],
\eeq
where the cut-off
$\Lambda$ applies to the
magnitude
$\sqrt{Q_E^2}$ of the Euclidean momentum, and 
$\Pi_0(Q_E;M_B,\Lambda)$
is given in App. A. The other
contribution,
$V^{(1)}_f$,
originates from the renormalization of the
coupling constant $g_B$ and the mass
$M_B$, at next-to-leading order, in the fermionic contribution, Eq.
(\ref{potUL}).  It will be examined
later in this section.

  We first address the problem of finding
a constant
$C(g_B,\Lambda)$ to eliminate the $M$-independent divergences of 
$V^{(1)}_{b,B}$.
  We
define:
\beq\label{CdeLambda}
C(g_B,\Lambda)=\frac{1}{2} \int^\Lambda \frac{d^2q}{(2\pi)^2}
\ln
\left[1+g_B^2\Pi_0(Q_E;M_\Lambda,\Lambda)\right]
 -\frac{M_\Lambda^2}{4\pi}\ln\frac{\Lambda^2}{M_\Lambda^2}\,
 -\frac{M_\Lambda^2}{4\pi}
\ln\ln\frac{\Lambda}{M_\Lambda} .
\eeq
with $M_\Lambda \equiv \Lambda {\rm e}^{-\pi/g_B^2}$, and we verify 
now that this subtraction fulfills our
requirements.  First, we express
$
V^{(1)}_{b,B} (M_B;g_B,\Lambda)-C(g_B,\Lambda)$ in terms of the 
renormalized mass and coupling
constant
according to Eqs.~(\ref{renormalizations}).
Note that since  $V^{(1)}_{b}$ is already of order
$1/N$, we need
only the leading order renormalization constants. That is,
we can set  $M_B=M$ and
$g^2_B=g^2/Z^{(0)}$, with
$Z^{(0)}$ given
by Eq.~(\ref{ZMf}) with $\bar Z^{(0)}=0$. Furthermore, we can also
replace
$M_\Lambda = \Lambda e^{-\pi/g_B^2}$ by $M_f$, using the relation 
(\ref{minimbare}) and the
identification at
leading order of
$M_{min}^{(0)}$ with $M_f$.
Thus, at this stage:
\beq\label{V1TzeroR}
V^{(1)}_b (M)= \frac{1}{2}
\int^\Lambda
\frac{d^2 q}{(2\pi)^2}\, \ln \left[
\frac{D^{-1}_0(Q_E;M)}{D^{-1}_0(Q_E;M_f)}
\right]
+\frac{M_f^2}{4\pi}\ln\frac{\Lambda^2}{M_f^2}\,
+
\frac{M_f^2}{4\pi}
\ln\ln\frac{\Lambda}{M_f} .
\eeq
where
$D^{-1}_0(Q_E;M)=Z^{(0)}+g^2\Pi_0(Q_E;M,\Lambda)$ is the
renormalized inverse $\sigma$ propagator given by Eq.~(\ref{Dinverse}).

The integrand in Eq.~(\ref{V1TzeroR})
vanishes   when $Q_E^2 \to \infty$, but not
fast enough to make the 
integral convergent as $\Lambda\to\infty$.
Indeed, from (\ref{logqdsD}) one
gets
\beq\label{devellog}
\frac{D^{-1}_0(Q_E;M)}{D^{-1}_0(Q_E;M_f)}\simeq
1+   \frac{2 {\cal A} (M)}{Q_E^2}
+\frac{2 {\cal B} (M)}{Q_E^2 \ln (Q_E^2/M_f^2)}
  + {\cal O} \left( \frac{M^4}{Q_E^4\ln (Q_E^2/M_f^2)} \right),\eeq
where:
\beq\label{divAB}
{\cal A} (M)=M^2-M_f^2 \, , \qquad  {\cal B} (M)=
M^2-M_f^2-M^2\ln(M^2/M_f^2) \, .
\eeq
The terms proportional to ${\cal A} $ and ${\cal B}$ yield the
following
contributions to the integral:
\beq\label{divM}
\frac{{\cal A}
(M)}{4\pi}\ln\frac{\Lambda^2}{M_f^2}\, ,\qquad\qquad
\frac{{\cal B}
(M)}{4\pi}\ln\ln\frac{\Lambda}{M_f},
\eeq
which contain $M$-dependent but also
$M$-independent
divergences. The latter
cancel exactly the last two terms of Eq.~(\ref{V1TzeroR}). The former
 contribute
$M$-dependent
terms to  Eq.~(\ref{V1TzeroR})
  which are of the same form as those of  the 
leading
order
  potential, $V^{(0)}$,  namely  $M^2$  and $M^2\ln M^2$ (see Eq. 
(\ref{potUR})), and they
 can  be
eliminated by an appropriate adjustment of the 
two renormalization constants $Z$
and
$Z'$. To see that,
we need to return to the  leading order contribution 
to the bare potential, (Eq. (\ref{potUL})),
and trade
then $M_B$ and $g_B$ for $M$ and $g$, using Eqs. 
(\ref{renormalizations}). We get:
\beq\label{Urenorm}
 \frac{M^2Z}{g^2} + \frac{M^2Z'}{4\pi}\left(\ln\left(\frac{M^2Z'}{\Lambda^2}
\right)-1\right) .
\eeq
By
expanding $Z$
and
$Z'$ up to order $1/N$ as in Eq.~(\ref{Z01}), one can write the above
expression
(\ref{Urenorm}) as $
  V^{(0)}+ ({1}/{N})V^{(1)}_f$, with
  $V^{(0)}$ given by Eq.
(\ref{renormU}) and, dropping terms of order $1/N^2$,
\beq\label{V10R}
V^{(1)}_f(M) =
M^2
\left(\frac{Z^{(1)}}{2g^2}+\frac{Z'^{(1)}}{4\pi}
\ln\left(\frac{M^2}{\Lambda^2}\right)\right) .
\eeq
As
mentioned at the beginning of this section,
$V^{(1)}_f(M)$ (from the equation above)
adds to $V^{(1)}_b(M)$
(from Eq.~(\ref{V1TzeroR})) to yield the complete
next-to-leading order
contribution to the renormalized
potential, 
$V^{(1)}(M)=V^{(1)}_f(M)+V^{(1)}_b(M)$.  The requirement that
$\Lambda$ disappears from
$V^{(1)}(M)$ then allows us to determine 
the $\Lambda$-dependence of
the renormalization  constants
$Z^{(1)}$ and $Z'^{(1)}$. One gets:
\beq\label{zeta}
Z^{(1)} =
\frac{g^2}{2\pi}\left[(Z'^{(1)}-1)
\ln
\left(\frac{\Lambda^2}{M_f^2}\right)-\ln\ln
\frac{\Lambda}{M_f}\right]
  + \bar
Z^{(1)}
\eeq
\beq\label{delta}
Z'^{(1)}=\ln\ln\left(\frac{\Lambda}{M_f}\right)+\bar
Z'^{(1)},
\eeq
where $\bar Z'^{(1)}$ and $\bar Z^{(1)}$ are finite constants to be fixed using
the
renormalization conditions
(\ref{renormalisation_conditions}). We postpone this determination till later.
Keeping
along $\bar Z'^{(1)}$ and $\bar Z^{(1)}$, one can then write  the  renormalized effective
potential
in the form:
\beq\label{renoV1}
V^{(1)} (M) =
\bar Z^{(1)}\frac{M^2}{2g^2}  + \bar Z'^{(1)}
 \frac{M^2}{4\pi} \ln \frac{M^2}{M_f^2}
-\frac{1}{2\pi} (M^2 - M_f^2) + \frac{M_f^2}{4\pi} F_0(x_M),
\eeq
with
$x_M=M^2/M_f^2  $ (see Eq.~(\ref{xMdef0})), and 
$F_0(x)$ is a finite function defined in
App. B.

The  renormalization conditions (\ref{renormalisation_conditions}) needed
to
  fix the
finite constants ${\bar Z}^{(1)}$ and ${\bar Z}'^{(1)}$  require  two
derivatives
of $V^{(1)}$.
The first
derivative  of $V^{(1)}$ with respect to $M^2$ is
easily obtained from
(\ref{renoV1}):
\beq\label{deriva}
\frac{dV^{(1)}}{dM^2}=\frac{\bar
Z^{(1)}}{2g^2} -
\frac{1}{2\pi}
+\frac{\bar Z'^{(1)}}{4\pi}(\ln x_M + 1)
+\frac{1}{4\pi}F_0'(x_M)
\eeq
where $F_0'={\rm
d}F_0/{\rm d}x$. The second derivative of $V^{(1)}$ may
be calculated as
\beq
\frac{d^2 V^{(1)}}{dM^2}= 2\frac{dV^{(1)}}{dM^2}+4x_M\frac{d}
{dx_M}\left(\frac{dV^{(1)}}{dM^2}\right).
\eeq
One thus easily gets:
\beq\label{deriv2}
\frac{d^2 V^{(1)}}{dM^2}
=
\frac{\bar Z^{(1)}}{g^2} + \frac{\bar Z'^{(1)}}{2\pi} (\ln x_M + 3)
-\frac{1}{\pi}
+ \frac{1}{2\pi}
F_0'(x_M) + \frac{x_M}{\pi} F_0''(x_M),
\eeq
where $F_0''={\rm d}^2F_0/{\rm d}x^2$. The derivatives of
$F_0(x)$ 
are given in App. B.

In this subsection we shall in fact only fix $\bar Z^{(1)}$ and
leave 
$\bar Z'^{(1)}$ undetermined. The reason
is that, as we shall see, it is not necessary to
specify
the value of $\bar Z'^{(1)}$ in order to
get a renormalization group
invariant expression for the effective
potential.
By imposing the second renormalization condition
(\ref{renormalisation_conditions}), viz 
$d^2V^{(1)}/dM^2=0$ at $M=M_0$,
we can express $\bar Z^{(1)}$ in terms of $\bar Z'^{(1)}$. Using
Eq.~(\ref{deriv2}), one gets:
\beq\label{Zbar}   
\bar Z^{(1)}= \frac{g^2}{2\pi} \left( 2- \bar
Z'^{(1)}\left(3+\ln x_0
\right)-   F'_0(x_0) -2 x_0 F''_0(x_0) \right) ,
\eeq
where the quantity $x_0$ is
the following function of $g$:
\beq\label{defx0}
x_0 \equiv {\rm e}^{2(\pi/g^2-1)} .
\eeq
At leading
order $x_0= {M_0^2}/{M_f^2}$, as can be easily deduced 
from Eq.~(\ref{running_g}). We can then write
the
next-to-leading order contribution to the renormalized potential  as
follows:
\beq\label{VcompR}
V^{(1)}(M) =  \frac{M^2}{4\pi} \left[
\bar Z'^{(1)} \left( \ln
\frac{M^2}{M_f^2}\!-\! 
3\!-\! \ln x_0\right)\! 
 - \! F'_0(x_0) \!-\!2 x_0 F''_0(x_0) \right]
\!+\!
\frac{M_f^2}{2\pi} + \frac{M_f^2}{4\pi} F_0(x_M).  
\eeq
The full renormalized potential up to 
next-to-leading order is thus $V=NV^{(0)}+V^{(1)}$ with $V^{(0)}$ and $V^{(1)}$
given by Eqs. (\ref{renormU}) and
(\ref{VcompR}), respectively.

The effective potential that we just constructed has an apparent
dependence on the scale $M_0$. However, as we did in the leading
order case, it is possible to exploit the
relation between the
  fermion mass $M_f$ and
the value $M_{min}$ of $M$ at the minimum of the effective
potential, 
in  order to get rid of this apparent
$M_0$ dependence. The situation is somewhat more
complicated here 
because,  as explained in subsection~\ref{Sec:gap_eqn},  the fermion mass
receives two 
contributions at order $1/N$,
$M_{min}$ and
$M_\Sigma$.
The latter is calculated in the
App. C and takes the form:
\beq\label{deltam}
M_\Sigma=M_f\left( \frac{\bar
Z'^{(1)}}{2N}+\frac{\varphi}{N}\right),
\eeq
where $\varphi$ is a numerical constant,  $\varphi=-0.126229$
(see Eq.~(\ref{varphiidef}))
. The value of
$M_{min}$ can be calculated by setting
$M_{min}=M_{min}^{(0)}+({1}/{N})M_{min}^{(1)}$ (see 
Eq.~(\ref{Mexpanded})), with
$M_{min}^{(0)}$
and
$M_{min}^{(1)}$ given by Eqs.~(\ref{Mmin0}) and (\ref{Mmin1gen}), respectively.
The required derivative
of $V^{(1)}$ can be obtained from Eq.~(\ref{deriva})
(together with (\ref{Zbar})). 
Note also that
$M_{min}^{(0)} = M_f + {\cal O}(1/N)$ and that, at the minimum, 
${\rm d}^2V^{(0)}/{\rm d}M^2=1/\pi$. One thus
gets:
\beq\label{Mmin1}
M_{min}^{(1)}=M_{min}^{(0)}\left(
\left(\frac{1}{2}\ln x_0 +1\right) \bar
Z'^{(1)}
+\frac{1}{2} F'_0(x_0) + x_0
F''_0(x_0) -\frac{1}{2} F'_0(1)
\right).
\eeq
Remembering that
$M_f=M_{min}+M_\Sigma,$ (see Eq.~(\ref{totalM})), and combining
Eqs.~(\ref{deltam}) and (\ref{Mmin1}), one can
therefore  write $M_f$ as:
\beq\label{new_mass}
M_f=M_0 \,\frac{1}{\sqrt x_0}\left[ 1+\frac{1}{N}
\left(
\frac{\bar Z'^{(1)}}{2} ( \ln x_0 +3)
\!+\!\frac{1}{2}  F'_0(x_0) + x_0
F''_0(x_0) \!-\!\frac{1}{2}
F'_0(1)
\!+\!\varphi\right)\right].
\eeq
For the purpose of
eliminating $M_0$, it is convenient to rewrite  Eq.~(\ref{new_mass})
as follows 
(neglecting terms of order
$1/N^2$):
\beq\label{logmass}
\ln \frac {M_f}{M_0} = -\frac{1}{2} \ln x_0
+\frac{1}{N} \left(
\frac{\bar Z'^{(1)}}{2} (\ln x_0 +3)
+\frac{1}{2}  F'_0(x_0) + x_0
F''_0(x_0)
-\frac{1}{2} F'_0(1)
+\varphi\right) .
\eeq
We shall return to this equation in subsection~\ref{sec:RGanal} below. For the moment we 
note that it allows us to easily eliminate the term $\sim \ln M_0$ in the leading
order potential
$V^{(0)}(M)$ given in Eq.~(\ref{renormU}). Combining the resulting
expression with Eq.~(\ref{VcompR}) and
droping again terms of order
$1/N^2$ (in particular, using $x_0=M_0^2/M_f^2+O(1/N)$), one finally
gets
\beq
V(M) &=& N V^{(0)}+ V^{(1)}
\nonumber \\
&=&\frac{NM^2}{4\pi}\left(-1+\ln\frac{M^2}{M_f^2}\right)
+\frac{M^2}{4\pi}\left(\xi+\bar Z'^{(1)}\ln\frac{M^2}{M_f^2}\right)
+\frac{M_f^2}{2\pi} + \frac{M_f^2}{4\pi} F_0(x_M),
\eeq
where $\xi\equiv 2\varphi-F'_0(1)=-0.8661636$. At this
point we observe
that
\beq\label{xm3}
\left(1+\frac{\bar
Z'^{(1)}}{N}\right)\ln\frac{M^2}{M_f^2}-1=\left(1+
\frac{\bar
Z'^{(1)}}{N}\right)
\left[\ln\left(\frac{M^2}{M_f^2}\left(1+\frac{\bar Z'^{(1)}}{N}\right)
\right)-1\right]
+{\cal O}(1/N^2).
\eeq
This allows us to express the full effective potential $V$ entirely in
terms of the redefined variable $x_M$:
\beq\label{xm2}
x_M\equiv \frac{M^2}{M^2_f}\left(1+\frac{\bar
Z'^{(1)}}{N}\right).
\eeq
The final result for the effective potential can then be put in the form (with a
slight abuse
in the notation):
\beq\label{renormVdexM}
V(x_M)   = \frac{M_f^2}{4\pi}\left[Nx_M (\ln x_M
-1)+\xi (x_M-1)
+F_0(x_M)\right].
\eeq
where we have explicitly  subtracted the constant
$2+\xi$ so
that $V^{(1)}(x_M=1) = 0$. 
Note  that the definition of $x_M$
 in Eq.~(\ref{xm2}) differs from that in
Eq.~(\ref{xMdef0})  by a term  of order
$1/N$. There is no contradiction however since $F_0(x)$ is
already of
order $1/N$,  so that using $x_M=M^2/M_f^2$ instead of the 
definition (\ref{xm2}) inside $F_0(x_M)$ is equivalent to
ignoring
terms of order
$1/N^2$. 
We shall see in the
next subsection that $x_M$ is  a renormalization group
invariant. The
effective potential $V(x_M)$ is thus explicitly independent of the scale
$M_0$, and for the
reasons discussed at the beginning of this section (after
Eq.~(\ref{renormM0})), also independent of
$g$.
Note that in terms of the redefined variable
$x_M$ (Eq.~(\ref{xm2})), the leading order  contribution in
Eq.~(\ref{renormVdexM}) is
identical to that of
Eq.~(\ref{URindepM0}).
\begin{figure}
\includegraphics[angle=180,width=13cm]{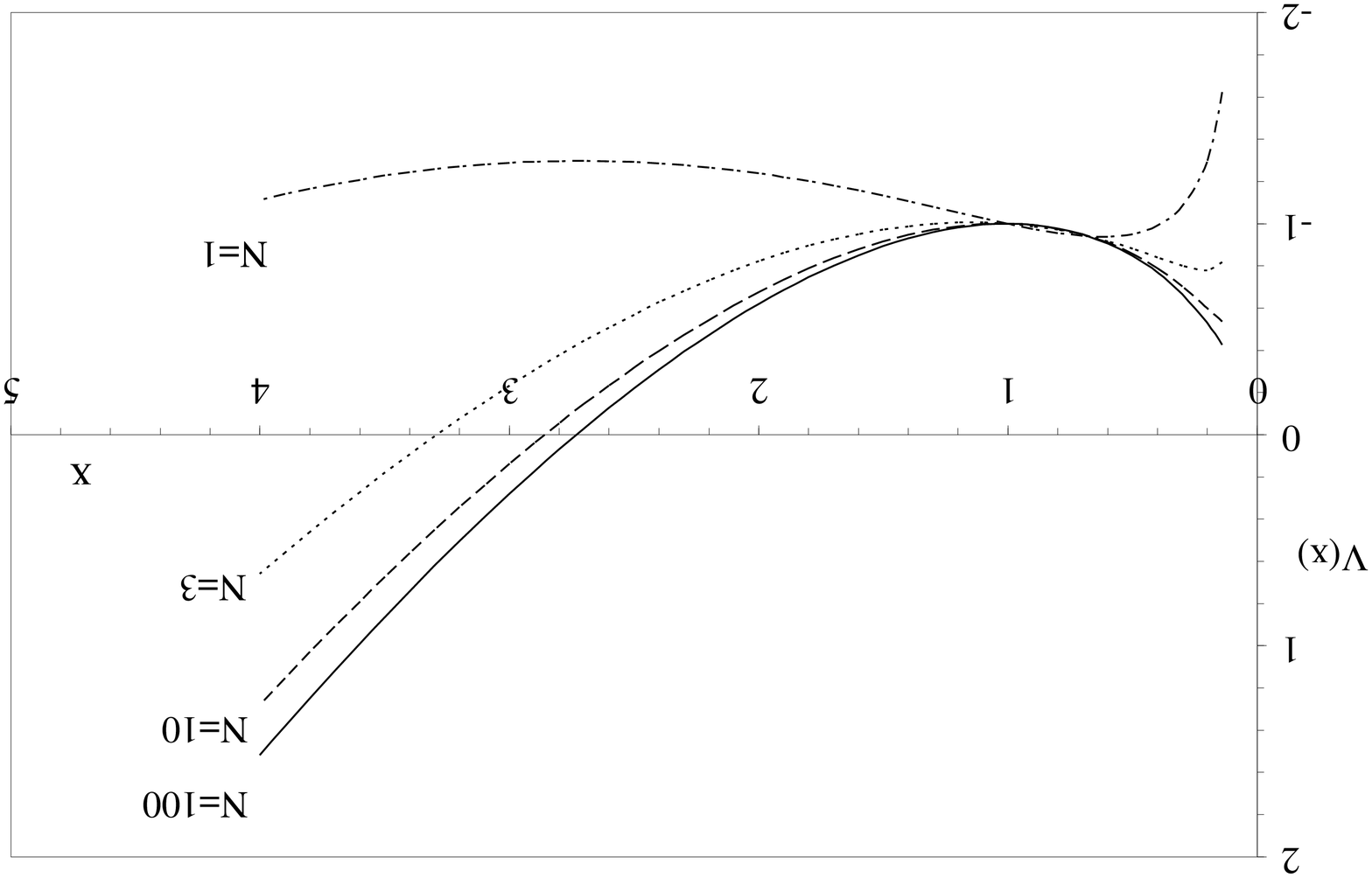}
\caption{\label{fig:potentiel_eff}The effective potential divided by
$NM_f^2/4\pi$ as a function of $x_M$ for the
values $N=1$, $N=3$, $N=10$ and $N=100$. The minimum occurs at
$x_{min}=2.721$ for $N=1$, $x_{min}=1.145$ for $N=3$, $x_{min}=1.068$
for $N=5$, $x_{min}=1.029$
for $N=10$, and  $x_{min}\approx 1-2\varphi/N \approx 1+0.25/N$ for
larger values of $N$. The  $1/N$ correction has been adjusted so as to
vanish at $x_M=1$; it also changes sign at this point.}
\end{figure}

The effective potential (\ref{renormVdexM}) is displayed in
Fig.~\ref{fig:potentiel_eff}
for various values of $N$. One sees that the $1/N$ correction
becomes
rapidly a small correction. For $N=1$, the approximation is of course
meaningless, but already
for
$N=3$, the shape of the potential is qualitatively the same as at leading
order.  The variation of
the
minimum with $N$ may be obtained as
explained in Sect.~\ref{sec:generalities} from the general equation (\ref{Mmin1gen}):
\beq\label{xMaccurate1surN}
x_{min}=1+\frac{x_{min}^{(1)}}{N}\qquad\qquad
x_{min}^{(1)}=-\frac{ 
\left.  \frac{ {\rm d}V^{(1)} }{ {\rm
d}x_M }
\right|_{x_M=1}   } {\left. \frac{
{\rm d}^2V^{(0)} }{ {\rm d}x_M^2 
}\right|_{x_M=1} }.
\eeq
 From the explicit
expression
(\ref{renormVdexM}) we get $\left.{ {\rm d}^2V^{(0)} }/{ 
{\rm d}x_M^2 }\right|_1=M_f^2/4\pi$, and
$\left.{
{\rm d}V^{(1)}}/{ {\rm d}x_M }\right|_1 = 
(M_f^2/4\pi)(\xi+F'_0(1))=\varphi M_f^2/2\pi$, so that,
at next-to-leading order,
$x_{min} = 1-2\varphi/N\approx 1+0.25/N$. The difference between this
value and that of the
``exact''  minimum found numerically from 
Eq.~(\ref{renormVdexM}) can be attributed to terms of order $1/N^2$.
As can be deduced
from the caption of Fig.~\ref{fig:potentiel_eff} for $N\simge 10$ these are
small,
confirming the previous
study of Ref.~\cite{Root:1974zr}. (Of course, such considerations say very
little about the
 magnitude of the true $1/N^2$ corrections.) It is interesting to recover the 
previous
result for
$x_{min}$ in a different way. Using the definition of $x_M$ in Eq.~(\ref{xm2}), we
can
write  
$M_{min}=\sqrt{x_{min}}\, M_f(1-\bar Z'^{(1)}/2N)$. The equation $M_f=M_{min}+M_\Sigma$
reads
then:
\beq
\label{eq89}
M_f=M_f\left[1+\frac{x_{min}^{(1)}}{2N}-\frac{\bar 
Z'^{(1)}}{2N}\right] +M_f\left[\frac{\bar
Z'^{(1)}}{2N}
+\frac{\varphi}{N}\right],
\eeq
where we have used Eq.~(\ref{deltam}) for
$M_\Sigma$ and the first equation in (\ref{xMaccurate1surN}).
Eq. (\ref{eq89}) gives back our previous result 
$x_{min}^{(1)}=-2\varphi$. It also clearly
illustrates how the
scale dependence cancels out between $M_{min}$ and $M_\Sigma$, 
whereas the magnitude of
each
individual contribution depends on $M_0$ (through $\bar Z'^{(1)}$). We come back on this issue in the next
subsection.

Note finally that the potential is plotted 
only for $x_M>1/e^2 \equiv x_{M_*}$. This is because for $x_{M}<x_{M_*}$ a Landau pole appears in the $\sigma$ propagator
(see subsection \ref{sec:T0LO}), making the function $F_0$ complex (see
Eq.~(\ref{V1TzeroR}) or, equivalently Eq.~(\ref{F0})). 
The influence of this Landau pole on the shape of the effective potential is clearly visible in Fig. \ref{fig:potentiel_eff}
on the plot of
$V(x_M)$ for $N=1$: not only does the potential becomes complex for $x_{M}<x_{M_*}$, but the ``$1/N$ correction'' is large
and negative for $x_M \simge x_{M_*}$. As soon as $N>3$, however, this peculiar behavior is no longer visible.

\subsection{\label{sec:RGanal}Renormalisation group
analysis}

In this subsection, we shall clarify the behavior under renormalization 
group
transformations of various quantities
that have been introduced in the previous subsections. In particular
we shall verify that the expression (\ref{new_mass}) of the fermion mass is invariant, and so
is the variable
$x_M$ defined in Eq.~(\ref{xm2}). But before we do  that, let
us verify that our calculation
reproduces known results for the renormalization group functions.

The
functions $\beta$  and $\gamma$ are most
easily obtained by differentiating the renormalization constants
with
respect to the cut-off $\Lambda$, according to Eq.~(\ref{betabare}).
Using Eqs.~(\ref{ZMf}),
(\ref{zeta}) and (\ref{delta}) one easily
gets:
\beq\label{expandbeta}
\beta(g_B)=\beta^{(0)}(g_B)+\frac{1}{N}\,\beta^{(1)}(g_B),
\eeq
with
\beq\label{beta0}
\beta^{(0)}(g_B)
=-\frac{g_B^3}{2\pi}, \qquad
\beta^{(1)}(g_B)=\frac{g_B^3}{2\pi} + \frac{g_B^5}{4\pi^2}.
\eeq
Eqs.
(\ref{expandbeta}) and (\ref{beta0})  show that,
for $N=1$, the leading term in a small $g$-expansion (i.e.,
that of
order $g^3$) in the $\beta$-function vanishes. This is in agreement with
earlier calculations and the
fact
that for $N=1$ the Gross-Neveu model reduces
to the Thirring model \cite{Thirring:in},
for which the $\beta$-function
vanishes identically. (The term of order $g^5$ does not vanish when
$N=1$ in Eq. (\ref{beta0})
because the $1/N$ correction does not contain all the terms of order $g^5$ in
the small $g$ limit). The
two-loop $\beta$-function has been 
calculated by Wetzel
\cite{Wetzel:1984nw} and
reads
$\beta(g)=-(1-1/N)(g^3/2\pi)+(1/N-1/N^2)(g^5/4\pi^2 )$). Note finally that $\beta^{(0)}(g_B)$
 follows
immediately from the explicit relation (\ref{running_gB}) between $g_B$ and the
cut-off $\Lambda$.

Following similar steps, one obtains the
function $\gamma\equiv \gamma^{(0)}+(1/N)\gamma^{(1)} $. Since at
leading order, $Z'=1$, $\gamma^{(0)}(g_B)=0$.
 Using Eq. (\ref{delta}) and also the
relation
(\ref{running_gB}) between the cut-off $\Lambda$ and the bare coupling
constant we
get:
\beq\label{gammabare}
\gamma^{(1)}(g_B)=-\frac{g_B^2}{2\pi}.
\eeq

It is instructive to
repeat the calculation of the renormalization 
group functions  by differentiating  with
respect to $M_0$.
The calculation is more involved because  $M_0$ 
enters the finite  parts of the renormalization constants.
The
calculation of $\beta^{(0)}(g)$ is trivial however and reproduces 
Eq.~(\ref{beta0}),
i.e.,
$\beta^{(0)}(g)=-g^3/2\pi$. Note that it also follows immediately from the
explicit relation between the
renormalized coupling and the scale $M_0$, Eq.~(\ref{running_g}). Using Eqs.~(\ref{zeta}), (\ref{delta}) 
and (\ref{Zbar}) we get for
$\beta^{(1)}(g)$:
\beq\label{beta1surN}
\beta^{(1)}(g)=-\frac{g^3}{2\pi}
\left[3x_0F''_0(x_0)+2x_0^2F'''_0(x_0)+\bar
Z'^{(1)}+
(\ln x_0 +3) x_0 \frac{\del \bar Z'^{(1)}}{\del x_0}
\right],
\eeq
where we used
\beq\label{derx0g}
\beta^{(0)}(g)\frac{\del }{\del g}= 2
  x_0 \,\, \frac{\del }{\del x_0}
\eeq
which follows from (\ref{defx0}). 
The calculation of $\gamma^{(1)}$ can be done similarly,
using  $Z'=1+Z'^{(1)}/N$:
\beq\label{gamma11}
\gamma^{(1)}(g)\equiv \frac{1}{2}\frac{{\rm d} Z'^{(1)}}{{\rm
d}\ln
M_0}=\frac{{\rm d} \bar Z'^{(1)}}{{\rm d}\ln x_0}.
\eeq

To complete these calculations, we now need 
to specify $\bar Z'^{(1)}$. This is done with the help of the first of the  renormalization
conditions (\ref{renormalisation_conditions}), which has not been used so far.
Using Eq.~(\ref{deriva}) for the first  derivative of
$V^{(1)}$, together with the expression (\ref{Zbar}) of $\bar Z^{(1)}$, we
get:
\beq\label{barzeta1}
\bar Z'^{(1)}=-x_0 F''_0(x_0).
\eeq
Eq.~(\ref{beta1surN}) becomes then:
\beq\label{beta1fix}
\beta^{(0)}(g)=-\frac{g^3}{2\pi},\qquad\qquad
\beta^{(1)}(g)=\frac{g^3}{2\pi}(\ln
x_0 +1)
\left( x_0F''_0(x_0) +x_0^2F'''_0(x_0)\right) ,
\eeq
where we have written also the leading order
contribution for future reference.
 In the small
$g$ limit, $x_0\sim 2\pi/g^2 \to \infty$, and from the
results quoted in App. B, we get
$(\ln x_0 +1) ( x_0F''_0(x_0) +x_0^2F'''_0(x_0)) \sim 1+g^2/2\pi$.
Thus,
the $\beta$-function (\ref{beta1fix}) coincides at small $g$ with
that obtained earlier,  Eq.~(\ref{beta0}), in
agreement with the usual
expectation that the first two terms in the expansion of the $\beta$-function
are
independent of the renormalization scheme provided the mapping between
the coupling constants corresponding to
the various schemes is analytic \cite{Gross:vu}. The
$\beta$-function given by (\ref{beta1fix}) is negative at small
$g$, reflecting  asymptotic  freedom;
this property   is 
not affected by the $1/N$ corrections:  $(\ln x_0 +1) \left(
x_0F''_0(x_0)
+x_0^2F'''_0(x_0)\right) <2$, so  
$\beta(g)$ remains negative as soon as  $N> 2$.

Finally, the function $\gamma^{(1)}$ is calculated by using
Eqs.~(\ref{gamma11}) and
(\ref{barzeta1}):
\beq\label{gamma1}
\gamma^{(1)}(g) = x_0 \, \frac{d\bar Z'^{(1)}}{d x_0}
=-
\left( x_0 F''_0(x_0) + x_0^2 F'''_0(x_0) \right)   .
\eeq
In the small $g$ limit, $x_0F''_0(x_0)
+x_0^2F'''_0(x_0) \sim 1/\ln x_0
\sim g^2/(2\pi)$. Thus, when $g$ is small this expression of
$\gamma^{(1)}$
coincides with that of Eq.~(\ref{gammabare}).

One may now verify explicitly that, as claimed in the
previous subsection, the expressions (\ref{new_mass}) 
of the fermion mass $M_f$ and (\ref{xm2}) of $x_M$ are renormalization
group invariant. Consider first $M_f$. In this case, the verification can be done by
a direct calculation, but it is more instructive to
proceed by considering first
the  variation of $M_{min}=M_{\rm min}^{(0)}+(1/N)M_{\rm
min}^{(1)}$, with $M_{\rm min}^{(0)}$ given by
Eq.~(\ref{Mmin0}) and 
$M_{\rm min}^{(1)}$ given  by Eq.
(\ref{Mmin1}). 
One finds:
\beq\label{variationdeMmin}
M_0\frac{{\rm d}M_{min}}{{\rm d}M_0}=-\frac{1}{N} M_{min}^{(0)}
  \,
x_0 \, \frac{d\bar Z'^{(1)}}{d x_0}.
\eeq
 Next, from Eq.~(\ref{deltam}) we get: 
\beq
M_0\frac{{\rm d}M_\Sigma}{{\rm d}M_0}=\frac{1}{N} M_f
 \, x_0 \, \frac{d\bar Z'^{(1)}}{d x_0}.
\eeq
 Note that we can replace $M_{min}^{(0)}$ by $M_f$ in
Eq.~(\ref{variationdeMmin}). It is then obvious that the  variation of
$M_{\rm min}$
  exactly compensates  that of $M_\Sigma$, showing that  $M_f=M_{min}+M_\Sigma$ is indeed independent of
$M_0$.

The relation (\ref{variationdeMmin}) can be viewed, quite 
generally, as a direct
consequence
of the invariance of the effective potential under renormalization group
transformations. Indeed,
consider such a transformation, in which
$M_0\to M_0'$, $M\to M'$, $g\to g'$ and 
$V(M';M_0',g')=V(M;M_0,g)$.
The relation between $M$ and $M'$ is determined, for an infinitesimal
variation 
${\rm d}M_0$,  by the renormalization group equation
(\ref{renormM0}). The same transformation relates the
values of $M$ at the
minimum before and after the transformation, that
is,
\beq
\frac{M_0}{M_{min}}\frac{{\rm d}M_{min}}{{\rm d}M_0}=-\gamma(g)  
\eeq
which, when expanded to
order $1/N$, is  
Eq.~(\ref{variationdeMmin}) (given the relation (\ref{gamma11}) for $\gamma$).

Consider now $x_M$. From the definition (\ref{xm2}) one gets
\beq
\frac{{\rm d}  x_M}{{\rm
d}\ln
M_0}=-2\frac{M^2}{M_f^2}\gamma+\frac{1}{N}\frac{M^2}{M_f^2}
2 x_0 \, \frac{d\bar Z'^{(1)}}{d x_0} \, ,
\eeq
which, according to Eq.~(\ref{gamma11}) vanishes at order $1/N$.
This may also be seen,
perhaps more directly, by writing
the relations between $M_B$ and $M$ as follows:
\beq
M_B=
\sqrt{Z'}M&=&\left(1+\frac{1}{2N}\ln\ln\frac{\Lambda}{M_f}+\frac{1}{2N} \bar
Z'^{(1)}\right)M\nonumber\\
&=&\left(  1+\frac{1}{2N}\ln\ln\frac{\Lambda}{M_f}\right)\left(
1+\frac{1}{2N} \bar Z'^{(1)}\right)M +{\cal
O}(1/N^2).
\eeq
This equation shows  that, up to terms of
order $1/N^2$, the quantity 
$\left(  1+ \bar
Z'^{(1)}/2N \right)M=M_f\sqrt{x_M}$ remains constant when $M_0$
varies, with $M_B$ 
and $\Lambda$ kept fixed.

\begin{figure}
\includegraphics[angle=90,width=13cm]{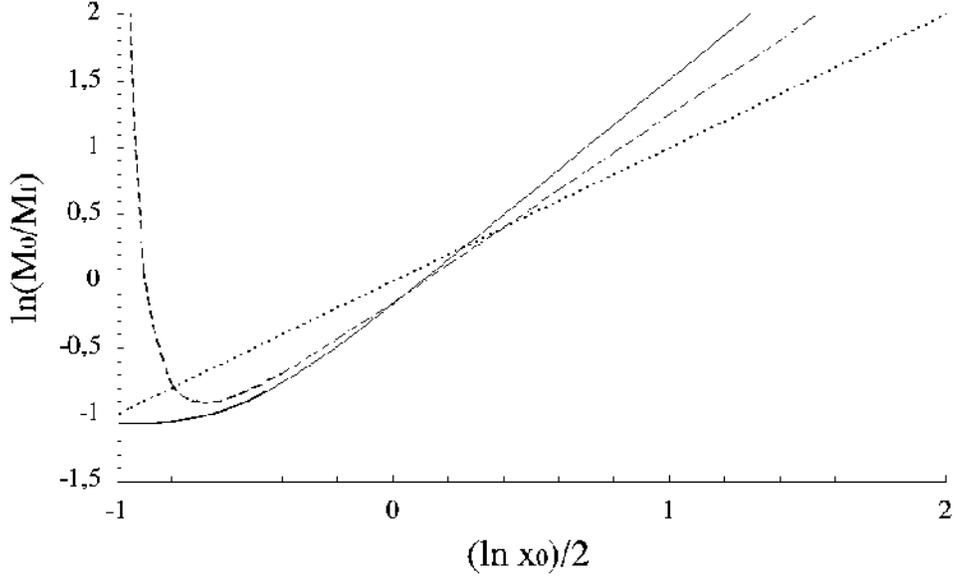}
\caption{The quantity $\ln(M_0/M_f)$ as a function of $\, \pi/g^2-1=(\ln x_0)/2 $, for N=3.
Full line: numerical integration;
dashed line: leading order relation (from Eq. (\ref{running_g}));
long dashed line: approximate integration according to equation
(\ref{relM0etg}).\label{fig:modeg}}
\end{figure}

Before closing this subsection, let us 
discuss the relation between the renormalized
coupling constant $g$ and the scale $M_0$. At leading order, this
is given by
Eq.~(\ref{running_g}).  To go beyond leading order, we use the relation  
${\rm d}g=\beta(g) {\rm
d}\ln M_0$ to write:
\beq\label{M0deg1}
\ln{ M_0(g)\over
M_0(g_1)}=\int_{g_1}^g
{dg'\over\beta(g')}.
\eeq
The result of the numerical integration is
plotted in Fig.~\ref{fig:modeg}, for the choice
$g_1=\sqrt \pi$. For this value of $g_1$, $M_0(g_1)=M_f$
in leading order; at order $1/N$, using Eq.~(\ref{new_mass}), we get
\beq
M_0(g_1)=M_f\left(1-\frac{1}{2N}\left(F_0'(1)-F_0''(1)+\xi\right)\right)\approx
M_f(1-\frac{0.48}{N}). 
\eeq
An approximate analytical evaluation can also be obtained  by
expanding the integrand in  (\ref{M0deg1}):
\beq\label{M0deg2}
\ln{ M_0(g)\over M_0(g_1)}\approx
\int_{g_1}^g
{dg'\over\beta^{(0)}(g')}\left(1-{1\over N}{\beta^{(1)}\over\beta^{(0)}}
\right).
\eeq
 By
using Eq.~(\ref{derx0g}), and $x_0(g_1)=1$, one then gets:
\beq\label{relM0etg}
M_0(g)=M_f
\exp\left\{
\frac{1}{2}\ln x_0-\frac{1}{N}\frac{1}{2}
\left( F'_0(x_0)-x_0(\ln x_0
+1)F''_0(x_0)+\xi\right)
\right\}.
\eeq
This expression is identical to Eq.~(\ref{logmass}) that has
been  used to eliminate the scale dependence in the effective potential. In contrast to the linearized
version, Eq.~(\ref{new_mass}), it holds even in cases where $M_0\gg M_f$, i.e., at weak coupling,
where  terms involving $\ln(M_0/M_f)$ may become of order $N$; such large logarithms do not enter the
$\beta$-function (as we have seen earlier, $\beta^{(1)}$ remains small at weak coupling), but can be
generated by the integration over a sufficiently large range of values of $g$ in Eq.~(\ref{M0deg2}).

The effective potential that we have obtained does not depend on the 
scale $M_0$ (at order $1/N$) , but
the
separation of the mass between a contribution coming from the minimum 
of the potential
$M_{min}\sim\langle
\bar\psi\psi\rangle$ and one coming from the self-energy $\Sigma$ 
depends on the scale
$M_0$. In particular,
since (see Eqs.~(\ref{gamma11}) and (\ref{derx0g}))
\beq
\bar Z'^{(1)}=2\int {\rm
d}g\frac{\gamma^{(1)}(g)}{\beta^{(0)}(g)}\sim \ln\frac{1}{g},
\eeq
$M_\Sigma/M_f $ may become of order unity if $g$ is too small (see Eq. (\ref{deltam})); this, however, occurs 
only for very
small $g$,  such that $g^2\simle {\rm
e}^{-N}$.  

Finally let us mention that while the choice of the scale for the effective potential is not an issue
at this point (since the effective potential does not depend on $g$), 
we shall see that at high
temperature the coupling constant effectively reappears in the calculation,
at a scale determined by the temperature. This will be discussed in subsection~\ref{sec:thermohighT}.

\section{\label{sec:renoreffpot} The renormalized effective potential at finite temperature}
We turn now to the computation of the effective potential at finite 
temperature. At leading order in
the $1/N$ expansion, the effective potential can be split into a zero 
temperature contribution and a finite
temperature one which is free of ultraviolet divergences. Thus, at 
leading order, the technical difficulties
associated with the elimination of divergences are localized in the 
contribution which has been studied in the
previous section. Beyond leading order however, the situation becomes 
more complicated and, at some
stage of the calculation, ultraviolet divergences appear in terms 
which depend on the temperature. As mentioned in the introduction, one expects
on general grounds \cite{LeBellac96,Landsman:uw,Collins84} that such contributions should 
eventually  cancel without the need for counterterms other than those introduced  at zero temperature.
We shall  verify explicitly in this section that this indeed happens.

\subsection{\label{sec:fTLO}The leading order contribution}

At leading order in the $1/N$ expansion, the effective potential is given
by the fermionic contribution, i.e., the first two terms in 
Eq.~(\ref{Vdesigmac}).
As we just mentioned, it can be written as the sum of a ``zero 
temperature'' contribution, $V^{(0)}(M)$, and a
finite temperature
one, $\tilde V^{(0)}(M,T)$, which vanishes as $T\to 0$. We shall 
denote the complete bare potential at leading
order by
$V^{(0)}_B(M_B,T)$, i.e., $V^{(0)}_B(M_B,T)=V^{(0)}_B(M_B)+\tilde 
V^{(0)}_B(M_B,T)$. To isolate the zero
temperature contribution, we rewrite the sum over Matsubara 
frequencies in  Eq.~(\ref{Vdesigmac}) as
\beq\label{Matsupotential}
-T\sum_{n, odd} \ln\left(1+\frac{M_B^2}{\omega_n^2+p^2}\right)=\oint
\frac{{\rm d}\omega}{2\pi i}\, n(\omega)\ln\left(1+\frac{M_B^2}{-\omega^2+p^2}
\right)
\eeq
where
\beq\label{nomega}
n(\omega)=\frac{1}{{\rm e}^{\beta\omega}+1}
\eeq
is the fermionic statistical factor, and the integration contour
is a set of circles surrounding each of the Matsubara frequencies. By 
deforming this contour into two
lines parallel to the imaginary axis, i.e., by writing
$\oint=\int_{-i\infty+\epsilon}^{+i\infty+\epsilon}+ 
\int_{i\infty-\epsilon}^{-i\infty-\epsilon}$, and using
the property $n(-\omega)=1-n(\omega)$,  one can rewrite the integral as
\beq\label{potenfermiocontour}
\int_{-i\infty+\epsilon}^{+i\infty+\epsilon}\frac{{\rm d}\omega}{2\pi i}
(1-2n(\omega)) \ln\left(1+\frac{M_B^2}{-\omega^2+p^2}\right).
\eeq
The term which does not contain the statistical factor gives the zero
temperature contribution  and it has been dealt with in subsection
\ref{sec:T0LO}. The term with the statistical factor is the finite 
temperature contribution
$\tilde V^{(0)}_B(M_B,T)$. It can be calculated by performing a 
further deformation of the
integration contour so as to pick up the singularities on the real 
$\omega$-axis. We get:
\beq\label{V0Tfini}
\tilde V^{(0)}_B(M_B,T)=-\frac{2}{\beta}\int_0^\infty \frac{{\rm d}p}{\pi}
\ln\left(1+{\rm e}^{-\beta E_p} \right).
\eeq
where $E_p=\sqrt{p^2+M_B^2}$.

\begin{figure}
\includegraphics[angle=90,width=13cm]{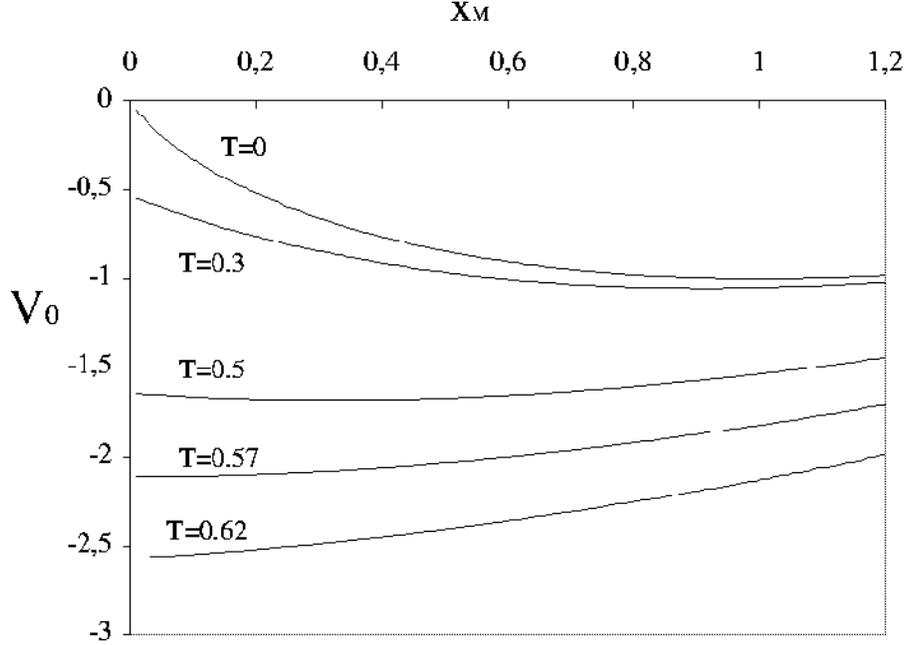}
\caption{\label{fig:fig-potV0_T}The leading order renormalized potential $V^{(0)}$ (divided by $NM_f^2/4\pi$) as
a function of $x_M$,  for various temperatures
$T$ (in units of the fermion mass
$M_f$). The temperature $T=0.57M_f$ is the temperature $T_c$ discussed in the
next section (see Eq. (\ref{Tc})),
and 
$0.62M_f=1.1T_c$.}
\end{figure}

At this order, the renormalized potential is immediately obtained by
substituting $M$ for $M_B$ in the equation above.  By combining  with 
the zero temperature contribution
(\ref{URindepM0}), one obtains  then the  leading order renormalized 
effective potential at finite temperature:
\beq\label{V0TR}
V^{(0)}(M,T)=\frac{M^2}{4\pi}\left(\ln\frac{M^2}{M_f^2}-1\right)-\frac{2}{\beta}\int_0^\infty 
\frac{{\rm
d}p}{\pi}
\ln\left(1+{\rm e}^{-\beta E_p} \right).
\eeq
A noteworthy feature of this expression is that it is analytic around $M=0$,
in contrast to the zero temperature contribution. To see that, consider the
derivative of
$V^{(0)}(M,T)$ with respect to
$M^2$:
\beq\label{deriveepot}
\frac{ \del V^{(0)}(M,T) }{ \del M^2}=\frac{1}{4\pi}\ln\frac{M^2}{M_f^2}+
\int_0^\infty \frac{ {\rm d}p }{ \pi E_p }
\,\frac{1}{{\rm e}^{\beta E_p}+1}.
\eeq
The integral in Eq.~(\ref{deriveepot}) is
infrared divergent when $M\to 0$. By isolating the (logarithmic) 
singularity one gets \cite{Dolan:qd}:
\beq\label{logintegral}
\int_0^\infty \frac{{\rm d}p}{ E_p}\,\frac{1}{{\rm e}^{\beta E_p}+1}\approx 
-\frac{1}{4}\ln\left(
\frac{\beta^2 M^2}{\pi^2}\right) -\frac{1}{2}\gamma_E+O(\beta^2 M^2),
\eeq
where $\gamma_E$ is Euler's constant. This formula shows that
the logarithms $\sim\ln M^2$, at the origin of the non-analyticity, 
cancel in Eqs.~(\ref{deriveepot}) and
(\ref{V0TR}).

As we did  in subsection~\ref{sec:T0LO}, it is convenient to express the 
effective potential in terms of the variable
$x_M=M^2/M_f^2$ (see Eq.~(\ref{URindepM0})). With a slight abuse in the notation, one can then write:
\beq\label{renormVT0}
V^{(0)}(x_M,T) = \frac{M_f^2}{4\pi}\left[x_M (\ln x_M -1)- 
\frac{8}{b}\;\int_0^\infty
{\rm d}{\rm u}
\ln ( 1+{\rm e}^{-b \sqrt{u^2+x_M}} )\right],
\eeq
where $b\equiv M_f/T$.

A plot of the leading order effective potential is given in Fig.~\ref{fig:fig-potV0_T}. As seen on this figure,
as $T$ increases, the minimum is shifted to lower and lower values of $x_M$ and eventually reaches $x_M=0$. We
come back to this in subsection~\ref{sec:chiral} when discussing the restoration of chiral symmetry.

\subsection{\label{sec:fTNLO}The $1/N$ contribution and the total effective potential}

As was the case already at T=0 (see subsection \ref{sec:T0NLO}), at 
$T\ne 0$ both the fermionic and the bosonic parts
in (\ref{Vdesigmac}) contribute   at next-leading-order. The 
fermionic contribution originates in the
next-to-leading order renormalization of $M_B$ and $g_B$ in $V^{(0)}_B(M_B,T)$.
The renormalization of $\tilde V^{(0)}_B(M_B,T)$ in 
Eq.~(\ref{V0Tfini}) amounts simply to the replacement $M_B^2\to
M^2(1+Z'^{(1)}/N)$, with $Z'^{(1)}$ given by Eq.~(\ref{delta}). The resulting expression
can be written as $\tilde V^{(0)}(M,T)+(1/N)\tilde V_f^{(1)}(M,T)$, with
$\tilde V^0{(0)}(M,T)$ given by Eq. (\ref{V0Tfini}) with $M_B \to M$ and
\beq\label{divtildeU}
\tilde V^{(1)}_f(M,T) = M^2\ln\ln\left(\frac{\Lambda}{M_f}\right)
\left(\int_{0}^\infty \frac{{\rm d}k}{ \pi}\frac{n_k}{E_k}\right)
+ \bar Z'^{(1)} M^2
\left(\int_{0}^\infty \frac{{\rm d}k}{ \pi}\frac{n_k}{E_k}\right).
\eeq
where $E_k=\sqrt{k^2+M^2}$ and $n_k=n(\omega=E_k)$ (see Eq. (\ref{nomega})).
The first term in Eq. (\ref{divtildeU}) introduces a new divergence,
not present at T=0, and which depends explicitly  on the temperature. 
From the discussion at the beginning of
this section, such a divergence should not remain in the final 
result, and indeed  we shall see shortly that
it is canceled by  a similar one coming  from the finite temperature
contribution of the bosonic part, to which we now turn.

The bosonic contribution  comes from the last term of Eq. (\ref{Vdesigmac}):
\beq\label{V1Tfini1}
  V^{(1)}_{b,B} (M_B,T)=\frac{1}{2}\int [d^2Q] \ln \left[1+ g_B^2
\Pi(\omega,q;M_B,T,\Lambda)\right].
\eeq
It is convenient to
write the sum over Matsubara frequencies in Eq.~(\ref{V1Tfini1}) as a
contour integral, as we did for the fermionic contribution in
Eq.~(\ref{Matsupotential}).  By exploiting the parity property
$\Pi(-\omega,q)=\Pi(\omega,q)$, one can then write:
\beq\label{potenbosocontour}
V^{(1)}_{b,B} (M_B,T)= \frac{1}{2} \int \frac{{\rm
d}q}{2\pi}\int_{-i\infty+\epsilon}^{i\infty+\epsilon}\,\frac{{\rm d}\omega}
{2\pi i}\,(1+2N(\omega))\,
\ln \left[1+g_B^2\Pi(\omega,q;M_B,T,\Lambda)\right],
\eeq
where
\beq\label{Nomega}
N(\omega)=\frac{1}{{\rm e}^{\beta\omega}-1}
\eeq
is the bosonic statistical factor.

\begin{figure}
\includegraphics[angle=90,width=13cm]{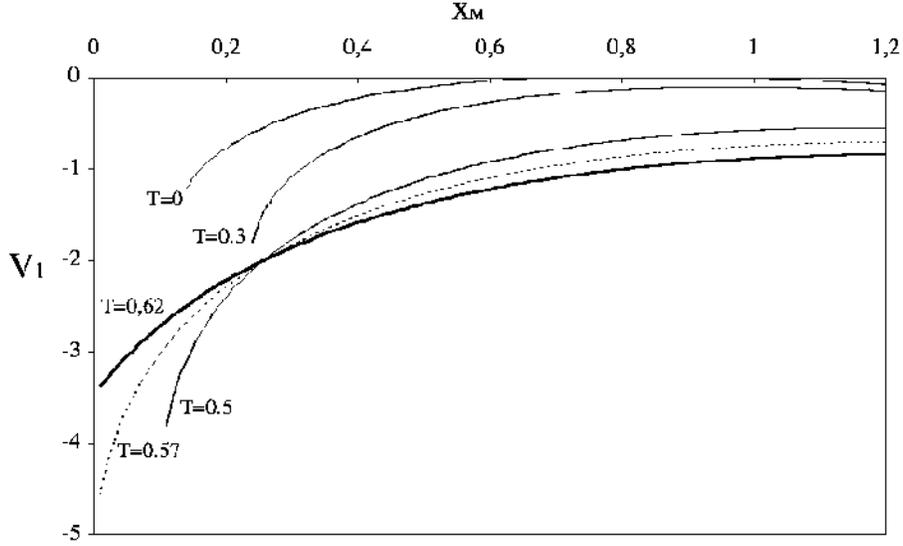}
\caption{\label{fig:fig-potV1_T} The next-to-leading order renormalized potential $V^{(1)}$ (divided by
$M_f^2/4\pi$) as a function of $x_M>x_{M_*}(T)$,  for various temperatures
$T$ (in units of the fermion mass
$M_f$). The temperature $T=0.57M_f$ is the temperature $T_c$ discussed in 
the next section (see Eq.\ref{Tc})), and 
$0.62M_f=1.1T_c$.}
\end{figure}

The expression (\ref{potenbosocontour}) of the bosonic part of the 
effective potential is
analogous to the corresponding one for the fermionic part, 
Eq.(\ref{potenfermiocontour}). However in contrast
to what happened in the fermionic case where 
Eq.(\ref{potenfermiocontour}) leads naturally to a separation
between zero temperature and non-zero temperature contributions, this is not quite so here: the
term containing the statistical factor $N(\omega)$ certainly vanishes 
as $T\to 0$, and as such can be
considered as a genuine finite temperature contribution; the term 
without the statistical factor reduces
to the zero temperature effective potential as $T\to 0$, but it does 
contain finite temperature
contributions (since $\Pi$ depends now on $T$). Nevertheless, the 
separation suggested by
Eq.(\ref{potenbosocontour}) is useful  as it allows us to clearly 
isolate the ultraviolet divergences: these
are contained entirely in the term without the statistical factor. We 
shall then add the zero temperature
counterterms (\ref{CdeLambda}) to that term, renormalize the mass and 
the coupling constant
with the leading order renormalization
constants ($Z^{(0)}$ from  Eq. (\ref{ZMf}) and $Z'^{(0)}=1$), 
and rewrite the
renormalized effective potential $V^{(1)}_{b}(M,T)$ as the sum of
two  contributions: $V^{(1)}_{b}(M,T)= V^{(1)}_{b,1}(M,T)+ 
V^{(1)}_{b,2}(M,T)$ with
\beq\label{V1_1}
  V^{(1)}_{b,1}(M,T)=\frac{1}{2} \int^\Lambda \frac{{\rm
d}^2q}{(2\pi)^2} \ln \left[
\frac{D^{-1}(iq_0,q;M,T)}{D_0^{-1}(Q_E;M_f)} 
\right]+\frac{M_f^2}{4\pi}\ln\frac{\Lambda^2}{M_f^2}\,
  +
\frac{M_f^2}{4\pi}
\ln\ln\frac{\Lambda}{M_f},
\eeq
\beq\label{V2}
  V^{(1)}_{b,2} (M,T)=\int_{-\infty}^{+\infty} \frac{{\rm
d}q}{2\pi}\int_{-i\infty+\epsilon}^{i\infty+\epsilon}\frac{{\rm d}\omega}{2\pi
i}\,N(\omega)\,
\ln
\left[D^{-1}(\omega,q;M,T)
\right] ,
&&
\eeq
where $D^{-1}(iq_0,q;M,T)\equiv D_0^{-1}(Q_E;M)+g^2\tilde{\Pi}(iq_0,q;M,T)$ is the renormalized
inverse $\sigma$-propagator studied in App. A. 

Clearly, when $T\to 0$,  $ V^{(1)}_{b,1}(M,T)$  becomes the zero 
temperature part of the bosonic
contribution to the potential analyzed in subsection~\ref{sec:T0NLO}; in 
the same limit $ V^{(1)}_{b,2}
(M,T)$ vanishes. The divergent contributions to $ V^{(1)}_{b,1}(M,T=0)$ are
  dealt with as in subsection~\ref{sec:T0NLO}.  When the temperature is non zero,
a new divergence occurs in $ V^{(1)}_{b,1} (M,T)$. This can be 
isolated  with the help  of the asymptotic
behaviors of $D_0^{-1}$ and $ \tilde\Pi$ given in App. A (Eqs. (\ref{logqdsD})
and (\ref{asymptotic_Pi}), respectively). One gets:
\beq\label{tildeV1}
  V^{(1)}_{b,1}(M,T)\sim -  M^2\ln\ln\left(\frac{\Lambda}{M}\right)
\left(\int_{0}^\infty
\frac{{\rm d}k}{ \pi}\frac{n_k}{E_k}\right).
\eeq
This divergence cancels  that of the
renormalized fermionic contribution $\tilde V^{(1)}_f$ of Eq. 
(\ref{divtildeU}), as anticipated (the two $\Lambda$-dependent terms in Eqs.~(\ref{divtildeU}) and
(\ref{tildeV1}) are identical to within terms which vanish as $\Lambda\to \infty$).

At this point, it is useful to go one step further and, in analogy 
with what we did in Eq.~(\ref{xm3}) of the
previous section, we absorb the finite part of the counterterm into 
the redefinition of the variable $x_M$
according to Eq.~(\ref{xm2}). Adding $\tilde V^{(0)}(M,T)$ and the second term in Eq. (\ref{divtildeU}) one can
write (discarding terms of order $1/N^2$):
\beq\label{fact}
\!\!-\frac{2}{\beta}\int_0^\infty
\frac{{\rm d}{\rm k}}{\pi}
\ln\left( 1+{\rm e}^{-\beta E_{\rm k}}\right)
\!+\!  \frac{{\bar Z'^{(1)}}}{N}\,M^2 \!\int_0^\infty \frac{{\rm 
d}{\rm k}}{\pi}
\frac{n_{\rm k}}{E_{\rm k}}
= -\frac{2M^2_f}{b}\!\int_0^\infty
\frac{{\rm d}{\rm u}}{\pi}
\ln\left( 1+{\rm e}^{-b \sqrt{u^2+x_M}}\right)
\eeq
with $x_M =M^2/M_f^2\;(1+\bar Z'^{(1)}/N)$, as defined in
Eq. (\ref{xm2}),  and $b= M_f/T$. Thus, in complete analogy with the
zero temperature case (see Eq.~(\ref{renormVdexM})), after the 
redefinition of the variable $x_M$, the leading
order contribution to the renormalized potential keeps the form of $V^{(0)}(x_M,T)$ 
given by Eq.~(\ref{renormVT0}).

By using the results just obtained and defining new functions $F(x_M,T)$, and $G(x_M,T)$, we can then write 
the
  complete effective potential at order $1/N$ in the form:
\beq\label{renormVT}
V(x_M,T) &=& \frac{M_f^2}{4\pi}\left[Nx_M (\ln x_M -1)-N\; 
\frac{8}{b}\;\int_0^\infty
{\rm d}{\rm u}
\ln ( 1+{\rm e}^{-b \sqrt{u^2+x_M}} )\right.
\nonumber
\\
&& \qquad\qquad\qquad\qquad
+\xi (x_M-1) +F(x_M,T)+G(x_M,T)\Big],
\eeq
where the functions $F$ and $G$ are given in App. B. The function 
$F(x,T)$ is an extension at finite
temperature of the function
$F_0(x)$ introduced in the previous section; to within an additive constant, 
it is proportional to the finite part of $ V^{(1)}_{b,1} (M,T)$.
The function $G(x,T)$  is proportional to
$ V^{(1)}_{b,2} (M,T)$:
\beq\label{V2G}
V^{(1)}_{b,2}= \frac{M^2_f}{4\pi}G(x_M,T).
\eeq

It  can be calculated by
deforming the integration
contour so that it goes around the singularities on the real axis. 
This operation
requires that grand circles at infinity generate no contribution, and that the
discontinuity of the integrand vanishes at $\omega=0$,  
properties that are both satisfied.
The resulting integral along the cut is then finite because of the statistical
factor. It can be written as:
\beq\label{V1phaseshift1}
  V^{(1)}_{b,2}(M)&=&\int_{-\infty}^{+\infty} \frac{{\rm
d}q}{2\pi}\int_{0}^{\infty}\frac{{\rm d}\omega}{2\pi\, 
i}\,N(\omega)\, \ln \left[
\frac{D^{-1}(\omega+i\epsilon,q;M,T)}
{D^{-1}(\omega-i\epsilon,q;M,T)}
\right] \, \nonumber\\
&=& \int_{-\infty}^{+\infty} \frac{{\rm
d}q}{2\pi}\int_{0}^{\infty}\frac{{\rm d}\omega}{2\pi}\,N(\omega)\, 2\,{\rm Im}\ln
D^{-1}(\omega+i\epsilon,q;M,T).
\eeq

\begin{figure}
\centering
\includegraphics[width=11.5cm]{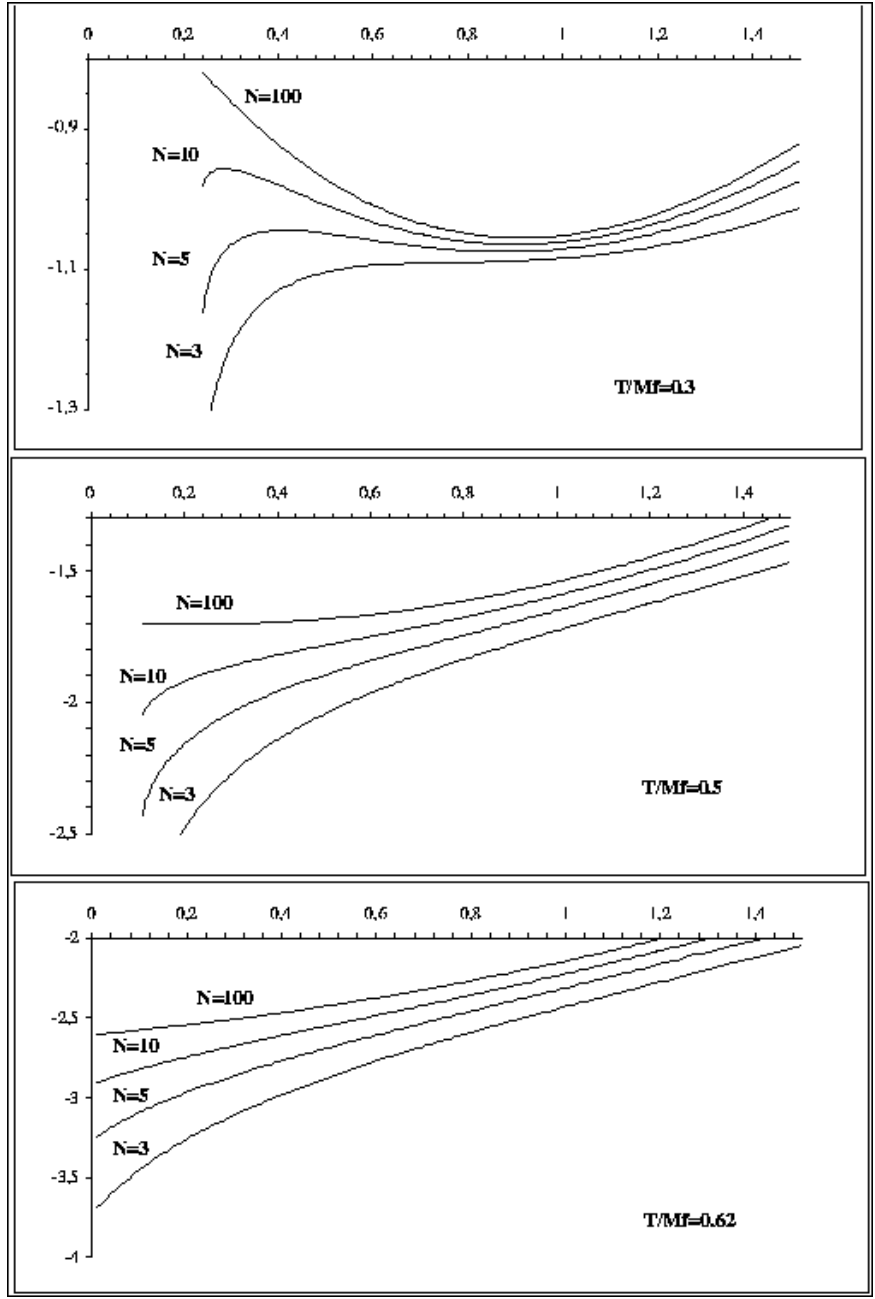}
\caption{\label{fig:veffT} The effective potential (divided by
$NM_f^2/4\pi$) as a function of $x_M>x_{M_*}(T)$,  for various values of $N$ 
and the temperature $T$. The temperature $T=0.57M_f$ is the temperature $T_c$ discussed in 
the next section (see Eq.\ref{Tc})), and $0.62M_f=1.1T_c$.}
\end{figure}

One may also express the argument of the logarithm in terms of the
phase shift $\delta$  of quark-quark scattering (see e.g. Eq. (\ref{GT})):
\beq\label{V1phaseshift2}
{\rm e}^{-2i\delta}=
\frac{D^{-1}(\omega+i\epsilon,q;M,T)}
{D^{-1}(\omega-i\epsilon,q;M,T)},
\eeq
with
\beq\label{fundelta}
\tan\delta(\omega,q;M,T)=-g^2\frac{{\rm Im}\Pi(\omega+i\epsilon,q;M,T)}
{{\rm Re}D^{-1}(\omega+i\epsilon,q;M,T)}.
\eeq
The latter are the expressions we use to calculate the function $G(x_M,T)$
numerically and to understand its behavior for small values of $x_M$ (see 
App. B).

To get an intuitive picture for what these formulae represent, let us first 
rewrite the expression (\ref{V1phaseshift1}) by integrating by parts,
\beq\label{V1phaseshift3}
 V^{(1)}_{b,2} (M,T)=\int_{-\infty}^{+\infty} \frac{{\rm
d}q}{2\pi}\int_{0}^{\infty}\frac{{\rm d}\omega}{\pi}\,\frac{1}{\beta}\ln
\left(1-{\rm e}^{-\beta\omega}\right) \frac{{\rm d} \delta}{{\rm d}\omega}.
\eeq
(The boundary term does not contribute because $\delta(\omega,q)$ 
vanishes at small $\omega$ in the same way as
${\rm Im}\Pi(\omega,q)$.)
Then, let us imagine that the
$\sigma$ excitation corresponds to a simple pole of the propagator, 
i.e., assume
$D^{-1}(\omega,q)=\omega^2-\omega_q^2$.  In this case, ${\rm d} 
\delta/{\rm d}\omega=\pi\delta(\omega-\omega_q)$,
and  $ V^{(1)}_{b,2}(M)$ is simply
\beq
 V^{(1)}_{b,2} (M)=\frac{1}{\beta}\int_{-\infty}^{+\infty} \frac{{\rm
d}q}{2\pi}\ln\left(1-{\rm
e}^{-\beta\omega_q}\right),
\eeq
which is the free energy density of noninterating bosons with energies $\omega_q$.
In the present case however, the sigma excitation does not correspond to 
a pole of the propagator (see subsection \ref{sigmaexcitation}),
and this simple picture does not quite 
hold. 

The $1/N$ contribution to the renormalized effective potential given by the second line in Eq. (\ref{renormVT}) is displayed in Fig.~\ref{fig:fig-potV1_T}. As
was the case at zero temperature (see the remark at the end of subsection~\ref{sec:T0NLO})), there is a minimum value
$x_{M_*}(T)$ of $x_M$  below which a Landau pole appears and the  function $F(x_M,T)$ becomes complex. 
For this reason, the 
next-to-leading order contribution to the potential is plotted in Fig.~\ref{fig:fig-potV1_T} only for
$x_M>x_{M_*}(T)$ (the way $x_{M_*}$ depends on the temperature is discussed in subsection~\ref{sigmaexcitation}).
In the same region
$x_M<x_{M_*}(T)$ where
$F$ is not real, the function
$G$ takes anomalously large and negative values (see Fig.~\ref{fig:funcion_G} and the discussion in App. B). For
$x_M \simge x_{M_*}$, both $G(x_M,T)$ and $F(x_M,T)$, and therefore also $V_1(x_M)$, have a large slope
as can be seen in Fig.~\ref{fig:fig-potV1_T}.

The complete effective potential is displayed in Fig.~\ref{fig:veffT} for various 
values of $N$ and the temperature $T$, as a function of $x_M$ (for $x_M>x_{M_*}(T)$). The curves labelled $N=100$ are
very close to  those corresponding to the leading order potential (not drawn). If we follow these curves, one
recovers the behavior of the leading order potential, already discussed: as the temperature increases, the
minimum of the potential shifts to lower values of
$x_M$ and becomes more and more shallow, until it reaches $x_M=0$ where it  becomes again a pronounced 
minimum. A striking feature  of the curves in Fig.~\ref{fig:veffT} is the large effect
of the $1/N$ corrections: already at moderate temperatures, one sees that the minimum disappears
for small values of $N$. This is to be related to the large slopes at small $x_M$ of the curves in
Fig.~\ref{fig:fig-potV1_T}, which drive quickly the minimum to low values of $x_M$, making this minimum
dangerously close to the threshold $x_{M_*}$ for the appearance of the Landau pole.
Note however that, at larger
temperatures, the potential has a well defined minimum at $x_M=0$ for 
all values of $N$. 

All these properties of the effective potential, 
and their physical interpretation, will be discussed at
length in the next section.

\section{Thermodynamics}
\label{sec:thermo}

We use now the results that have been established in the previous 
sections to discuss the
thermodynamical properties of the system. We start, in the next subsection, by reviewing known leading order results
concerning the restoration of chiral symmetry at a finite temperature $T_c$. In this one-dimensional
system, this phenomenon is specific of the mean field, or large $N$, 
approximation based on uniform
quark condensates, i.e., on constant solutions of the gap equation (\ref{gap}). 
(As argued in Ref.~\cite{Dashen:xz},  kink configurations of the 
condensate, if taken into account, would presumably prevent chiral symmetry breaking at any
non-zero temperature). At next-to-leading order in the 
$1/N$ expansion the fluctuations around the uniform condensate provide a small correction to the mean field picture as
 long as the temperature remains small. But as the temperature increases  these fluctuations
 become large, eventually making  the behavior of the system pathological;  as we shall see
 in subsection~\ref{quarkcondensate}, this results in a breakdown of 
the $1/N$ expansion  at some temperature below the mean field transition temperature $T_c$. We shall see
that this breakdown is related to the presence of a Landau pole which is discussed in
subsection \ref{sigmaexcitation}.

The last two subsections are devoted to the  high temperature limit  ($T\gg T_c$).
There chiral symmetry is realized, and the
$1/N$ expansion remains a good approximation scheme.  It provides a  description of the system as  weakly interacting
fermions, as expected from asymptotic freedom.
Quite remarkably, while the effective potential  obtained in the previous section  
does not depend explicitly on the coupling
constant, we shall see that the pressure can nevertheless 
be expanded in terms of an effective coupling whose magnitude 
decreases logarithmically with the temperature. To better understand this result, a
direct perturbative calculation of the pressure is presented in the last subsection.

\subsection{\label{sec:chiral}Restoration of chiral symmetry in leading order}

The finite temperature effective potential at leading order may be 
written (see Eq.~(\ref{renormVT0})):
\beq\label{calUdeT}
V^{(0)}(M,T)=\frac{M^2}{4\pi}\left( \ln \frac{M^2}{M_f^2}-1 \right)
-\frac{2}{\pi\beta}\int_0^\infty {\rm d}k\ln\left( 1+{\rm e}^{-\beta 
E_k} \right),
\eeq
with $E_k=\sqrt{k^2+M^2}$. As discussed in subsection IIC, at this order the minimum of Eq. (\ref{calUdeT})
is simply the quark mass
$M_f(T)$. It is a decreasing function of $T$. A simple analysis shows that, at small 
$T$, this decrease  is
exponentially small:
\beq
\frac{M_f(T)-M_f}{M_f}\approx \sqrt{\frac{2\pi T}{M_f}}\,{\rm e}^{-M_f/T}.
\eeq
\begin{figure}
\includegraphics[angle=90,width=12cm]{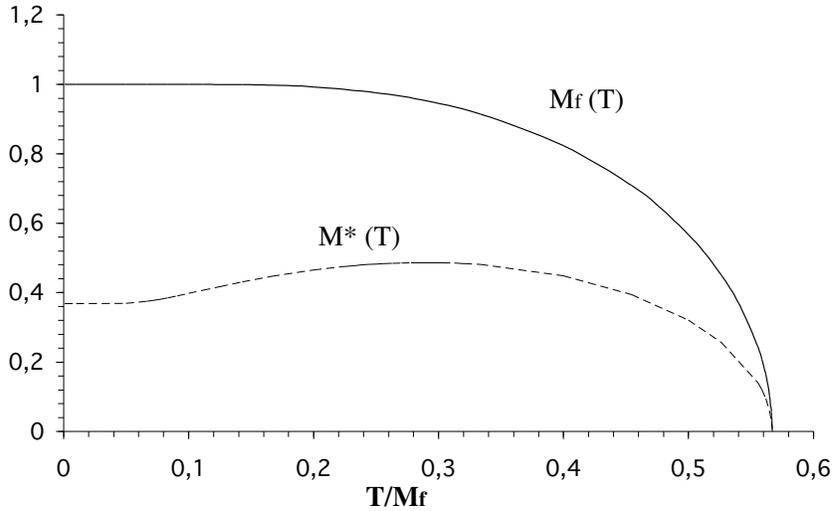}
\caption{\label{fig:MfdeT}The fermion mass $M_f(T)$ in leading order 
in the $1/N$ expansion as a function
of $T/M_f$. Also indicated is
the value
$M_*(T)$ of $M$ below which a Landau pole appears (see
subsection~\ref{sigmaexcitation} and App. A).}
\end{figure}
As the temperature increases further, however, $M_f(T)$ becomes 
smaller and smaller and
eventually vanishes, as can be seen from the plot in 
Fig.~\ref{fig:MfdeT} which displays the
function $M_f(T)$ obtained numerically. One may use the expansion in 
Eq.~(\ref{logintegral}) to
calculate the temperature
$T_c$ at which
$M_f(T)$ vanishes \cite{Jacobs:ys,Harrington:tf}. One gets:
\beq\label{Tc}
T_c=M_f\frac{{\rm e}^{\gamma_E}}{\pi}\simeq 0.567 M_f.
\eeq

  As $T\to T_c$  the system undergoes a second order
phase transition with mean field exponents. To verify this 
explicitly,  we expand
the effective potential around $M=0$:
\beq\label{Landaupot}
V^{(0)}(M,T)\simeq V^{(0)}(0,T)+ a \left(\frac{M}{M_f}\right)^2 +b
\left(\frac{M}{M_f}\right)^4
\eeq
with
\beq
a=\frac{M_f^2}{2\pi}\ln \left(\frac{T}{T_c}\right)
\qquad\qquad b=\frac{7\zeta(3) M_f^4}{32\pi^3 T^2}
\eeq
where $\zeta(3)\approx 1.202$. The coefficient $a$ follows easily 
form the expansion
(\ref{logintegral}). The coefficient
$b$ is obtained by using the expression (\ref{Matsupotential}) of the 
potential as a sum over
the Matsubara frequencies. By taking the second derivative of this 
expression with respect to
$M^2$ one obtains a convergent sum which is proportional to $b$:
\beq
\frac{T}{4}\sum_{n}\frac{1}{[|2n+1|\pi T]^3}=\frac{7\zeta(3)}{16\pi^3 
T^2}=\frac{2b}{M_f^4}.
\eeq
(Note that the expansion (\ref{Landaupot}) of the potential is
meaningful only at finite temperature, as is obvious on the form of 
the coefficients $a$ and
$b$ which are singular in the limit $T\to 0$.) The form
(\ref{Landaupot}) of the effective potential controls the behavior  of
$M_f(T)/M_f$ in the vicinity of $T_c$. A simple calculation gives:
\beq
\frac{M_f(T)}{M_f}\sim \sqrt{\frac{-a}{2b}}\simeq
\sqrt{\frac{8\pi^2}{7\zeta(3)}}\frac{T_c}{M_f}\sqrt{\frac{T_c-T}{T_c}}.
\eeq

Above $T_c$, the fermion mass $M_f(T)$ vanishes, and so does the 
(renormalized) quark condensate:  $\langle
\bar\psi
\psi\rangle=-NM_f(T)/g^2$ (see Eq.~(\ref{condensatedef2})). Thus, at $T_c$   the discrete chiral 
symmetry is restored.

The pressure is simply related to the effective potential at its 
minimum: $P(T)=NP^{(0)}=-NV^{(0)}(M=M_f(T),T)$.
The entropy density can be easily calculated from $s(T)={{\rm 
d}P}/{{\rm d}T}$. Since $V^{(0)}(M,T)$ is stationary for variations of $M$ around the 
value $M=M_f(T)$, the derivative can be simply 
taken at fixed
$M$. The result is that the entropy density is just that of free particles of 
mass $M_f(T)$:
\beq\label{sdeT}
s(T)=Ns^{(0)}(T)=- 2N\int_{-\infty}^\infty \frac{dk}{2\pi}\left[(1-n_k)\ln 
(1-n_k)-n_k\ln n_k\right],
\eeq
where $n_k=n(E_k)$ is the fermion statistical factor (see 
Eq.~(\ref{nomega})),  and $E_k=\sqrt{k^2+M_f^2(T)}$.
When $T>T_c$, the fermion mass vanishes and we have
($\int_0^\infty dx \ln(1+{\rm e}^{-x})=\pi^2/12$):
\beq\label{pressure0}
P^{(0)}=\frac{\pi T^2}{6}\qquad\qquad s^{(0)}=\frac{{\rm d}P^{(0)}}{{\rm d}T}= \frac{\pi T}{3}.
\eeq
Thus for $T>T_c$, interaction effects completely disappear, and the 
system behaves as a system of free massless
fermions.

We show in the last two subsections that the $1/N$ corrections do not 
alter much this picture at high
temperature where, due to asymptotic freedom, the effect of the interactions remains small.
But the $1/N$ corrections do affect very strongly the behavior of 
the system below $T_c$, as we shall see shortly. Before turning to these $1/N$ corrections,
 we shall, in the next 
subsection, continue to discuss physical
properties of the system in the mean field approximation; some of these will be useful in the interpretation
of the next-to-leading order results.

\subsection{\label{sigmaexcitation}The $\sigma$ excitation and 
quark-quark scattering amplitude}

At zero temperature, the propagator $D_0(Q;M_f)$, obtained by setting $M=M_f$ in 
Eq.~(\ref{Dinverse}), describes $\sigma$-meson
excitations with mass
$M_\sigma=2M_f$. This is easily seen by solving the equation 
$D_0^{-1}(Q_E^2=-M_\sigma^2;M_f)=0$ for
$M_\sigma$.  Note however that this point where $D_0^{-1}$ vanishes 
does not correspond to a simple pole in the
$\sigma$ propagator, but to a branch point limiting the region of 
phase space where
the $\sigma$  excitation can decay into
quark-antiquark pairs.

\begin{figure}
\includegraphics[angle=90,width=10cm]{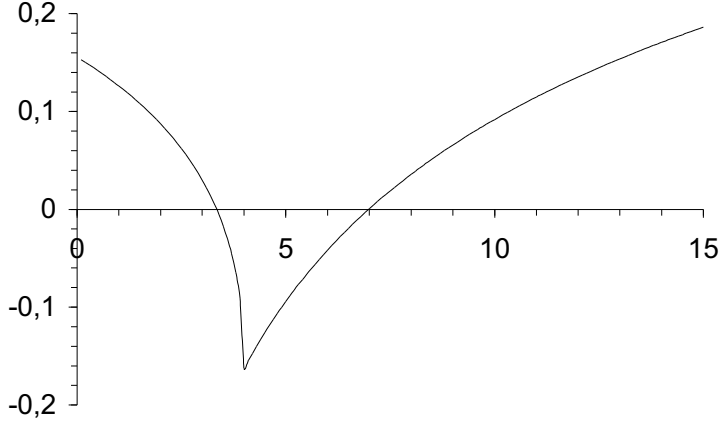}
\caption{\label{fig:do}The real part of $D_0^{-1}(\omega,q;M)$ as a 
function of $(\omega^2-q^2)/M^2$
for
$M<M_f$. For $M=M_f$, ${\rm Re} D_0^{-1}$ vanishes at the single 
point $(\omega^2-q^2)/M_f^2=4$, beyond which the
imaginary part starts to grow. For an arbitrary $M$, Re$D_0^{-1}(\omega,q;M)$ has a
minimum at this point, with value
$(g^2/\pi)\ln(M/M_f)$ (see App. A). When $M=M_*=M_f/e$, Re $D_0^{-1}$ vanishes at $\omega^2=q^2$.}
\end{figure}

In order to see that we need to
perform an analytic continuation from the Euclidean momentum 
$Q_E=(q_0,q)$ to the Minkowski one
$Q_M=(\omega,q)$, with $q_0\to -i\omega$. From Eq.~(\ref{Dinverse}), one gets:
\beq\label{D0inverse}
D^{-1}_0(\omega,q;M)=\frac{g^2}{2\pi}\left[\ln\left(\frac{M^2}{M_f^2}\right)
+B_0\left(\frac{-\omega^2+q^2}{M^2}\right)\right].
\eeq
Let us set $\omega_q=\sqrt{q^2+4M_f^2}$. The inverse propagator 
$D^{-1}_0(\omega,q;M_f)$ has cuts on the
real $\omega$ axis for $\omega>\omega_q$ and $\omega<-\omega_q$; the 
corresponding imaginary part
is (for $\omega>0$):
\beq\label{ImPi0}
{\rm Im} D^{-1}_0(\omega+i\epsilon,q;M_f)= 
-\frac{g^2}{2}\sqrt{\frac{\omega^2-\omega_q^2}{\omega^2-q^2} }
\,\Theta(\omega-\omega_q).
\eeq
The point $\omega=\omega_q$, is the branch point corresponding to the threshold
mentioned above. A plot of ${\rm Re} D_0^{-1}(\omega+i\epsilon,q;M)={\rm Re} D_0^{-1}(\omega+i\epsilon,q;M_f)+
(g^2/\pi)\ln(M/M_f)$  is displayed in Fig.~(\ref{fig:do}).

At finite temperature the threshold remains  located at
$2M_f(T)$, as long as $T<T_c$.  To see that, consider the equation 
which defines $M_\sigma$;
\beq
D^{-1}(\omega=M_\sigma,q=0;M_f(T))=0,
\eeq
where $D^{-1} = D_0^{-1}+g^2\tilde\Pi$ is the finite temperature $\sigma$ 
propagator studied in App. A.
Let us set
$M_\sigma=2M_f(T)$ and verify that indeed this is the solution. Using 
the explicit
expressions for $\tilde\Pi(\omega,q)$ given in App. A, we get:
\beq\label{gapsigma}
D^{-1}(\omega=M_\sigma,q=0;M_f(T))=\frac{g^2}{\pi}
\ln\left(\frac{M_f(T)}{M_f}\right)+2 g^2\int_0^\infty\frac{{\rm
d}k}{\pi}\frac{n_k}{E_k}=0,
\eeq
with $n_k$ as in Eq.~(\ref{sdeT}) above. 
This equation is precisely that which determines $M_f(T)$, i.e., the
gap equation obtained by equating the right hand side of 
Eq.~(\ref{deriveepot}) to zero, which proves the point.

\begin{figure}
\includegraphics[width=12.5cm]{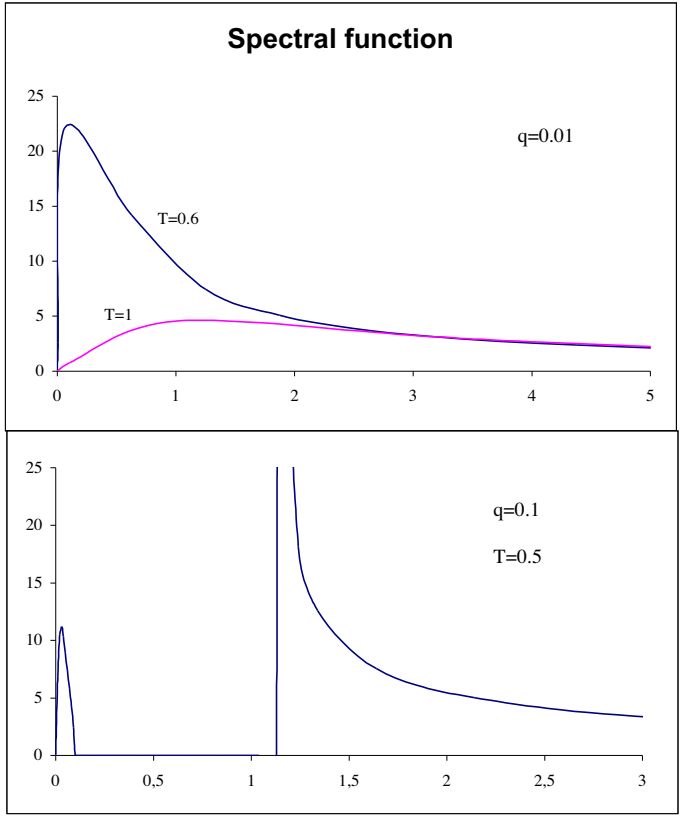}
\caption{\label{fig:spec}The spectral function of the $\sigma$ 
excitation above $T_c$ (top, $q=0.01$) and below
$T_c$ (bottom,
$q=0.1$), as a function of $\omega$ expressed in units of $M_f$ 
(upper curves) or $M_f(T)$ lower curves.}
\end{figure}

A more complete description of the $\sigma$ excitation is obtained from the
corresponding spectral function $\rho(\omega,q)$. At zero 
temperature, we define
($\omega$ real):
\beq\label{spetral0}
\rho_0(\omega,q)=\frac{1}{i}[D_0(\omega+i\epsilon, 
q;M_f)-D_0(\omega-i\epsilon,q;M_f)]=2{\rm
Im}D_0(\omega+i\epsilon, q;M_f).
\eeq
This is an odd function of $\omega$.
Using the formulae given above, one easily obtains (for $\omega>0$):
\beq\label{specT0}
\rho_0(\omega,q)=\frac{4}{g^2}\sqrt{\frac{\omega^2-q^2}{\omega^2-\omega_q^2}}
\frac{\Theta(\omega-\omega_q)}{ 1+\frac{4}{\pi^2}
\left({\rm 
Arctanh}\sqrt{\frac{\omega^2-\omega_q^2}{\omega^2-q^2}}\right)^2   }.
\eeq
 At fixed $q$, this is a decreasing 
function of $\omega$, peaked at
$\omega\simeq\omega_q$; when
$\omega\to\infty$,
$\rho_0(\omega,q)$ vanishes as
$\rho_0(\omega,q)\sim 4\pi^2/(g\ln\omega^2)^2$.  By replacing, in 
Eq.~(\ref{spetral0}),  $D_0$ by the finite
temperature propagator $D(\omega,q;M_f(T))$, one obtains 
the spectral function at finite temperature
$\rho(\omega,q;T)$.  The typical behavior of
$\rho(\omega,q;T)$ for temperatures below and above
$T_c$ is displayed in Fig.~\ref{fig:spec}.  We note, below $T_c$, 
the threshold at
$\omega=\sqrt{q^2+4M_f(T)^2}$, beyond which the spectral function 
resembles $\rho_0(\omega,q)$ given by 
Eq.~(\ref{specT0}). For $0<\omega<q$, a typical finite temperature 
contribution appears, that of the scattering
of quark-antiquark pairs on quarks or antiquarks present in the heat 
bath (see App. A for more details).  As
$T$ approaches
$T_c$ from below,
$M_f(T)\to 0$, and the two contributions merge, leaving a sharp peak 
at small $\omega$. The peak
structure survives just above $T_c$ (see the curve labelled $T=0.6$ 
in Fig.~\ref{fig:spec}), but as one raises the
temperature the spectral weight becomes spread over a larger and 
larger energy range, as can be seen in
Fig.~\ref{fig:spec}. The behavior of the spectral function is then 
governed  by the high temperature regime of the
fermion loop discussed in App. A. In this regime and for large values of
$q$ ($q\gg T_c$), the spectral function becomes  a step function at
$\omega=q$ followed by a long logarithmic tail.

Consider finally the quark-quark scattering amplitude 
${\cal T}(\omega,q;M,T)$, where the mass $M$ is treated  as an independent parameter.  This is the  
generalization at finite temperature of the quantity introduced in Eq.~(\ref{scatteringamplitude01}), i.e.:
${\cal T}(\omega,q;M,T)= g^2 D(\omega,q;M,T).$
At zero momentum, we have (see Eqs. (\ref{V0andD})): 
\beq\label{sumrule1}
{\cal T}(0,0;M,T) = g^2\lim_{q\to 0} D(0,q;M) =
\left(\frac{{\rm d}^2V^{(0)}(M,T)}{{\rm d}M^2}\right)^{-1},
\eeq
This relation may be verified by calculating the second derivative  from the expression (\ref{V0TR}) of the
potential and using the results given 
in App. A to obtain:
\beq\label{veffdeM}
\lim_{q\to 0} \frac{1}{{\cal T}(0,q;M,T)}=\frac{1}{\pi}\left(
1+\ln\frac{M}{M_f}\right)+2\int_0^\infty \frac{{\rm
d}k}{\pi}\left(\frac{M^2}{E_k^2}\frac{{\rm d}n}{{\rm 
d}E_k}+\frac{k^2}{E_k^2}\frac{n}{E_k}\right).
\eeq
The second term in the r.h.s. of Eq. (\ref{veffdeM}) is 
$\lim_{q\to 0} \tilde\Pi(0,q;M,T)$ (where we have used Eq.~(\ref{tildePi})).
The two terms in the integral correspond 
respectively to the  contributions of scattering processes and
pair creations (see App. A).  As mentioned already, at finite temperature, and when
$M\ne 0$, the double  limit $\omega\to 0,
q\to 0$ of $D(\omega,q)$ is   not regular  and the result depends on the order in which 
the two limits are taken. If the limit $q\to 0$ had been taken first, the scattering term, 
proportional to ${\rm d}n/{\rm d}E$, would be absent in Eq.~(\ref{veffdeM}), as may be seen from
Eq.~(\ref{tildePi}). 

When
$M\to 0$,  the scattering term vanishes and the limit $\omega\to 0,q\to 0$
of ${\cal T}(\omega,q)$ becomes regular. In that same limit where $M\to 0$,  the second term in the integral of
Eq.~(\ref{veffdeM})  (i.e., the pair creation term)  develops a logarithmic divergence, which in fact cancels 
the logarithm of the  $T=0$ contribution
 (the first term of 
Eq.~(\ref{veffdeM})).
Thus one has (see Eq.~(\ref{limit1pitilde})):
\beq\label{TM0}
\lim_{M\to 0}{\cal T} (0,0;M,T) = \frac{\pi}{\ln (T/T_c)},
\eeq
which is nothing but the square of the renormalized coupling constant evaluated at a scale $M_0\sim T$ 
(see Eq.~(\ref{running_g})).  Thus at high temperature, i.e.,
when $T\gg T_c$, the scattering amplitude becomes small, as expected from asymptotic freedom. It can be also
verified that no Landau pole occurs in this regime.

Below $T_c$ the situation is complicated by the presence of a Landau pole in the $\sigma$ propagator.
The threshold $M_*(T)$ for the appearance of the Landau pole, which coincides with the value of $M$ at which the second derivative 
of $V^{(0)}$ vanishes, is plotted as a function of the temperature in Fig.~\ref{fig:MfdeT}. As the 
temperature increases, $M_*$ is shifted first to slightly larger values: 
this is because at
moderate temperatures the finite temperature contribution in 
Eq.~(\ref{veffdeM}) is dominated
by the first, negative, term in the integral. For temperatures beyond 
approximately $T_c/2$ however the
second, positive, contribution becomes dominant, resulting in a shift 
of $M_*$ to lower and lower values, reaching $M_*=0$ at $T_c$. 
That $M_*$ vanishes precisely at
$T_c$ follows from the simple fact that $M_*$ is located between 
two extrema of the potential, which merge at
$T_c$. Thus, as $T$ approaches $T_c$, the fermion mass
$M_f(T)$ becomes close to $M_*(T)$, and the calculation of the next-to-leading contribution to the effective
potential becomes ill defined in the vicinity of its minimum: in particular $F(x_M;T)$ becomes complex for $x_M<x_{M_*}$
($x_{M_*}=(M_*/M_f)^2$) and both $F(x_M;T)$ and $G(x_M;T)$ present large negative slopes for $x_M\simge x_{M_*}$ (see discussion in
subsection \ref{sec:fTNLO} and App. B). The latter has important consequences which will become clear in the next subsection.

\subsection{\label{quarkcondensate}The quark condensate at order 1/N}

In this subsection, we   examine how the physics of chiral symmetry breaking and its restoration, 
which we have just 
discussed, is modified by the $1/N$ corrections. Chiral symmetry breaking is characterized by a non-vanishing
value of the quark-condensate, and the latter 
is proportional to the value $M_{min}$ of the  minimum of the effective
 potential (see Eq.~(\ref{condensatedef2})). The effective potential
 has been obtained in subsection~\ref{sec:fTNLO}
to order $1/N$  in terms of the renormalization group invariant 
$x_M$ defined in Eq.~(\ref{xm2}).
We shall analyze here the variation with the temperature of  $x_{min}\equiv x_{M_{min}}\propto M_{min}^2$
obtained  at next-to-leading-order.

The quantity $x_{min}(T)$ may be determined to accuracy $1/N$ by
following the strategy exposed in subsection~\ref{sec:generalities} B.    
Generalizing Eq.~(\ref{xMaccurate1surN}) at finite temperature,  we set
\beq\label{xmin0a}
x_{min}(T)=x_{min}^{(0)}(T)+ \frac{1}{N} x_{min}^{(1)}(T),
\eeq
where $x_{min}^{(0)}(T)$ is the minimum of $V^{(0)}(x_M,T)$ in 
Eq.~(\ref{renormVT0}).
Clearly, $x_{min}^{(0)}(T) M_f^2=M_f^2(T)$, where $ M_f(T)$ is the 
temperature dependent quark
  mass introduced in subsection~\ref{sec:chiral}, and plotted in 
Fig.~\ref{fig:MfdeT}. The correction
$x_{min}^{(1)}(T)$ is given by  (see Eq.~(\ref{xMaccurate1surN})):
\beq\label{xMaccurate1surN2}
x_{min}^{(1)}=-\frac{   \left.  \frac{ {\rm d}V^{(1)}(x_M,T) }{ {\rm
d}x_M }
\right|_{x_{min}^{(0)}(T)}   } {\left. \frac{ {\rm d}^2V^{(0)}(x_M,T) 
}{ {\rm d}x_M^2 }\right|_{x_{min}^{(0)}(T)} },
\eeq
with
\beq\label{dV1dxM}
\frac{ {\rm d}V^{(1)}(x_M,T) }{ {\rm
d}x_M }  =\xi
+F'({x_{min}^{(0)}(T)},T)+G'({x_{min}^{(0)}(T)},T),
\eeq
and ${ {\rm d}^2V^{(0)}(x_M,T) }/{ {\rm d}x_M^2 }$ can be 
determined from the results of subsection \ref{sec:chiral}. 
In Eq.~(\ref{dV1dxM}),
$F'$ and $G'$ are the derivatives of the functions $F$ and $G$ with 
respect to $x_M$; they are calculated numerically.

\begin{figure}
\includegraphics[angle=90,width=12cm]{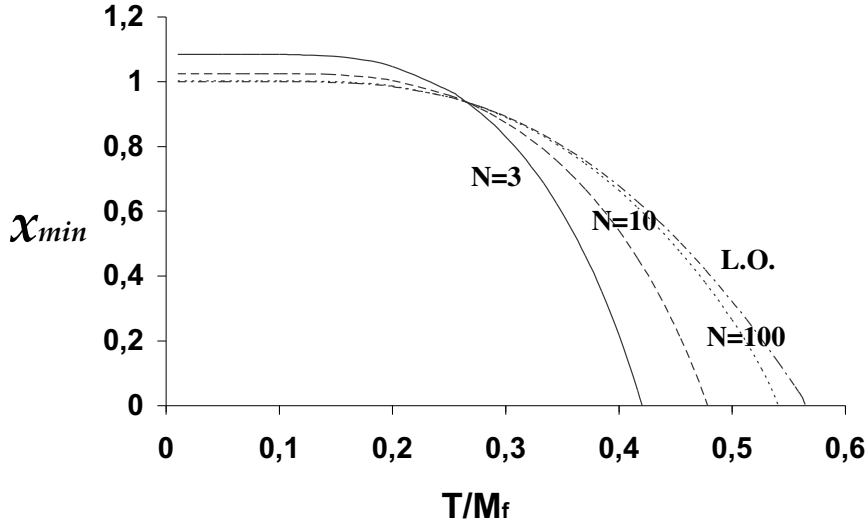}
\caption{\label{fig:xminun}The value of $x_{min}$ calculated from 
Eqs.~(\ref{xmin0a}) and
(\ref{xMaccurate1surN2}), as a function of the temperature $T$ (in 
units of $M_f$), and for various values of $N$. The curve labelled L.O. corresponds
to the leading order potential. }
\end{figure}

A plot of $x_{min}(T)$ is given in Fig.~\ref{fig:xminun}. At small
$T$, $x_{min}>1$, in agreement with the study of the effective potential 
at zero temperature (see the caption of
Fig.~\ref{fig:potentiel_eff}).  However, when the temperature $T$
increases, the $1/N$ correction eventually turns negative at $T \sim T_c/2$ and becomes large when
the temperature increases further. Thus, for instance,
$|x_{min}^{(1)}/x_{min}^{(0)}|\simeq 0.1$ when $T/T_c \sim 0.55$,  $|x_{min}^{(1)}/x_{min}^{(0)}|\simeq 1$ when
$T/T_c \sim 0.65$,  
$|x_{min}^{(1)}/x_{min}^{(0)}|\simeq 10$ when $T/T_c \sim 0.85$   and 
$|x_{min}^{(1)}/x_{min}^{(0)}|\simeq 30$ when $T/T_c \sim 0.92$.  Thus, when $T/T_c \simge 0.65$
the $1/N$ correction becomes too large to be really considered as a ``correction'', 
and the $1/N$ expansion breaks down. This is to be contrasted with the situation at zero temperature where 
$|x_{min}^{(1)}/x_{min}^{(0)}|\simeq 0.25$ (see Eq.~(\ref{xMaccurate1surN}) and the discussion after). Note that
what makes $|x_{min}^{(1)}/x_{min}^{(0)}|$ large as $T$ approaches $T_c$ is the numerator
${\rm d}V^{(1)}/{\rm d}x_M$ (the denominator ${\rm d}^2V^{(0)}/{\rm d}x_M^2 $ goes to a constant as $T\to
T_c$); as we have already discussed, both $F'(x_M)$ and $G'(x_M)$ become large and negative as $x_M$ approaches $x_{M_*}$,
and $x_{min}$ does approach $x_{M_*}$ as the temperature increases.

This pathological result is confirmed by a direct (numerical) minimization of the potential in Eq.~(\ref{renormVT}). For all values of $N$ and $T$,
one finds values of  $x_{min}(T)$ very close to those deduced from Eqs. (\ref{xmin0a})-(\ref{xMaccurate1surN2})
whenever the minimum exists (the two evaluations of $x_{min}(T)$  may 
differ a priori by terms of order  $1/N^2$). However,  as anticipated in subsection \ref{sec:fTNLO}, the
potential  no longer  has a minimum 
when $T$ exceeds a certain value, which depends on $N$: for $N=100$,
the minimum ceases to exist when $T \geq 0.93T_c$, for $N=10$ when 
$T\geq 0.63T_c$ and for
$N=3$ when $T\geq 0.51T_c$. Again the disappearance of the minimum may be related to the rapid drop of the functions
$F(x_M,T)$ and $G(x_M,T)$ in the vicinity of $x_{M_*}$, the value of $x_M$ below which a Landau pole
appears, as discussed in App. B. 

At this point one could speculate and try to relate the breakdown of the $1/N$ expansion to the fact that 
the true transition temperature of the model 
is presumably  $T_c=0$ \cite{Dashen:xz}. The mean field approximation which ignores part of the important
degrees of freedom (e.g. the kink configurations) gives a poor
representation of the physics of the system at finite temperature. This mean field physics  persists
for moderate temperatures: then the fluctuations around
the uniform condensate do not change significantly the state of the system. But beyond some value of 
the temperature the fluctuations become dominant and perhaps mimic the effect of degrees of freedom left-out in the mean field
approximation: these fluctuations are responsible for the rapid decrease of $x_{min}$ for $T \simge T_c/2$ which drives
the system towards its true equilibrium state, where chiral symmetry is restored.

\begin{figure}
\includegraphics[angle=90,width=14cm]{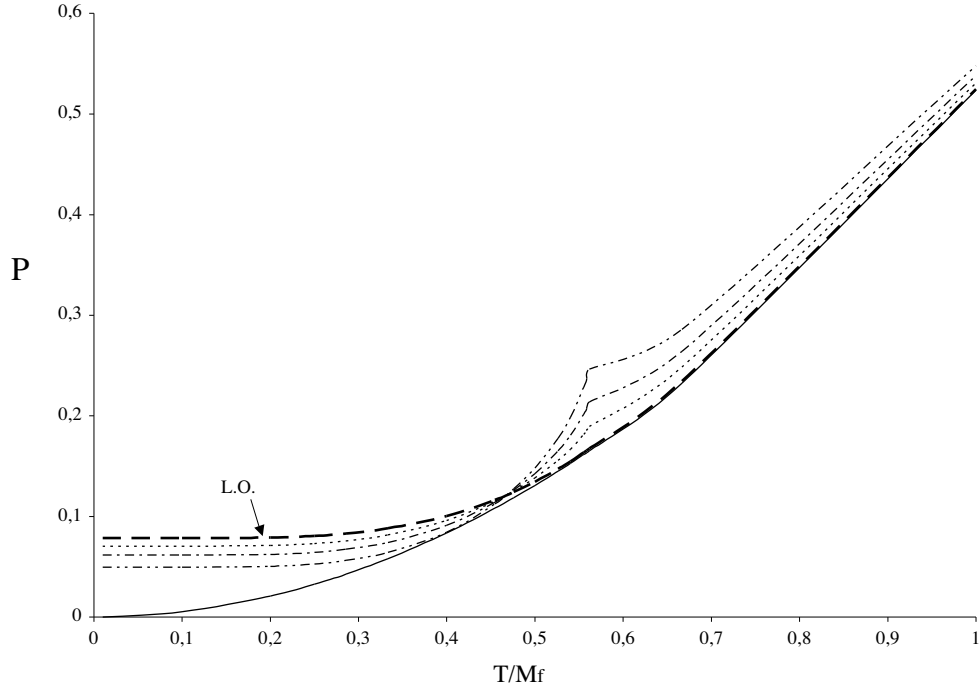}
\caption{\label{fig:pressure} The pressure as a function of $T/M_f$. 
The full line is the pressure of free massless
fermions, Eq.~(\ref{pressure0}). The thick dashed line is the 
pressure at leading order (L.O.). The other curves include
the $1/N$ corrections, for $N=10$, $N=5$ and $N=3$.  For small 
temperatures, the pressure is an increasing function
of $N$; the curves corresponding to the various values of $N$ cross each other for $T\simle 0.5 M_f$, and  for
larger temperature, the pressure decreases with increasing
$N$.  }
\end{figure}

Another 
signal of the inadequacy of the $1/N$ expansion for temperatures $T\simle T_c$ is provided by the study of the
pressure. 
A plot of this quantity  is given in Fig.~\ref{fig:pressure}. Note that although the determination of the minimum
of the effective potential becomes inaccurate in the vicinity of $T_c$ (and may not even exist), this is of little
consequence for the present calculation:
since the potential is flat in the vicinity of $T_c$, its value, i.e. the pressure, is well estimated. One sees
in particular that the entropy density,
i.e., the  derivative of the pressure with respect to $T$, smoothly increases with the
temperature below $T_c$, but decreases suddenly at $T_c$, suggesting a first order transition. This is clearly
unphysical. 
In fact the plot reveals two regimes: at low temperature the system exhibits unphysical mean field behavior;
this stops abruptly around $T_c$ where a new regime sets in. This regime is that of high temperature with
restored chiral symmetry: there, the  $1/N$ approximation appears to be a good approximation;
it allows us to study the physics of the system in conditions where, because of asymptotic freedom, it is
expected to behave as a gas of weakly interacting massless fermions. 
We shall verify in the last two subsections that this is indeed true.

\subsection{\label{sec:thermohighT}Thermodynamics at high temperature}

We discuss now the limit of high temperature, $T\gg T_c$, where the 
condensate vanishes and, in leading
order, the quarks are massless.  
As we shall see, the thermodynamical functions
can then be expanded in powers of $\pi/\ln(T/T_c)$ which we shall interpret as
the running coupling constant at a scale of the order of the temperature.
This provides a nice, and non trivial, illustration of the behavior expected from
asymptotic freedom.
Let us consider the limit of the $1/N$ contribution to the effective 
potential $V^{(1)}$, i.e., the second line in Eq. (\ref{renormVT}) when $T\gg T_c$. 
It is obtained from the high temperature limit of $F(0,T) + G(0,T)$, the constant $\xi$ playing no role in 
this limit. The function $F(0,T)$ is given by the first line of 
Eq. (\ref{FTx0}) (the second line is a finite constant and plays no role 
at high temperature). It can be written as
\beq\label{V1highT2}
  F(0,T)=\frac{T^2}{M_f^2} \frac{2}{\pi} \int_0^\infty {\rm d}s\,\int_0^\infty {\rm d}r
\,\,\ln\left[ \frac{ 1+\frac{\tilde
g^2}{2\pi}B_T(is,r)   }  {  1+\frac{\tilde 
g^2}{2\pi}\ln \left( T_c^2/M_f^2 \sqrt{(s^2+r^2)^2+M_f^4/T^4}  \right)   }
\right].
\eeq
where the quantity $\tilde g$ that we have 
introduced in this equation is
\beq\label{gtilde}
\tilde g^2\equiv \frac{\pi}{\ln(T/T_c)}.
\eeq
which coincides with the scattering amplitude at zero momentum for massless particles (see Eq. (\ref{TM0})).
According to
Eq.~(\ref{running_g}), this quantity may also be interpreted as the effective coupling constant at the scale $M_0
\sim T$, i.e.,  $\tilde
g^2=g^2(M_0=\pi T/{\rm e}^{\gamma_E+1})$. We shall come back to this identification in the next subsection. As
is clear on Eq.~(\ref{V1highT2}), aside from the overall factor
$T^2$ in front of the integral, all the temperature dependence of the 
effective potential is entirely contained in $\tilde g^2$ (the extra temperature dependence in the
denominator in Eq. (\ref{V1highT2}) is numerically negligible).

Since $\tilde g$ becomes small as
$T$ increases, one may attempt to expand $V^{(1)}_{b,1}(T)$ in powers 
of $\tilde g$. In order to do
so, we note that $B_T(is,r)- \ln\left((s^2+r^2)T_c^2/M_f^2\right)$ decreases rapidly as 
$r,s\gg~1$ (see
App. A), and the contribution of this region to the integrals is negligeable.
The integrand is also regular in the $r,s\to 0$ limit, and correspondingly the contribution to
integrals of the region $r,s\ll M_f^2/T^2$ is negligeable. We conclude that the only region of the
$s,r$ plane which contributes to the
integrals in the limit $T\gg T_c$ is the region $r,s\sim 1$. There, both 
$B_T(is,r)- \ln\left((s^2+r^2)T_c^2/M_f^2\right)$ and
$\ln(s^2+r^2)$ are small compared with $\ln(T/T_c)$ when $T\gg T_c$. 
One can then proceed to an expansion of $F(0,T)$ to get
\beq
F(0,T)= \frac{T^2}{M_f^2}\left(\frac{\tilde g^2}{2\pi} F_2+ 
\frac{\tilde g^4}{4\pi^2} F_4+\cdots\right),
\eeq
with
\beq\label{F2}
F_2=\frac{2}{\pi}\int_0^\infty {\rm d}s\,\int_0^\infty {\rm d}r
\,\,\left[B_T(is,r)- \ln\left((s^2+r^2)T_c^2/M_f^2\right)
\right],
\eeq
and
\beq
F_4=-\frac{1}{\pi}\int_0^\infty {\rm d}s\,\int_0^\infty {\rm d}r
\,\,\left(\left[B_T(is,r)\right]^2- \left[\ln\left((s^2+r^2)T_c^2/M_f^2\right)
\right]^2\right).
\label{F4ref}
\eeq
$F_2$ can be calculated analytically by noting that $B_T(is,r)- \ln\left((s^2+r^2)T_c^2/M_f^2\right)
=2\pi \tilde{\Pi}(iTs,Tr;M=0,T)$, that is,
$F_2$ is proportional to the $(q_0,q)$ integral of $\tilde{\Pi}(iq_0,q;M=0,T)$.
It can be calculated, by using Eq. (\ref{tildePi}) for $\tilde{\Pi}$, performing first
the $q_0$ integral, changing to the variables $k_{\pm}=k\pm q/2$ and
doing the $k_{\pm}$ integrals, and finally using $\int_0^{\infty}dy \ln (1+e^{-y})=\pi^2/12$ for the remaining integration.
$F_4$ has been calculated numerically. We obtain then:
\beq\label{F2F4}
F_2=\frac{2}{3}{\pi^2}\approx 6.5797 \qquad
F_4\approx -4.322
\eeq

Let us now turn to the evaluation of the function $G(0,T)$, given by Eq. (\ref{GT}).
It can be written as:
\beq\label{V1phaseshift124}
G(0,T)=-\frac{T^2}{M_f^2} \frac{4}{\pi} \int_0^{\infty}{\rm d}s 
\int_0^{\infty}{\rm d}r\,\,N(s)\,  \delta(s,r;T) \, ,
\eeq
where $s=\omega/T$, $r=q/T$, $N(s)=1/({\rm e}^s-1)$, and:
\beq\label{tandelta}
\tan\delta=-\frac{ {\rm Im} B_T(s,r) }{\ln(T^2/T_c^2)+{\rm 
Re} B_T(s,r)  }.
\eeq

To obtain the limit of $G(0,T)$ for$T\gg T_c$ one observes that the statistical factor
$N(s)$ limits the $s$-integral to $s\simle 2$. For these values of 
$s$, the imaginary part is
significant only in a finite range of values of $r$ (e.g. $r\simle 
10$ for $s\sim 1$), and vanishes exponentially at
larger
$r$. In this domain of values of $s$ and $r$,  the factor ${\rm 
Re}B_T(s,r) $ in the denominator remains of order 1
while the term
$\ln(T/T_c)$ becomes large when $T\gg T_c$. One can then expand:
\beq
  \tan\delta=-\frac{ \frac{\tilde g^2}{2\pi}{\rm Im} B_T(s,r) 
}{1+\frac{\tilde g^2}{2\pi}B_T(s,r)  }\approx
-\frac{\tilde g^2}{2\pi}{\rm Im} B_T(s,r) \left[1-\frac{\tilde 
g^2}{2\pi}{\rm Re}B_T(s,r) +\cdots\right]\simeq\delta
\eeq
The expansion of $G(0,T)$ follows:
\beq\label{develG}
G(0,T)=\frac{T^2}{M_f^2} \left(\frac{\tilde g^2}{2\pi}\, \,G_2 + 
\frac{\tilde g^4}{4\pi^2}
\, G_4+\cdots\right),
\eeq
where
\beq\label{G2}
G_2=\frac{4}{\pi}\int_0^\infty dr \int_0^\infty ds \,N(s)\,{\rm Im} 
B_T(s,r),
\eeq
and
\beq\label{G4}
G_4=-\frac{4}{\pi}\int_0^\infty dr \int_0^\infty ds \,N(s)\,{\rm Im} 
B_T(s,r)\, {\rm Re}B_T(s,r).
\eeq
$G_2$ can be calculated analytically, by noting that $\mathrm{Im} B_T(s,r)=2\pi \mathrm{Im}(\Pi(Ts,Tr,T))$.
Using then Eq. (\ref{imaginaryasymp}), we can perform the two integrals using
$\int_0^{\infty}dy\ln(1-e^{-y})=\pi^2/6$. $G_4$ has been calculated numerically. We obtain:
\beq
G_2= -\frac{2}{3}\pi^2\approx -6.5797, \qquad G_4\approx 6.281 \,.
\eeq

At this stage, several comments are in order. Recall that, at 
leading order, the pressure is that of non interacting
massless particles as soon as $T>T_c$: $P^{(0)}=\pi T^2/6$ 
(see Eq.~(\ref{pressure0})); thus, when $T>T_c$ all the
effects of the interactions are contained in $V^{(1)}$. 
Asymptotic freedom leads us to expect that, at high temperature, 
these interactions should
be weak. Since the 
coupling constant has completely disappeared in the final expression of $V^{(1)}$ this
property is not immediately obvious, but the  analysis of this subsection indicates how this happens. 
The next-to-leading-order contribution to 
the pressure can be written as $P^{(1)}(T)=T^2
f(\tilde g^2)$ where, aside from the explicit $T^2$ which carries 
the dimension, (almost) all temperature dependence is
contained in the effective coupling $\tilde g$. This effective constant is
nothing but the running coupling at a scale $\sim T$, and it 
becomes small at large $T$.  The function $f(\tilde g^2)$, which vanishes for $ \tilde g^2=0$, can be 
expanded in powers of $\tilde g$, and we have given
above the leading contributions of this expansion. A noteworthy feature of this expansion is the
vanishing of the  term of order
$\tilde g^2$: $F_2+G_2=0$ (see Eqs.~(\ref{F2F4}) and (\ref{G2})). 
We shall clarify in the next subsection the origin of this cancellation and, more generally, verify that the
properties of the pressure at high temperature that emerge from the $1/N$ expansion are 
those one expects form perturbation theory.

\subsection{\label{perturbation}Perturbation theory at high temperature}

\begin{figure}
\includegraphics[width=8cm]{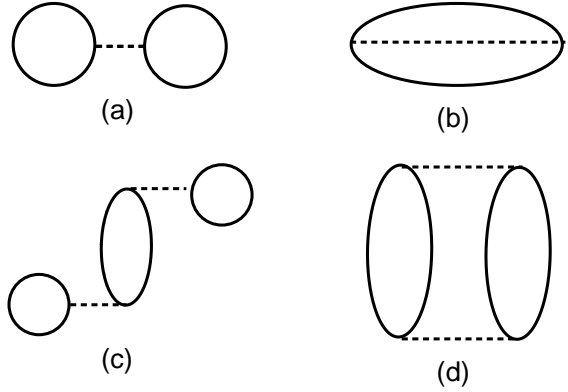}
\caption{\label{fig:pt}The diagrams of perturbation theory contributing to the pressure to order $g^2$ ((a) and (b)) and
$g^4$ ((c) and (d)). The dotted
line represents the fermionic interaction (proportional to
$g^2/N$), not the full  $\sigma$-propagator. The diagrams (a) and (c) are of order $N$,
the diagrams (b) and (d) of order 1.   }
\end{figure}

We  shall now reconsider the high temperature limit within ordinary
perturbation theory, i.e. within an expansion in powers of the coupling constant $g^2$ starting directly
from the lagrangian (\ref{GrossNeveu}). We shall then be able to recover the results of the previous subsection, and understand the origin of the properties of the pressure that we
have just discussed. Note that in contrast to what happens at zero temperature,
 where perturbation theory about
the symmetric vacuum is meaningless because of infrared divergences, at finite temperature no such
divergences occur because the fermions are effectively massive (because of their non
vanishing Matsubara frequencies).
Consider then  the diagrams contributing
to the pressure up to order $g^4$ (recall that the pressure is minus the effective
potential). These are displayed in Fig.~\ref{fig:pt}. Note that these are all included in the
calculation of the pressure at order $1/N$. That is, when expanding the result of our $1/N$ calculation in powers of
$g$, one should reproduce the perturbative expansion up to order $g^4$. Let us call
$P_a$,
$P_b$,
$P_c$,
$P_d$ their respective contributions. To handle their ultraviolet divergences, we shall follow
the standard procedure of perturbation theory. This implies using a  renormalization
scheme somewhat different from that of the rest of this paper, but this has no consequence
for the final results. In the first part of our analysis,  we shall assume that the quarks have a
small constant mass
$m$, i.e. a term $m \bar\psi \psi$ is added to the lagrangian (\ref{GrossNeveu}). This mass will serve
both to avoid spurious infrared divergences in intermediate calculations, and also to keep finite expressions which would vanish if $m$
were zero to start with (in fact, all contributions except $P_d$ vanish when $m\to0$): keeping $m$
finite is then useful to illustrate the systematics of the
cancellation of ultraviolet divergences and of the various renormalizations involved. Once this will be understood we shall
take the limit $m \to 0$.

Consider then the first contribution:
\beq
P_a= \frac{g^2 N}{2} \left[\int\frac{{\rm d}k}{2\pi}\frac{m}{E_k}(1-2n_k)  \right]^2
\eeq
where $E_k=\sqrt{k^2+m^2}$ and $n_k=n(E_k)$ is the fermion statistical factor (\ref{nomega}).
Note that $P_a$ is proportional to $\langle\bar\psi \psi\rangle^2$ (given by Eq. (\ref{G(0)}) with $M \to m$), from which one could guess
that $P_a$ vanishes in the chiral limit $m=0$. However, as it stands, $P_a$ is ill defined and needs renormalisation. To do
that, we rewrite $P_a$ as follows:
\beq\label{Pa}
P_a=\frac{g^2N}{2} 
\left\{\left[\int\frac{{\rm d}k}{2\pi}\frac{m}{E_k} \right]^2
-2\left[\int\frac{{\rm d}k}{2\pi}\frac{m}{E_k} \right]\left[\int\frac{{\rm d}k}{2\pi}\frac{m}{E_k}2n_k  
\right]+
\left[\int\frac{{\rm d}k}{2\pi}\frac{m}{E_k}2n_k  \right]^2\right\}, 
\eeq
The first term in the second line of Eq.~(\ref{Pa}) is a vacuum contribution and it can be
discarded.  The last term is finite. The potential difficulty comes from the second term which is
divergent and temperature dependent. However such a divergence can be eliminated by a mass
renormalization, i.e., by adding to the Lagrangian the counterterm $\delta m_2\, \bar\psi\psi$ with $\delta m_2$
chosen so as to cancel the correction of the zero temperature mass at order $g^2$, i.e., 
\beq
\delta m_2 ={g^2}N \int\frac{{\rm d}k}{2\pi}\frac{m}{E_k}.
\eeq
With the mass counterterm included, $P_a$ becomes
\beq
P_a\longrightarrow P'_a=P_a+{g^2}N \int\frac{{\rm d}k}{2\pi}\frac{m}{E_k} \int\frac{{\rm
d}k}{2\pi}\frac{m}{E_k}2n_k,
\eeq 
and the unwanted contribution disappears in $P'_a$, as expected.

Consider next the ``exchange'' term
\beq\label{Pb0}
P_b=\frac{g^2}{2} \Pi(\tau\!=\!0,x\!=\!0;m,T),
\eeq
One can notice that $\Pi(\tau\!=\!0,x\!=\!0;m,T)=\lim_{\tau\to
0}{\rm Tr} S(\tau,x\!=\!0;m,T)S(-\tau,x\!=\!0;m,T)$, where $S$ is the fermion propagator (see
Eq.~(\ref{Sdex})), to write Eq.~(\ref{Pb0}) as: 
\beq\label{integrale1}
\Pi(\tau\!=\!0,x\!=\!0;m,T)\!=\!-\int \frac{{\rm d}k}{2\pi}\int \frac{{\rm
d}q}{2\pi}\left\{
\!-\frac{1}{2} \!-\!\frac{m^2+k\,q}{2E_k E_q}(1\!-\!2n_{k})(1\!-\!2n_{q})\right\}.
\eeq
Note that before taking the limit $\tau\to 0$, the integrand in Eq.~(\ref{integrale1}) contains exponential
factors
${\rm e}^{\pm E_{k,q}\tau}$  which, together with statistical factors, insure the convergence of the integrals over $k$ and $q$.   In the
limit $\tau=0$, the first term in Eq.~(\ref{integrale1}) is a divergent constant independent of the
temperature and it can be ignored. The contribution proportional to $k\,q$ can be written 
as a product of two integrals which vanish by parity (at any finite $\tau$). What remains then in
Eq.~(\ref{Pb0}) is:
\beq\label{integrale3}
-\frac{g^2}{4}\left(\int\frac{{\rm d}{k }}{2\pi} \frac{m}{E_k}(1-2n_{k})\right)\, \left(\int\frac{{\rm
d}{q}}{2\pi}
\frac{m}{E_q}(1-2n_{q})\right)=-\frac{P_a}{2N}.
\eeq
Thus the exchange term is just
proportional to the direct contribution $P_a$ calculated above. It can be renormalized in the same way. We note now
that all the contributions considered so far remain, after renormalization,  proportional to $m$. It follows
that all the contributions of order
$g^2$ to the pressure cancel in the limit $m\to 0$: such contributions are proportional to
$\langle\bar\psi\psi\rangle^2$, a quantity  which vanishes when
$m=0$.

Let us turn now to the first of the order $g^4$ contributions. 
\beq\label{Pc}
P_c=-\frac{g^4N}{2}  \left[\int\frac{{\rm d}k}{2\pi}\frac{m}{E_k}(1-2n_k)  \right]^2 \Pi(0,0;m,T)=-g^2P_a
\Pi(0,0;m,T)
\eeq
where $\Pi(0,0;m,T)$ stands for $\Pi(\omega\!=\!0,q\!=\!0;m,T)$. One divergence contained in this quantity is
eliminated by coupling constant renormalization. The needed counterterm   is obtained form $P_a$ of Eq.~(\ref{Pa})
in which one  replaces 
$g^2$ by
$g^2+g^4\Pi_0(0,0;M_0)$. One gets then:
\beq
P_a''=\frac{g^2N}{2}  \left[\int\frac{{\rm d} k}{2\pi}\frac{m}{E_k}(1-2n_k)  \right]^2 (1+g^2\Pi_0(0,0;M_0))
\eeq
By combining $P_c$ and the order $g^4$ of $P_a''$, one gets
\beq\label{Pc'} 
P_c'=-\frac{g^4N}{2}  \left[\int\frac{{\rm d} k}{2\pi}\frac{m}{E_k}(1-2n_k)  \right]^2
(\Pi(0,0;m,T)-\Pi_0(0,0;M_0))
\eeq
where the quantity $\Pi(0,0;m,T)-\Pi_0(0,0;M_0)$ is finite. The contribution $P_c'$
is still divergent, but the divergences can be eliminated by the same procedure as that used above for $P_a$, namely
by the mass renormalization counterterm. The result obtained at the end of this procedure is again proportional to
$m$ and vanishes when $m\to 0$. 

Note that the combination $\Pi(0,0;m,T)-\Pi_0(0,0;M_0)$  in Eq.~(\ref{Pc'}) enters also the
calculation of the renormalized  quark-quark scattering amplitude to order
$g^4$ in perturbation theory:
\beq\label{calTg4}
{\cal T}(0,0)=g^2-g^4\left[\Pi(0,0;T)-\Pi_0(0,0;M_0)\right]=g^2-\frac{g^4}{\pi}\ln\frac{\pi T}{{\rm
e}^{\gamma_E+1} M_0},
\eeq
where the limit $m\to 0$ has already been taken (we used Eqs. (\ref{A2}), (\ref{A4}) and (\ref{Ax})). It can easily be verified that this expression is invariant under
renormalization group transformations (to order $g^4$) when $g$ runs according to the lowest order
$\beta$-function  (\ref{beta1fix}). Now, as usual in perturbation theory, one can choose the scale
$M_0$ so as to minimize higher order corrections. In the present case, the order
$g^4$ contribution to the scattering amplitude vanishes for $M_0=\pi T/{\rm e}^{\gamma_E+1}$. Note
that for this choice of $M_0$, 
\beq\label{gtilde2}
g^2(M_0=\pi T/{\rm e}^{\gamma_E+1})=\frac{\pi}{\ln(T/T_c)},
\eeq
which, is the coupling constant $\tilde g^2$ introduced above, in Eq.~(\ref{gtilde}). Note also that, for this
choice, the quark-quark scattering amplitude in Eq. (\ref{calTg4}) coincides with that calculated within the
non-perturbative scheme (see Eq. (\ref{TM0})).

Before we move on to the last order $g^4$ contribution (the only one to survive the limit $m\to 0$
which we implicitly assume from now on), let us return to the contribution
$P_b$ given in Eq.~(\ref{Pb0}), in order to make closer contact with the
$1/N$ calculation. One can rewrite $P_b$ as
\beq\label{Pb1}
P_b=\frac{g^2}{2} \int \frac{{\rm
d}q}{2\pi}\int_{-i\infty+\epsilon}^{i\infty+\epsilon}\,\frac{{\rm d}\omega}
{2\pi i}\,(1+2N(\omega))\, 
\Pi(\omega,q;T)=\Pi(\tau=0,x=0;T).
\eeq
(By performing the $\omega$-integral and suitably rearranging the statistical factors, one can recover, form
this expression, Eq.~(\ref{integrale1}) above.) The quantity $\Pi(\omega,q;T)$ is ill defined, since
it contains an infinite constant contribution, independent of the temperature. We can absorb this
constant into 
$\Pi_0(0,0;M_0)$ and discard the corresponding vacuum contribution. Next we write: 
\beq
\Pi(\omega,q;T)-\Pi_0(0,0;M_0)=\left[\Pi(\omega,q;T)-\Pi(0,0;T)\right]+
\left[\Pi(0,0;T)-\Pi_0(0,0;M_0)\right].
\eeq
The last term within brackets is $(1/\pi)\ln\left(\pi T/M_0{\rm e}^{\gamma_E+1}\right)$ (see
Eq.~(\ref{calTg4}) above). The first term within brackets is $(1/2\pi) B_T(Q/T)$
which, at large Q, behaves as $(1/2\pi)\ln(Q^2T_c^2/(T^2M_f^2))$ (see App. A).  This suggests
the following temperature independent subtraction which transforms $P_b$ into
\beq
P_b''=\frac{g^2}{4\pi}\int [{\rm d}^2Q] \left(B_T(Q/T)+\ln\left(\frac{\pi^2T^2}{M_0^2{\rm
e}^{2(\gamma_E+1)}}\right)\right)- \frac{g^2}{4\pi} \int \frac{{\rm d}^2q}{(2\pi)^2}
\ln\left(\frac{Q_E^2}{M_0^2{\rm e}^2}\right).
\eeq
At this stage, it is necessary to separate the integral into the part which contains the statistical factor, and
the other which does not. The part with the statistical factor is calculated by deforming the contour in the
usual way (see subsection~\ref{sec:fTNLO}) and yields:
\beq\label{PbG}
P_b^G= \frac{g^2}{2\pi^3}\int_0^\infty {\rm d}q\,\int_0^\infty {\rm d}\omega
 \,N(\omega)\,{\rm Im}B_T(\omega/T, q/T).
\eeq
This contribution is
independent of the vacuum subtraction.
Changing to the dimensionless variables $s=\omega/T$, $r=q/T$, fixing the scale $M_0$ at the value $M_0=\pi T/{\rm e}^{\gamma_E+1}$ and identifying $g^2$ with the running coupling at that scale (i.e., replacing $g^2$ by
$\tilde g^2$ in Eq.~(\ref{PbG})), and finally multiplying by
$(4\pi/M_f^2)$ (see Eq. (\ref{V2G})) and dividing by $(T^2/M_f^2)(\tilde g^2/2\pi)$ (see Eq. (\ref{develG})), one recovers
the expression (\ref{G2}) of $G_2$. The part without the statistical factor in $P_b''$ reads
\beq\label{PbF}
P_b^F=\frac{g^2}{4\pi}\int\frac{{\rm d}^2q}{(2\pi)^2} \left(B_T(Q/T)+\ln\left(\frac{\pi^2T^2}{M_0^2{\rm
e}^{2(\gamma_E+1)}}\right)\right)- \frac{g^2}{4\pi} \int\frac{{\rm d}^2q}{(2\pi)^2}
\ln\left(\frac{Q_E^2}{M_0^2{\rm e}^{2}}\right),
\eeq
and it can be verified that  it is finite. By choosing again 
$M_0=\pi T/{\rm e}^{\gamma_E+1}$, and substituting  
$\tilde g^2$ for $g^2$ in Eq.~(\ref{PbF}), one gets:
\beq
P_b^F=\frac{\tilde g^2}{4\pi}\int\frac{{\rm d}^2q}{(2\pi)^2}  B_T(Q/T) - \frac{\tilde g^2}{4\pi} \int\frac{{\rm
d}^2q}{(2\pi)^2}
\ln\left(\frac{Q_E^2T_c^2}{T^2M_f^2}\right).
\eeq
The same manipulations as done before for $P_b^G$ allow us to recover the expression
(\ref{F2}) of $F_2$. 

At this point, we have all the necessary ingredients to understand the origin of the cancellation $F_2+G_2=0$ observed in the
previous subsection. We have just seen that $P_b=P_b^G+P_b^F \propto F_2+G_2$; but from the calculation of $P_b$ done above
with Eq. (\ref{Pb0}) we have $P_b=0$. Hence the cancellation.
Consider finally 
\beq
P_d=\frac{g^4}{4}\int [{\rm d}^2Q] \left[\Pi(\omega,q;T)\right]^2.
\eeq
We can proceed as we did for $P_c$ and first  combine $P_d$ with the contribution coming from the
coupling constant renormalization of 
$P_b$ which transforms $P_b$ into $P_b'$:
\beq
P_b'=P_b(1+g^2\Pi_0(0,0;M_0))=P_b-\frac{g^4}{2} \int [{\rm d}^2Q] \Pi(\omega,q;T) \Pi_0(0,0;M_0).
\eeq
(Of course, according to the previous discussion, this is a vanishing correction,
 but it is convenient to carry it along as it makes the
procedure more transparent.)  By adding to $P_d$ the order $g^4$ contribution from $P_b'$, we get:
\beq\label{Pdprime}
P_d'=\frac{g^4}{4}\int [{\rm d}^2Q]
\left[\Pi(\omega,q;T)-\Pi_0(0,0;M_0)\right]^2
\eeq
where we have also added an irrelevant constant $\sim (\Pi_0(0,0;M_0))^2$. The divergence in
Eq.~(\ref{Pdprime}) is dealt with in analogy with what we just did for $P_b$. Thus we write
\beq
P_d''=\!\frac{g^4}{4} \!\int [{\rm d}^2Q]
\left[\frac{1}{2\pi}B_T(Q/T)\!+\!\frac{1}{ \pi}\ln\left(\frac{\pi T}{M_0 {\rm e}^{(\gamma_E+1)}}\right)\right]^2\!-\!
\frac{g^4}{4} \!\int\frac{{\rm d}^2q}{(2\pi)^2} 
\left[\frac{1}{2\pi}\!\ln\!\left(\frac{Q_E^2}{M_0^2{\rm e}^{2}}\right)\right]^2.
\eeq
and it can be verified that  $P_d''$ is finite. 

At this stage, one may choose the scale to be $M_0=\pi T/{\rm e}^{\gamma_E+1}$. Then $P_d''$ simplifies
 into:
\beq
P_d''=\frac{\tilde g^4}{4}\int [{\rm d}^2 Q] \left[\frac{1}{2\pi}B_T(Q/T)\right]^2- \frac{\tilde g^4}{4} 
\int\frac{{\rm d}^2q}{(2\pi)^2} 
\left[\frac{1}{2\pi}\ln\left(\frac{Q_E^2T_c^2}{M_f^2T^2}\right)\right]^2.
\eeq
We can now perform the
$\omega$-integral as we did earlier and separate the  contributions containing one or no statistical factors. We obtain
in this way the contribution of order $g^4$ to the functions $F$ and $G$. These  are:
\beq
F_4=-\frac{1}{\pi}\int_0^\infty {\rm d}s\int_0^\infty{\rm d}r 
\left(\left[B_T(is,r)\right]^2-\left[\ln\left(T_c^2/M_f^2(s^2+r^2)\right)\right]^2\right),
\eeq
\beq
G_4=-\frac{4}{\pi}\int_0^\infty  {\rm d}s\int_0^\infty {\rm d}r N(s)
{\rm Im} B_T(s,r) {\rm Re} B_T(s,r),
\eeq
which agrees with the expressions given above (Eqs. (\ref{F4ref}) and (\ref{G4})).

To summarize, the perturbative calculation performed in this last subsection has allowed us to recover the high
temperature limit obtained previously within the $1/N$ expansion. The calculation involves a running coupling
at a scale $M_0$ which can be chosen, as usual in perturbation theory, so as to minimize the high order
corrections. Here the choice $M_0=\pi T/{\rm e}^{\gamma_E+1}$ allows us to identify the running coupling with
$\tilde g$ which naturally emerges in the high temperature expansion of the order $1/N$
calculation. We have seen that the second order terms vanish, because they are proportional to the quark
condensate which vanish in the high temperature phase; this explains the cancellation of $F_2+G_2$ shown
in the previous subsection. Finally, we have also been able to reproduce explicitly the analytical
formulae giving the first few terms of the high temperature expansion of the pressure. This analysis confirms
that, as could be expected from asymptotic freedom, the behavior of the system at high temperature can be
understood in terms of perturbation theory.

\section{\label{ref:Summary} Conclusions}
The Gross-Neveu model in 1+1 dimension has proven to be a very useful playground to study 
various aspects of renormalization at  finite temperature. We have obtained, at next-to-leading order in the $1/N$
expansion,  a new expression for the effective potential, which is explicitly invariant under renormalization group transformations,
both at zero and at finite temperature. We have verified, in a non-perturbative context,
that the temperature dependent ultraviolet divergences cancel, as expected from general arguments: the same counterterms which make
finite the effective potential at zero temperature also make it finite at non zero temperature.
There is presently a lot of interest in exploring non
perturbative techniques to study the thermodynamics of gauge fields, in particular of QCD \cite{vanHees:2002js,
Blaizot:2000fc,Andersen:2000yj} and the present investigations
contribute to this general effort,  in a much simpler context than that of gauge theories.  In contrast to other methods, such as
for instance those based on $\Phi$ derivable approximations \cite{vanHees:2002js,Blaizot:2000fc}, the
$1/N$ expansion has the advantage of being at the same time non perturbative and  of providing an expansion parameter: the latter is
helpful to understand the systematics of the cancellation of ultraviolet divergences. Within this framework, there are obvious
generalizations of the model studied in this paper, for example considering continuous symmetries and higher dimensions, which would
be worth exploring. 

Part of the motivation for studying  the Gross-Neveu model specifically in 1+1 dimension is that it shares with three dimensional non
abelian gauge theories like QCD the property of asymptotic freedom. This  leads us to expect that at high temperature, the
system behaves as a weakly interacting gas of the original constituents, somewhat analogous to the quark-gluon
plasma of QCD. The detailed analysis of the high temperature limit is an important part of the present study.
We have  verified that, as expected from asymptotic freedom, 
the pressure can be expanded in terms of an effective coupling which decreases with increasing
temperature. In fact, we have shown explicitly that, in this high temperature regime the prediction of the $1/N$ expansion is identical to that
of ordinary perturbation theory. We find that the pressure
goes slowly towards that of the ideal gas as the temperature increases. Compared to QCD where a similar
phenomenon occurs, one finds that in the present calculation the pressure is higher than that of the free gas, while it is lower in QCD. In QCD, the
dominant correction to the free gas behavior is of second order
in the coupling strength, and the sign of the effect is easily understood from the corresponding expression of the entropy density
\cite{Blaizot:2000fc}. Here the contribution of the order $g^2$ cancels and the leading correction to the free gas behavior is of order $g^4$.
Even if the basic mechanisms at work are different here and in QCD, it would be interesting to explore further the
properties of the high temperature phase of the Gross-Neveu model along the lines developed in Ref.~\cite{Blaizot:2000fc}.

We find that the $1/N$ expansion provides a sensible approximation scheme at both low and high temperature: At
low temperature, this is because it produces only small corrections to the mean field physics; at high
temperature,  it reproduces essentially perturbative calculations. But, for temperatures of the order of the
mean field transition temperature the approximation breaks down: a
Landau pole appears in the calculation and fluctuations become too large to be treated as corrections. (The Landau pole is harmless
at low temperature and   disappears at high temperature.)
It would be interesting to explore further this breakdown of the $1/N$ expansion, in particular to
study to which extent it
is a consequence of the one-dimensional character of the system, by using different techniques,
such as Exact Renormalization Group Equations, with which  fermionic models have already been extensively studied   at zero
\cite{Ellwanger1994} and finite
temperature \cite{Berges1999} (for a review see \cite{Berges2002}). In this context, bosonic fluctuations can be included in a
simple way, but the $1/N$ expansion has not been explored beyond the leading term.  

\newpage
\appendix
\section{Renormalized $\sigma$ propagator}

 The bare fermion loop 
is defined in Eq.~(\ref{Pi_oneloop}) and reads:
\beq\label{Pi_oneloop1}
\Pi(Q;M,T) = 2 \int \{d^2{K}\} \frac{ M^2+{K}\cdot {K}'}{(-{K}^2+M^2)(-{K}'^2+M^2)}.
\eeq
where the notation $\int \{d^2{K}\}$
is that of Eq.~(\ref{measure_fermion}) and  ${K}'={K}+{Q}$.   In this
equation, and throughout this 
Appendix, 
$M$  is considered as a given parameter, not necessarily equal to the physical fermion mass $M_f$. Note that
$\Pi(Q)$ is dimensionless.

One may  perform the sum  over the Matsubara frequencies (see Eq.~(\ref{tildePi}) below) to verify that 
$\Pi$ can be written as a sum of a zero temperature contribution $\Pi_0$ and a finite
temperature contribution
$\tilde \Pi$ which goes to zero when $T\to 0$. This separation is well defined only when 
$M\ne 0$: indeed the zero temperature contribution, $\Pi_0(Q,M)$ is divergent when $M\to 0$. At
finite temperature, the fermionic Matsubara frequencies provide an infrared cut-off which makes
$\Pi(Q;M\to 0,T)$ well defined. 

We consider first the zero temperature contribution $\Pi_0(Q;M)$. This may be obtained directly 
from  Eq.~(\ref{Pi_oneloop1}) as an Euclidean integral, function of 
 $Q_E^2=q_0^2+q^2$ only. Once regularized  with  a  cut-off
$\Lambda$, it yields:
\beq
\label{A2}
\Pi_0(Q_E;M) = \frac{1}{2\pi}\left[  \ln\left(\frac{M^2}{\Lambda^2} \right) +
B_0\left(\frac{Q_E^2}{M^2}\right) \right],
\eeq
where
\beq
 B_0(z)\equiv  
\sqrt{\frac{4+z}{z}} \ln \left[ \frac{ \sqrt{4+z}+\sqrt{z} }
{ \sqrt{4+z}-\sqrt{z} }\right]
\eeq
 is an analytic function of $z$ with a cut on the negative real axis. Some properties of
$B_0(z)$ for $z=x$ real and positive are useful. For small $x$
\beq
\label{A4}
B_0(x)=2+\frac{x}{6}-\frac{ x^2}{60}  +{\cal O}(x^3),
\eeq
while for large $x$
\beq\label{A5}
B_0(x)= \ln x+\frac{2}{x}(1+\ln x) +{\cal O}\left( \frac{a+b\ln x}{x^2}\right).
\eeq

The expression $1+g^2_B \Pi_0(Q;M)$ may be interpreted as the bare inverse propagator for
the $\sigma$  field  in leading order of the $1/N$ expansion (see Eq.~(\ref{integrale_gaussienne})). The
renormalized inverse propagator is obtained by multiplying it
 by
$Z$ (coming from the renormalization of $\tilde \sigma$ in Eq.~(\ref{integrale_gaussienne})), and replacing 
$g^2_B$ by
$g^2/Z$, with $Z=Z^{(0)}=(g^2/\pi)\ln(\Lambda/M_f)$ (see Eq.~(\ref{ZMf}) with $\bar Z^{(0)}=0$). One then
obtains:
\beq\label{Dinverse}
 D_0^{-1}(Q_E;M)&=&Z\left(1+\frac{g^2}{Z}\Pi_0(Q_E;M)\right)= \frac{g^2}{2\pi}\left[ 
\ln\left(\frac{M^2}{M_f^2} \right) + B_0\left(\frac{Q_E^2}{M^2}\right) \right] \nonumber\\ 
& =& 
\frac{g^2}{\pi}\ln\left(\frac{M}{M_f}\right)+
\frac{g^2}{2\pi}  \sqrt{1+\frac{4M^2}{Q_E^2}} \ln \left[ 
\frac{Q_E^2}{4M^2} \left(1+\sqrt{1+\frac{4M^2}{Q_E^2}}\right)^2 \right] .
\eeq
This is an increasing function of $Q_E^2$, whose minimum, at 
$Q_E^2=0$, coincides, up to a factor $g^2$ (see Eq. (\ref{V0andD})), with the second derivative with respect to
$M$ of the leading order contribution to the renormalized effective potential (Eq.~(\ref{URindepM0})):
\beq\label{Dmoins1}
D^{-1}_0(Q_E=0;M)=\frac{g^2}{\pi}\left( 1+\ln\frac{M}{M_f}\right)= 
g^2\frac{{\rm d}^2V^{(0)}}{{\rm
d} M^2}.
\eeq
Note that $D^{-1}_0(Q_E=0;M)$ vanishes at the  value $M_*=M_f/e$. For $M>M_*$,  $D^{-1}_0(Q_E;M)$ is
always positive; for $M<M_*$, $D^{-1}_0(Q_E;M)$ vanishes at some positive value of $Q_E^2$ and is
negative for smaller values 
of $Q_E^2$. That is, when $M\leq M_*$ the $\sigma$ propagator has an unphysical
pole, whose role in the next-to-leading order calculation of the effective potential is discussed
in the main text.

For large values of $Q_E^2$, we obtain from Eq.~(\ref{A5}):  
\beq\label{logqdsD}
D^{-1}_0(Q_E;M)\simeq \frac{g^2}{2\pi} \left[\ln \left(\frac{Q_E^2}{M^2_f}\right)+
\frac{2M^2}{Q_E^2}\left(1+\ln \frac{Q_E^2}{M^2}\right)\right] +{\cal O}\left(\frac{M^3}{Q_E^3}\ln
\frac{Q_E^2}{M^2}\right).
\eeq

We  now turn to 
the finite temperature contribution, $\tilde\Pi(\omega,q;M,T)$. This is obtained from
Eq.~(\ref{Pi_oneloop1}) by performing the sum over Matsubara frequencies. The resulting expression for
the total $\Pi$ is: 
\beq\label{tildePi}
\lefteqn{ \Pi_0(\omega,q;M)+\tilde\Pi(\omega,q;M,T)=-\int_{-\infty}^{\infty} \frac{{\rm d}k}{2\pi} 
}\nonumber\\ & &
\left[
\frac{E_+\,E_-+M^2-k^2+q^2/4}{2E_+\,E_-}\left(\frac{n_+-n_-}{\omega-E_++E_-}-
\frac{n_+-n_-}
{\omega+E_+-E_-}\right)\right.\nonumber\\ & & 
+\left.\frac{E_+\,E_--M^2+k^2-q^2/4}{2E_+\,E_-}\left(\frac{n_++n_--1}{\omega-E_+-E_-}-
\frac{n_++n_--1}
{\omega+E_++E_-}\right)\right],
\eeq
where 
\beq
E_\pm=\sqrt{( k\pm q/2)^2+M^2},\qquad\qquad n_\pm\equiv n(E_\pm)=\frac{1}{{\rm e}^{\beta E_\pm}+1}.
\eeq
The finite temperature contribution proper, i.e., $\tilde\Pi(\omega,q;M,T)$, is obtained by keeping only
terms proportional to statistical factors (i.e., by  dropping the 1's in
the third line of Eq.~(\ref{tildePi})).  One can then verify that 
\beq\label{symPi}
\tilde\Pi(\omega,q)=\tilde\Pi(-\omega,q)=\tilde\Pi(\omega,-q).
\eeq
Thus we need to calculate $\tilde\Pi$ only in the region $\omega>0$, $q>0$.  Note also
that changing $k$ into $-k$ in the integrand of Eq.~(\ref{tildePi}) is equivalent to
changing
$q$ into $-q$ which leaves the integrand invariant; we can therefore limit the
$k$-integration to the range $[0,\infty[$.

The imaginary part is defined as usual as the imaginary part of
$\tilde\Pi(\omega+i\epsilon,q)$ with $\omega$ real. In the region of interest, this is non vanishing  for 
$k$ values such that (a):
$E_+-E_-=\omega$,   or  (b):
$E_++E_-=\omega$. The case (a)
is typical of finite temperature and corresponds to scattering processes 
in the heat bath. The  case (b) corresponds to processes which are analogous to those
occurring at zero temperature (decay of the sigma into quark-antiquark  pairs; see
Sect.~\ref{sigmaexcitation}).  For each value of $\omega$ and $q$, the ``phase-space'' is limited to a
single value of the momentum and the energy of the quark. This is given by ($\omega_q=\sqrt{q^2+4M^2}$):
\beq
(a) \qquad
 0<\omega<q
\qquad\qquad {\rm (scattering \, processes)}\qquad\qquad\qquad\qquad
\qquad\qquad\nonumber\\\qquad\nonumber\\
k^a=\frac{\omega}{2}\sqrt{\frac{\omega_q^2-\omega^2}{q^2-\omega^2}}\qquad E^a_\pm
=\frac{1}{2}\left(\pm\omega+q\sqrt{\frac{\omega_q^2-\omega^2}{q^2-\omega^2}}\right)\qquad
\eeq
\beq
(b) \qquad
 \omega>\omega_q
\qquad\qquad{\rm (pair \, creation)}\qquad\qquad\qquad\qquad\qquad\qquad\nonumber\\
\qquad\nonumber\\
k^b=\frac{\omega}{2}\sqrt{\frac{\omega^2-\omega_q^2}{\omega^2-q^2}}=k^a\qquad E^b_\pm
=\frac{1}{2}\left(\omega\pm q\sqrt{\frac{\omega^2-\omega_q^2}{\omega^2-q^2}}\right)\qquad
\eeq
The corresponding imaginary parts are given by:
\beq
(a) \qquad0<\omega<q\qquad\qquad{\rm
Im}\tilde\Pi=\frac{1}{2}\sqrt{\frac{\omega_q^2-\omega^2}{q^2-\omega^2}}\,\left(
n(E_+^a)-n(E_-^a)\right)
\eeq
\beq
(b)  \qquad\omega>\omega_q\qquad {\rm
Im}\tilde\Pi=\frac{1}{2}\sqrt{\frac{\omega^2-\omega_q^2}{\omega^2-q^2}}\,\left(
n(E_+^b)+n(E_-^b)\right)
\eeq
For $\omega>0$ outside the regions $[0,q]$ and $[\omega_q,\infty[$, ${\rm Im}\tilde\Pi=0$. 
The real parts are determined as  principal value  integrals. 

Note that the contribution of the scattering processes to the imaginary part is  negative, while
that of the pair creation processes is positive. In fact, in order to get a physically
meaningfull result, the latter should be added to the (negative) zero temperature contribution (see
Eq.~(\ref{ImPi0})):  the finite temperature contribution can then be interpreted as a ``suppression'' of
pair creation due to Pauli blocking.

Consider now $\Pi$ in Eq.~(\ref{tildePi}), and take the limit $M\to 0$. We get:
\beq\label{PiMzero}
\Pi(\omega,q;T)=-\!\!\int_0^{\Lambda/2}\frac{{\rm d}k}{\pi}\,\frac{4k}{\omega^2-4k^2}
\left[ \frac{1}{{\rm e}^{\beta(k+q/2)}+1}-\frac{1}{{\rm e}^{\beta(q/2-k)}+1} \right],
\eeq
where we have used  the property
\beq
\frac{1}{{\rm e}^{\beta(k-q/2)}+1}-1=-\frac{1}{{\rm e}^{\beta(q/2-k)}+1} 
\eeq
to rearrange terms. 
The ultraviolet cut-off $\Lambda/2$ (rather than $\Lambda$) is chosen so that $Z+g^2\Pi$ coincides with the
previously  defined renormalized inverse propagator in the limit
$M\to 0$. Indeed, we have
\beq
\label{Ax}
\Pi(0,0;T)=\!\!\int_0^{\Lambda/2}\frac{{\rm
d}k}{\pi}\,\frac{1}{k}
\left[ \frac{1}{{\rm e}^{\beta k}+1}-\frac{1}{{\rm e}^{-\beta k}+1}
\right] =\frac{1}{\pi}\ln\left(\frac{\pi T}{{\rm
e}^{\gamma_E}\Lambda}\right),
\eeq
where the second equality holds up to terms which vanish when $\Lambda\to \infty$. Combining this
result with the expression (\ref{ZMf}) of the renormalization constant $Z$ (with $\bar Z^{(0)}=0$), and
using Eq.~(\ref{Tc}), we get 
\beq\label{limit1pitilde}
Z+g^2\Pi(0,0;T)=\frac{g^2}{\pi}\ln\left(\frac{T}{T_c}\right).
\eeq
As expected, up to the factor $g^2$, this is the second derivative of the effective
potential at
$M=0$ (see Eqs.~(\ref{sumrule1}) and (\ref{TM0})). 

The inverse propagator at $M=0$ is then:
\beq
D^{-1}(\omega,q;T)= Z+g^2\Pi(\omega,q;T)= Z+g^2\Pi(0,0;T)+ g^2[\Pi(\omega,q;T)-\Pi(0,0;T)]
\eeq
where the combination $\Pi(\omega,q;T)-\Pi(0,0;T)\equiv (1/2\pi)B_T(Q/T)$ is finite.
Thus, in analogy with Eq.~(\ref{D0inverse}), we write
\beq\label{DTetBT}
D^{-1}(\omega,q;T)=\frac{g^2}{2\pi}\left[  \ln\left(\frac{T^2}{T_c^2}\right)+
B_T\left(\frac{\omega}{T}, \frac{q}{T}\right)\right].
\eeq
Thus defined, the dimensionless function $B_T(\omega/T,q/T)$  behaves as
$\ln(T_c^2/M_f^2(-\omega^2+q^2)/T^2)+\mathcal{O}(T^4/(q^2-\omega^2)^2)$ when
$q\gg T$ or
$\omega\gg T$. At small
$\omega,q$, $B_T$ is regular ($B_T(\omega/T,q/T)\to 0$ as $\omega,q\to 0$). 

\begin{figure}
\includegraphics[width=12cm]{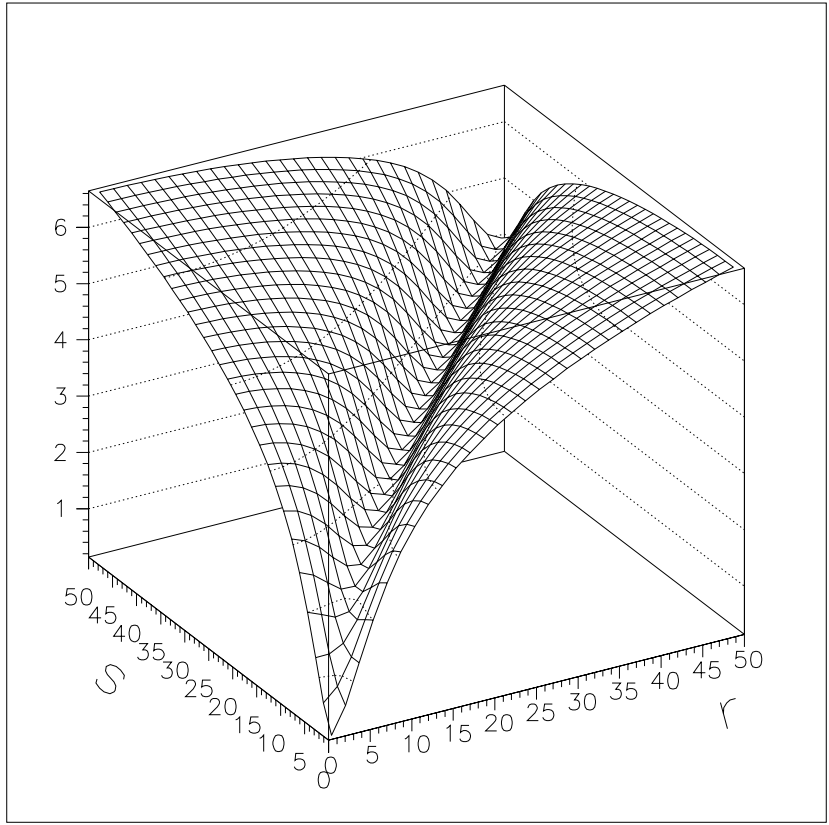}
\caption{\label{fig:BT}The function ${\rm Re}B_T(s,r)$}
\end{figure}

The imaginary part  of $D^{-1}(\omega,q;T)$ is finite and given simply by ($\omega$ real and positive):
\beq\label{imaginaryasymp}
 {\rm Im}\Pi(\omega,q;T)=\frac{1}{2}\,\left(
\frac{1}{{\rm e}^{\beta(q+\omega)/2}+1}-\frac{1}{{\rm e}^{\beta(q-\omega)/2}+1}  \right).
\eeq
At small $\omega$, and as long as $q\simle T$, ${\rm Im}\Pi(\omega,q;T)$ is a  linear  function
of $\omega$, with a slope
$-q/4T$. When $q>T$ however, the slope vanishes exponentially, as $(q/T)\exp(-q/T)$.
For large values of $q/T$, the imaginary part becomes a step function of height -1/2 located at
$\omega=q$. 

The real part is
obtained numerically from the principal value of the integral (\ref{PiMzero}), by combining with the
renormalization constant to eliminate the cut-off. It can be also expressed in  terms of the real part of
the function $B_T$:
\beq\label{functionh}
{\rm Re} B_T(s,r)={\rm PP}\int_0^\infty {\rm d}x\left[ 
\frac{x}{x^2-r^2/4}(n_+-n_-)-\frac{1}{x}(f_+-f_-)\right],
\eeq
where we have set  $x\equiv k/T$, $s\equiv \omega/T$, $r\equiv q/T$,
and
\beq
f_\pm=\frac{1}{{\rm e}^{\pm x} +1}\qquad\qquad n_\pm=\frac{1}{{\rm e}^{{s\over 2}\pm x}+1}.
\eeq
A plot of ${\rm Re}B_T(s,r)$ is given in Fig.~\ref{fig:BT}.

Let us finally examine the asymptotic behavior of $\Pi(\omega,q)$ when
$q$ (or $\omega$) is large. This is dominated by the zero temperature contribution and the
leading term is given by Eq.~(\ref{logqdsD}). However, at next-to-leading order, there is a
relevant contribution coming from 
$\tilde\Pi$. To estimate it, we go back to Eq.~(\ref{tildePi}), change
$\omega\to iq_0$ and to go to polar coordinates:
$q_0=r\cos\theta$,
$q=r\sin\theta$. Next,  one performs changes of variables such that 
 all the statistical factors in  Eq.~(\ref{tildePi}) have the same energy argument. Then it
is not hard to get:
\beq\label{asymptotic_Pi1}
\tilde\Pi(iq_0,q)\sim \int_{-\infty}^\infty \frac{{\rm d}k}{2\pi}\frac{n_k}{E_k}\left[
8k^2-8\sin^2\theta (E_k^2+k^2)\right]\frac{1}{r^2}.
\eeq
In particular,
\beq\label{asymptotic_Pi}
\int_0^{2\pi} \frac{{\rm d}\theta}{2\pi} \tilde\Pi(iq_0,q)\sim  -\frac{4
M^2}{r^2}\int_{-\infty}^\infty\frac{{\rm d}k}{2\pi}\frac{n_k}{E_k}.
\eeq

\section{The functions $F_0(x)$, $F(x,T)$ and $G(x,T)$}

The function $F_0(x)$ introduced in  Eq.~(\ref{renoV1}) of the main 
text is defined by:
\beq\label{F0}
F_0(x) \equiv \frac{1}{2} \int_{e^2}^\infty du \left\{ \ln
\left[\frac{\ln x +B_0(u/x)}{B_0(u)}\right]
  - 2 \frac{x-1}{u}
  - 2 \frac{x-1-x\ln x}{u\ln u} \right\} \nonumber
  \\
+ \frac{1}{2} \int_0^{e^2} du
  \ln \left[\frac{\ln x +B_0(u/x)}{B_0(u)}\right] \; .
  \eeq
The last two terms in the first line of this equation correspond to 
the
$\Lambda$-dependent terms in Eq.~(\ref{V1TzeroR}), which we have 
rewritten as:
\beq\label{infint}
\frac{1}{4\pi}\ln\frac{\Lambda^2}{M_f^2} & = &\int_{eM_f}^\Lambda
\frac{1}{(2\pi)^2}\frac{d^2Q}{Q^2} + \frac{1}{2\pi},
\nonumber \\
\frac{1}{4\pi} \ln\ln\frac{\Lambda}{M_f}& = &\int_{eM_f}^\Lambda
\frac{1}{(2\pi)^2}\frac{d^2Q}{Q^2\ln(Q^2/M_f^2)} .
\eeq
Thus, by combining the integrand as we did in Eq.~(\ref{F0}), we have 
obtained 
convergent integrals
and the limit $\Lambda\to\infty$ could be taken. Note that, as
stated in subsection~\ref{sec:T0NLO}, the function  $F_0(x)$ is 
defined only for $x>x_{M_*}=1/{\rm e}^2$.

To find the first derivative $F_0'$, it is convenient to separate
the contribution of the last two terms of the first line of 
Eq.~(\ref{F0}), which we call $F'_{\rm
CT}$. Since $F'_{\rm
CT}$ is divergent, we re-introduce a cut-off
$\Lambda$, and get:
\beq\label{CT}
F'_{\rm CT} =  \int_{e^2}^{\Lambda^2} \frac{{\rm d}u}{u} \left( -1 +
\frac{\ln x}{\ln u} \right) .
\eeq
In order to calculate the remaining contribution to $F_0'(x)$, we use 
the change
of variables
\beq
v=\frac{u}{4x}\left(1+\sqrt{\frac{4x}{u}+1}\right)^2
\eeq
to get
\beq\label{der1}
F'_0(x)-F'_{\rm CT}(x)=
\int_1^{\alpha}\frac{{\rm d}v}{v}\frac{\ln v}{\ln 
x+\frac{v+1}{v-1}\ln v},
\eeq
where $\alpha \equiv v(u=\Lambda^2) \sim \Lambda^2/x$ for large 
$\Lambda$.
In order to recombine both contributions (\ref{CT}) and (\ref{der1}), 
and let $\Lambda \to \infty$,
one observes that
\beq\label{der2}
\int_{e^2}^{\Lambda^2} \frac{{\rm d}u}{u} \left( -1 +
\frac{\ln x}{\ln u} \right) =
\int_{e^2}^{\Lambda^2/x} \frac{{\rm d}u}{u} \left( -1 +
\frac{\ln x}{\ln u} \right) - \ln x + {\cal O}(\ln x/\ln \Lambda^2).
\eeq
Thus, adding (\ref{der1}) and (\ref{der2}) and now letting $\Lambda 
\to
\infty$,
one gets after simple algebra:
\beq\label{F'0}
F'_0(x) &=& \ln x \int_{e^2}^\infty \frac{{\rm d}u}{u}
\left( \frac{1}{\ln u} - \frac{(u-1)^2}{(u^2-1)\ln x + (u+1)^2\ln u}
\right) \nonumber \\
&+& \int_1^{e^2}\frac{{\rm d}u}{u}\frac{\ln u}
{\ln x+\frac{u+1}{u-1}\ln u} -\ln x -2\ln (e^2+1) + 4.
\eeq
The second derivative $F''_0$  follows easily from the previous 
expression.
One has:
\beq
x F''_0(x) +1 = \int_{e^2}^\infty \frac{{\rm d}u}{u} \left(
\frac{1}{\ln u} -
\frac{\ln u}{(\ln x + \frac{u+1}{u-1}\ln u)^2} \right)
-\int_1^{e^2}\frac{{\rm d}u}{u}\frac{\ln u}
{(\ln x + \frac{u+1}{u-1}\ln u)^2}
\eeq
Finally, the third derivative $F_0'''$,  needed for instance to 
calculate the
$\beta$-function at next-to-leading order, is obtained by taking the 
derivative of the expression
above:
\beq
x^2F'''_0(x) +xF''_0(x) = 2 \int_1^\infty
\frac{{\rm d}u}{u}\frac{\ln u}
{(\ln x + \frac{u+1}{u-1}\ln u)^3}
\eeq
Note that $F'_0$, $xF''_0$ and $x^2F'''_0 +xF''_0$ are actually 
functions
of the variable $\ln x$.
In the main text we use the limit of these expressions as
  $\ln x \to \infty$. This limit is easily obtained from the equations
above. On finds:
\beq
F'_0(x) \sim (\ln x) (\ln \ln x ) \hskip 1 cm xF''_0(x) \sim \ln \ln x
\hskip 0.7 cm {\rm and } \hskip 0.7 cm
xF''_0 + x^2 F'''_0(x) \sim 1/ \ln x.
\eeq

\begin{figure}
\includegraphics[angle=90,width=13cm]{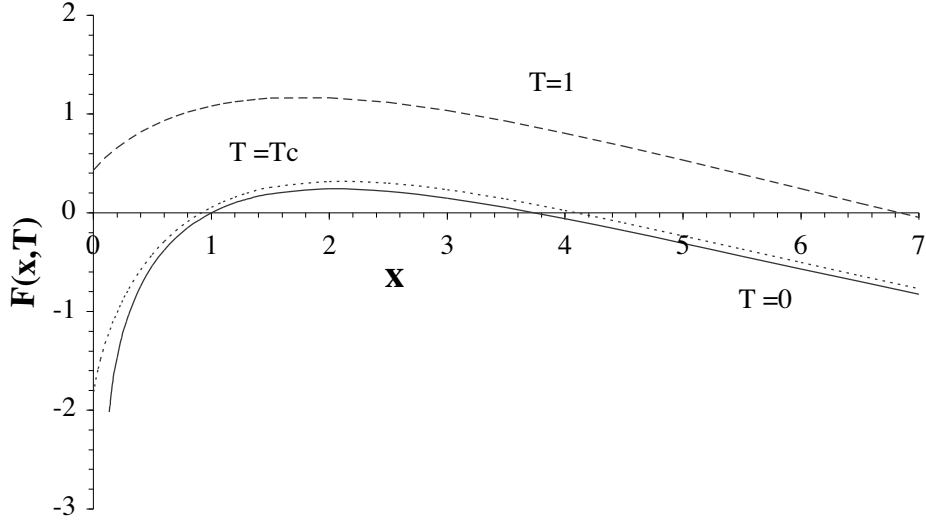}
\caption{\label{fig:funcion_F}The function $F(x,T)$ for various 
temperatures. Full line:
$T=0$, i.e., $F_0(x)$; dotted line: $T=0.565M_f=T_c$; dashed line: 
$T=M_f$}
\end{figure}

We turn now to the function $F(x,T)$ which
is obtained when we regroup what remains of $V^{(1)}_{b,1}$ after 
elimination
of the temperature dependent
divergence in Eq.~(\ref{tildeV1})  with the function
$F_0$ introduced above, Eq.~(\ref{F0}). We define:
\beq\label{extra2}
\lefteqn {
\frac{M^2_f}{4\pi} F(x_M,T)
\equiv \frac{M^2_f}{4\pi} F_0(x_M)
+ \frac{1}{2} \int^{eM_f}_{0} \frac{r dr}{2\pi} \int^{2\pi}_0
\frac{d\theta}{2\pi}\ln \left[
\frac{D^{-1}(r,\theta;M)}{D^{-1}_0(r^2;M_f)} \right]
}
\qquad \qquad \qquad \qquad \qquad \qquad \qquad \qquad \qquad
\qquad \qquad \qquad \qquad \qquad \qquad \qquad \qquad \qquad \qquad
\nonumber \\
+\frac{1}{2} \int^{\infty}_{eM_f}
\frac{r dr}{2\pi} \left\{ \int^{2\pi}_0 \frac{d\theta}{2\pi}\ln \left[
\frac{D^{-1}(r,\theta;M)}{D^{-1}_0(r^2;M_f)} \right]
+ \frac{8M^2}{r^2 \ln (r^2/M^2_f)}
\left( \int_0^{\infty} du \frac{n_u}{\sqrt{u^2+x_M}} \right) 
\right\}.\nonumber\\
\eeq
In this equation we have set
$\omega =i q_0$ and changed to polar coordinates: $q_0=r\cos\theta$,
$q=r\sin\theta$.
Also, we use abusively the notation 
$D^{-1}(r,\theta)$ for $D^{-1}(\omega=ir\cos\theta,q=r\sin\theta)$, 
and similarly 
for $D_0^{-1}$. To get 
the last term in
the expression above we used the second equation in (\ref{infint}) 
and the  fact that $\ln \ln
\Lambda/M_f = \ln \ln \Lambda/M+ {\cal O}(1 /\ln (\Lambda/M_f))$.
Note finally that, since $F$ contributes to a term of order $1/N$, we 
have set
$x_M=M^2/M^2_f$, ignoring the factor of order $1/N$ in 
Eq.~(\ref{xm2}).
One has:
\begin{eqnarray}\label{FT}
F(x,T) & = &  \int^{\infty}_{e} 2\,v\,dv \;\; \left\{
\int^{\pi/2}_0 \frac{d\theta}{\pi} \ln ( A(v,\theta;x,T))
  \right.
\nonumber
\\
& - & \left. \frac{x -1}{v^2} - \frac{1}{v^2\ln v^2}
\left( x-1-x\ln x \;
-4x \; \int_0^{\infty} du \frac{n_u}{\sqrt{u^2+x}}
\right) \right\}
\nonumber
\\
&  +&  \int^e_0 2\,v\,dv
\int^{\pi/2}_0 \frac{d\theta}{\pi} \ln (A(v,\theta;x,T) )
\eeq
where the function $A(v,\theta;x,T) $ is
\beq
A=\frac{\ln x +B_0(v^2/x) + 2\pi \tilde\Pi(v,\theta;x,T)}
{B_0(v^2)}.
\end{eqnarray}
In (\ref{FT}) we have used the symmetry properties (\ref{symPi}) 
to limit the integration over $\theta$ to the interval $[0,\pi/2]$. 
Note  finally that $F(x,T=0)=F_0(x)$, with $F_0$ given by 
Eq.~(\ref{F0}).
The function $F(x,T)$,  evaluated numerically, is shown in 
Fig.~\ref{fig:funcion_F}, for various values of $T$. Note that, when 
$T<T_c$, $F(x_M,T)$ 
is defined only for $x_M>x_{M_*}(T)$ (see  
subsection~\ref{sec:fTNLO}).

To study the high temperature limit of the 
the function $F(x_M,T)$ (needed in subsection \ref{sec:thermohighT}) it is useful to write
$F(x_M=0,T)$ in a different form. Adding and subtracting the expression
\beq
\int_0^\infty 2v dv \int_0^{\pi/2} \frac{d \theta}{\pi} 
\ln \left( \frac{B_0(v^2)}{\ln (\sqrt{1+v^4})} \right)
\eeq
we can rewrite $F(0,T)$ as
\begin{eqnarray}
\label{FTx0}
&&F(0,T)= \frac{T^2}{M_f^2}\frac{2}{\pi} \int_0^\infty ds \int_0^\infty dr
\ln \left( \frac{\ln(T^2/T_c^2) + B_T(is,r)}
{\ln (\sqrt{1+(s^2+r^2)T^4/M_f^4})} \right)
\nonumber
\\
&+& \int_e^\infty 2 v dv \left\lbrace \int_0^{\pi/2} \frac{d\theta}{\pi}
\ln \left( \frac{\sqrt{1+v^4}}{B_0(v^2)} \right)
+ \frac{1}{v^2} + \frac{1}{v^2\ln v^2}
\right\rbrace
+  \int_0^e 2 v dv  \int_0^{\pi/2} \frac{d\theta}{\pi}
\ln \left( \frac{\sqrt{1+v^4}}{B_0(v^2)} \right) \nonumber \\
\end{eqnarray}
In the first line we introduced the function $B_T$ 
(defined in App. A) and made the change of variables 
$s=(M_f/T)v\cos \theta$, $r=(M_f/T)v\sin \theta$. It is a finite expression 
(see the comment after Eq. (\ref{DTetBT})). The second
line is also finite (see Eq. (\ref{A5})) and it is 
independent of the temperature.

\begin{figure}
\includegraphics[angle=90,width=12cm]{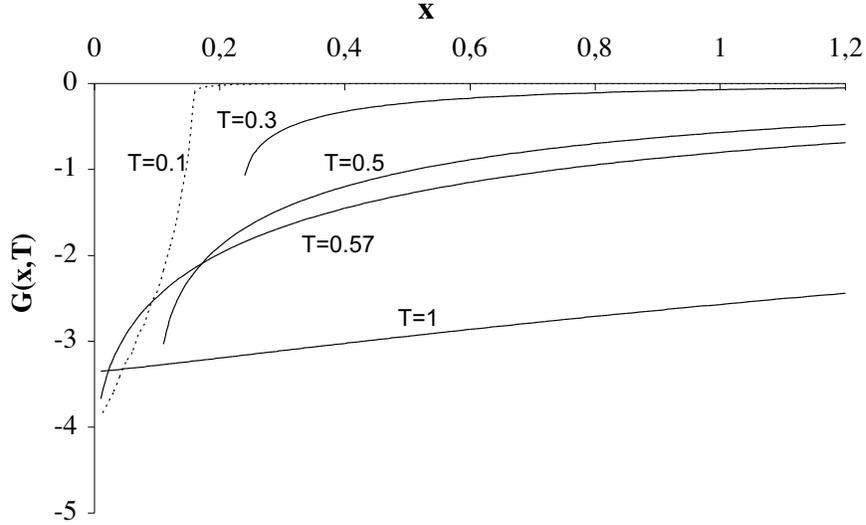}
\caption{\label{fig:funcion_G}The function $G(x,T)$ versus $x$ for 
various temperatures $T$ 
(in units of the fermion mass $M_f$). The dotted line corresponds to 
$T/M_f=0.1$
 and indicates the typical behavior of the function for small values 
of $x$ when $T<T_c$.
  For
 the other temperatures  the function is plotted only for  
$x>x_{M_*}(T)$. }
\end{figure}

Finally the function  $G$ is
\beq\label{GT}
G(x,T) = -\frac{4}{\pi} \int_0^{\infty} dr \int_0^{\infty} ds
N(s) \,\delta(s,r;x,T).
\eeq
where $\delta$ is given by Eq.~(\ref{fundelta}). 
We have set here $s=\omega/M_f$, $r=q/M_f$ and 
$N(s)=1/(e^{b s}-1)$.
The function $G(x,T)$, evaluated numerically, is presented in
Fig.~\ref{fig:funcion_G} for various temperatures.

One can understand the behavior of $G$ at small $x_M$ from the 
behavior of the phase shift. 
At moderate temperatures it is enough to consider the zero temperature
phase-shift in evaluating the integral in Eq.~(\ref{GT}). For 
$M=M_f$, using the expression of the
zero temperature propagator
$D_0^{-1}(\omega,q;M_f)$ given in App. A,  we find 
($\omega_q=\sqrt{q^2+4M^2_f}$):
\beq
\tan(\delta)=\frac{\pi}{2}\frac{\Theta(\omega-\omega_q)}{{\rm
Arctanh}\sqrt{(\omega^2-\omega_q^2)/(\omega^2-q^2)}}.
\eeq
Consider this expression for fixed $q$, as a function of $\omega$. 
For $\omega\simge \omega_q$ this behaves as 
$\delta(\omega,q)\sim 
{\pi}/{2}-({2}/{\pi})\sqrt{(\omega^2-\omega_q^2)/(\omega^2-q^2)}$
while at large $\omega$, $\delta(\omega,q)\sim \pi/\ln 
(\omega^2-q^2)$. For $M\ne M_f$,
$D_0^{-1}(\omega,q;M)=D_0^{-1}(\omega,q;M_f)+(g^2/\pi)\ln(M/M_f)$. A 
plot of 
${\rm Re }D_0^{-1}(\omega,q;M)$ is 
given in Fig.~\ref{fig:do}. For $M<M_f$, the real part of
$D_0^{-1}(\omega,q;M)$ vanishes at points $\omega_a$ and $\omega_b$ 
and is minimum at the point $\omega_q$ where
it vanishes for $M=M_f$ ($\omega_a\le \omega_q\le \omega_b$). For
$\omega<\omega_a$,
$\delta$ is small and positive; it jumps to a value close to $\pi$ 
when $\omega$ crosses the value $\omega_a$,
and remains approximately constant until $\omega\simeq \omega_q$, at 
which point it decreases rapidly to reach
the value $-\pi/2$ at $\omega_b$;  it then goes slowly to zero as 
$\omega\to \infty$. When $M\to M_*(T)$ from
above, $\omega_a\to 0$, and the region $\omega<\omega_q$, where 
$\delta\simeq \pi$, gives a large contribution to
$G$, amplified by the large value of the statistical factor at small 
$\omega$, 
where $N(\omega)\simeq T/\omega$. This is the origin of the large 
negative contribution at small $x_M$ that
is visible in Fig.~\ref{fig:funcion_G}, particularly in the curve 
labelled $T=0.1 M_f$.

\section{The fermion self-energy}

Here we calculate the fermion self-energy defined in Eq.~(\ref{sigmadef}). We have:
\beq
\Sigma(K)=-\frac{g^2}{N}\int [{\rm d}^2 Q]\,
D(Q;M)\,\frac{(\slashchar{K}-\slashchar{Q})+M}{-(K-Q)^2+M^2}\equiv a \slashchar{K} +
b,
\eeq
where $M$ is the renormalized mass, and  $g$ the
renormalized coupling.
The functions $a(K)$ and $b(K)$ can be obtained easily by projection:
\beq
a(K)=\frac{1}{2K^2}{\rm Tr}\slashchar{K} \Sigma(K)=-\frac{g^2}{NK^2}\int [{\rm d}^2 Q]\,
D(Q;M) \,\frac{K^2-K\cdot Q}{-(K-Q)^2+M^2},
\eeq
\beq
b(K)= \frac{1}{2} {\rm Tr} \Sigma(K)=-\frac{g^2}{N}\int [{\rm d}^2 Q]\, D(Q;M)\,
\frac{M}{-(K-Q)^2+M^2}.
\eeq

One can separate  the self-energy $\Sigma$ into a zero temperature 
contribution and a finite temperature one, as we did for the fermion loop in App. A. Accordingly we
write:
\beq
\Sigma(K)=\Sigma_0(K)+\tilde\Sigma(K)
\eeq
and similarly for the functions 
$a(K)$ and
$b(k)$. 
It will be verified shortly that $a(K)$ is finite, while $b(K)$ is divergent. The 
divergence is a mass correction which is eliminated by the mass counterterm 
$ Z'^{(1)}$. Introducing a cut-off
$\Lambda$, one can write:
\beq
b(K)=-\frac{M}{2N}\ln\ln\left(\frac{\Lambda^2}{M_f^2}\right)+\frac{M}{N} b'(K),
\eeq
where, as we shall verify,  $b'(K)$ is  finite. 
The full fermion two-point function takes then the form:
\beq
S^{-1}(K)&=&-\slashchar{K}(1-a)+M\left(1+\frac{ Z'^{(1)}}{2N}\right)+b(K)\nonumber\\
 &=&-\slashchar{K}\left(1-a(K)\right)+M\left(1+ 
 \frac{\bar Z'^{(1)}}{2N}+b'(K) \right)
\eeq
where we have used $M_B=\sqrt{Z'} M$, $\sqrt{Z'}=1+Z'^{(1)}/2N$ and 
$Z'^{(1)}=\ln\ln(\Lambda/M_f)+\bar Z'^{(1)}$. 

We now proceed to the explicit
calculation of the zero temperature contribution to $\Sigma$. The corresponding functions
$a_0$ and
$b_0$ are functions of $K^2$ only. By going over to Euclidean momenta, and performing the
angular integrals, one obtains:
\beq
b_0'(K_E^2)=-\frac{g^2 }{N}\int_0^{\Lambda^2} \frac{{\rm d}Q_E^2}{4\pi}
\frac{D_0(Q_E^2;M)}{\sqrt{(K_E^2+M^2+Q_E^2)^2-4K_E^2Q_E^2}}+
\frac{1}{2N}\ln\ln\frac{\Lambda}{M_f},
\eeq
and 
\beq
a_0(K_E^2)=-\frac{g^2 }{N} \int \frac{{\rm d}Q_E^2}{8\pi K_E^2}\, D_0(Q_E^2;M)
\left(1+\frac{K_E^2-Q_E^2-M^2}{\sqrt{(K_E^2+M^2+Q_E^2)^2-4K_E^2Q_E^2}}\right).
\eeq
One can verify on these explicit expressions that both $a_0(K^2)$ and $b_0'(K^2)$ are finite.

 In fact we shall need these functions only for $M=M_f$ and $K_E^2=-M_f^2$, and we shall set
$a_0\equiv a_0(K_E^2=-M_f^2)$, $b_0'\equiv
b_0'(K_E^2=-M_f^2)$. 
 We have:
\beq
a_0=\frac{g^2 M_f^2}{N}\int_0^\infty\frac{{\rm
d}Q^2}{2\pi}\frac{D_0(Q_E^2;M_f)}{(Q^2+2M_f^2)\sqrt{Q^4+4M_f^2 Q^2}+Q^4+4Q^2M_f^2}
\eeq
\beq
b_0'=\frac{1}{2N}\int_{e^2M_f^2}^{\Lambda^2}\frac{{\rm
d}Q^2}{Q^2}\frac{1}{\ln(Q^2/M_f^2)}-\frac{g^2}{N}\int_0^{\Lambda^2}
\frac{{\rm d}Q^2}{4\pi}\frac{D_0(Q_E^2;M_f)}{\sqrt{Q^4+4M_f^2 Q^2}}.
\eeq
By using the explicit expression (\ref{Dinverse}) of the propagator together with the change
of variables:
\beq
u=\sqrt{\frac{Q^2}{4M_f^2}}+\sqrt{\frac{Q^2}{4M_f^2}+1},
\eeq
one can simplifies these expressions, and get:
\beq
a_0=\int_1^\infty\frac{{\rm d}u}{u^3}\,\frac{u^2-1}{u^2+1}\,\frac{1}{\ln
u^2}.
\eeq
\beq
b_0'=-\int_1^e \frac{{\rm
d}u}{u}\,\frac{u^2-1}{u^2+1}\,\frac{1}{\ln u^2}+\int_e^\infty \frac{{\rm
d}u}{u}\,\frac{2}{u^2+1}\,\frac{1}{\ln u^2}.
\eeq

We are now in a position to calculate the change in the fermion mass coming from  $\Sigma$.
We have, in the vicinity of the mass shell, to within terms of order $1/N^2$:
\beq
0=-M_f(1-a)+M_{min}\left(1+\frac{\bar Z'^{(1)}}{2N}
+b'\right),
\eeq
where we have replaced $M$ by the value $M_{min}$ it takes at the minimum of the effective
potential. Writing $M_f=M_{min}+M_\Sigma$, and ignoring terms of order $1/N^2$, we get:
\beq
 \frac{M_\Sigma}{M_f}=a+b'+\frac{\bar Z'^{(1)}}{2N}= \frac{\bar Z'^{(1)}}{2N}+\frac{\varphi}{N},
\eeq
where 
\beq\label{varphiidef}
\varphi\equiv\left[\int_e^\infty\frac{{\rm
d}u}{u^3}\,\frac{3u^2-1}{u^2+1}\,\frac{1}{\ln u^2}-\int_1^e \frac{{\rm
d}u}{u^3}\,\frac{(u^2-1)^2}{u^2+1}\,\frac{1}{\ln u^2}\right]
\eeq
Numerical integration gives $\varphi=-0.126229$.

Finally, we use in the main text the expression  of the propagator in coordinate space. This is easily
obtained from the following ``mixed'' representation
\beq\label{Sdex}
S(\tau>0,p)=\Lambda_+\gamma_0{\rm e}^{-E_p\tau}(1-n_p)+\Lambda_-\gamma_0{\rm e}^{E_p\tau}n_p\nonumber\\
S(\tau<0,p)=-\Lambda_+\gamma_0{\rm e}^{-E_p\tau}n_p-\Lambda_-\gamma_0{\rm e}^{E_p\tau}(1-n_p)
\eeq
where $\Lambda_\pm=(E_p\pm\gamma_0(\gamma^1p+m))/2E_p$ and $E_p=\sqrt{p^2+m^2}$.

\end{document}